\documentclass[twocolumn,aps,prc,showpacs,superscriptaddress]{revtex4}
\usepackage{epsfig}
\newcommand {\bperi}	{$5<b<11$~fm}
\newcommand {\bcent}	{$0<b<5$~fm}
\newcommand {\ampt}	{{\sc ampt}}
\newcommand {\hijing}	{{\sc hijing}}
\newcommand {\Npart}	{N_{\rm part}}
\newcommand {\pt}	{p_T}
\newcommand {\bw}	{\begin{widetext}}
\newcommand {\ew}	{\end{widetext}}
\newcommand {\be}	{\begin{equation}}
\newcommand {\ee}	{\end{equation}}
\newcommand {\bea}	{\begin{eqnarray}}
\newcommand {\eea}	{\end{eqnarray}}
\newcommand {\mn}[1]	{\langle#1\rangle}
\newcommand {\mnmn}[1]	{\langle\langle#1\rangle\rangle}
\newcommand {\etaa}	{\eta_{\alpha}}
\newcommand {\etab}	{\eta_{\beta}}
\newcommand {\etaref}	{\eta_{\rm ref}}
\newcommand {\deta}	{\Delta\eta}
\newcommand {\detaa}	{\Delta\eta_1}
\newcommand {\detab}	{\Delta\eta_2}
\newcommand {\detamax}	{\Delta\eta_{\rm max}}
\newcommand {\mneta}	{\mn{\eta}}
\newcommand {\mnetaa}	{\mn{\eta_1}}
\newcommand {\mnetab}	{\mn{\eta_2}}
\newcommand {\psiPP}	{\psi_{\rm PP}}
\newcommand {\psiRP}	{\psi_{\rm RP}}
\newcommand {\fl}	{\sigma'}
\newcommand {\nf}	{\delta'}
\newcommand {\chisq}	{\chi^2/{\rm ndf}}
\newcommand {\N}[1]	{N_{#1}}
\newcommand {\Q}[3]	{Q_{#1}^{#2}(#3)}
\newcommand {\Vn}[2]	{V_{#1}\{{\rm #2}\}}
\newcommand {\vn}[2]	{v_{#1}\{{\rm #2}\}}
\newcommand {\Vt}[2]	{\tilde{V}_{#1}\{{\rm #2}\}}
\newcommand {\vt}[2]	{\tilde{v}_{#1}\{{\rm #2}\}}
\newcommand {\tV}	{\tilde{V}}
\newcommand {\tv}	{\tilde{v}}
\newcommand {\vnsq}[2]	{v^2_{#1}\{{\rm #2}\}}
\newcommand {\vtsq}[2]	{\tilde{v}^2_{#1}\{{\rm #2}\}}

\newcommand {\note}[1]	{}

\begin{document}

\title{Decomposition of flow and nonflow in relativistic heavy-ion collisions}
\author{Lingshan Xu}
\affiliation{Department of Physics, Purdue University, West Lafayette, Indiana 47907, USA}
\author{Li Yi}
\affiliation{Department of Physics, Purdue University, West Lafayette, Indiana 47907, USA}
\author{Daniel Kikola}
\affiliation{Department of Physics, Purdue University, West Lafayette, Indiana 47907, USA}
\author{Joshua Konzer}
\affiliation{Department of Physics, Purdue University, West Lafayette, Indiana 47907, USA}
\author{Fuqiang Wang}
\affiliation{Department of Physics, Purdue University, West Lafayette, Indiana 47907, USA}
\author{Wei Xie}
\affiliation{Department of Physics, Purdue University, West Lafayette, Indiana 47907, USA}

\begin{abstract}
We propose a method to separate $\deta$-dependent and $\deta$-independent azimuthal correlations using two- and four-particle cumulants between pseudo-rapidity ($\eta$) bins in symmetric heavy-ion collisions. The $\deta$-independent correlation may be dominated by harmonic flows, a global correlation resulting from the common collision geometry. The $\deta$-dependent correlation can be identified as nonflow, particle correlations unrelated to the common geometry. Our method exploits the $\eta$ symmetry of the average harmonic flows and is ``data-driven.'' 
We use the \ampt\ and \hijing\ event generators to illustrate our method. We discuss the decomposed $\deta$-independent and $\deta$-dependent correlations regarding flow and nonflow in the models.
\end{abstract}


\maketitle

\section{Introduction}

A long standing issue in relativistic heavy-ion collisions is the entangled flow and nonflow in particle angular correlation measurements~\cite{flowreview}\note{[better ref]}. Flow is a global property of particle distributions being anisotropic with respect to a common symmetric direction in the transverse plane of a collision event (e.g. the reaction plane or the participant plane). Flow is thought to be composed of at least two contributions, the hydrodynamic type global flow dominant at low transverse momentum ($\pt$)~\cite{Ollitrault_ecc} and the pathlength dependent jet-quenching effect dominant at high $\pt$~\cite{jet_quenching}. Nonflow, on the other hand, is caused by intrinsic particle correlations, such as from jet-correlations and resonance decays~\cite{Borghini_nonflow}, and is not related to the common symmetric direction. In general nonflow is restricted to few-body correlations while flow is a global correlation of the entire event.

The common symmetric direction of the overlap geometry in the configuration space is unknown a priori, and is estimated experimentally using particle correlations in momentum space~\cite{SWang,v2method}. The resultant anisotropy measurement is therefore contaminated by nonflow~\cite{Borghini_nonflow,Miller_fluc,STAR_nonflow,Ollitrault_fluc_nonflow}. 
There are several methods to analyze azimuthal anisotropy. Broadly speaking they can be categorized into two classes: the two-particle cumulant method and the multi-particle cumulant method~\cite{v2method}. For example, the event-plane method is effectively a two-particle correlation method because a test particle is correlated with the event-plane reconstructed from the rest particles~\cite{v2method,trainor}. 
The two-particle cumulant method is contaminated by nonflow, and pseudo-rapidity ($\eta$) gaps, $\deta$, are often applied between the particle pair to reduce nonflow contributions. 
However, how much nonflow still remains is hard to quantify. In addition, back-to-back inter-jet correlations can still contribute considerably at large $\deta$.

On the other hand, nonflow contributions in the four-particle cumulant method are suppressed by a high order in event multiplicity and are therefore negligible in heavy-ion collisions~\cite{Borghini_multipart,Borghini_multipart2}. 
However, the four-particle cumulant method is affected by flow fluctuations differently from the two-particle cumulant; the fluctuation effect is negative in four-particle cumulant while positive in two-particle cumulant~\cite{v2method,Miller_fluc,Ollitrault_fluc_nonflow}. The difference between two- and four-particle cumulants is therefore a net effect of flow fluctuations and nonflow. So far one cannot separate flow and nonflow effects from the combination of two- and four-particle cumulants without relying on model assumptions. In fact, a great deal of effort has been invested in studying flow fluctuations and nonflow based on various model assumptions~\cite{GangWang,Alver_nonGaus,PHOBOS_nonflow,Sorensen1,Sorensen2,Bessel,Wang_nonflow,ALICE,LiYi,Kikola}.

In this paper we propose a data-driven method to separate $\deta$-dependent and independent correlations by analyzing two- and four-particle cumulants of particles from different pseudo-rapidity bins. The $\deta$-dependent correlation can be identified as nonflow, and the $\deta$-independent correlation may be dominated by flow. We lay out our method in Sec.~\ref{method}. We illustrate our method using the \ampt\ event generator in Sec.~\ref{AMPT}. We argue that our method is successful in separating, to a large extent, the flow and nonflow in \ampt\ based on expected features for flow. We employ the \hijing\ model to render additional support to our method.




\section{Method\label{method}}

When analyzing flow, one often applies $\eta$-gaps in two-particle cumulant measurements. This reduces, but does not completely eliminate, nonflow contributions~\cite{FTPC_v2,ALICE,CMS,ATLAS}. When analyzing jet-like correlations, one often subtracts large-$\deta$ azimuthal correlations from small-$\deta$ ones to obtain the near-side intra-jet correlation~\cite{jetcorr_RP,CMS}. This assumes that flow is $\eta$-independent. In this paper we extend these ideas systematically, to all possible $\eta$-gaps, to hopefully extract the $\deta$-dependent nonflow in entirety without any assumptions on the $\eta$-dependence of flow.

We propose to analyze the two- and four-particle cumulants between two $\eta$ bins, $\etaa$ and $\etab$. The azimuthal cumulants can be obtained from the moments
\bea
\mn{2}_n(\etaa,\etab)&=&\mn{e^{in(\phi_{1,\etaa}-\phi_{2,\etab})}}\,,\\
\mn{4}_n(\etaa,\etab)&=&\mn{e^{in(\phi_{1,\etaa}+\phi_{2,\etaa}-\phi_{3,\etab}-\phi_{4,\etab})}}\,.
\eea
 For four-particle cumulant, two particles are from $\etaa$ and the other two from $\etab$. 

\note{
Technically the azimuthal moments can be calculated efficiently using the $Q$-vector
\be\Q{n}{}{\eta}=\sum_{i=1}^{\N{\eta}}e^{in\phi_i}\,,\ee
where $\N{\eta}$ is the number of particles in the $\eta$ bin.
For particles from the same $\eta$ bin, the $n^{\rm th}$-harmonic moments are given by~\cite{Bilandzic,Wang_formula}
\bea
\mn{2}_n(\eta)&=&\frac{|\Q{n}{}{\eta}|^2-\N{\eta}}{\N{\eta}(\N{\eta}-1)}\,,\\
\mn{4}_n(\eta)&=&\frac{|\Q{n}{}{\eta}|^4+|\Q{2n}{}{\eta}|^2-2{\rm Re}(\Q{2n}{}{\eta}\Q{n}{*2}{\eta})}{\N{\eta}(\N{\eta}-1)(\N{\eta}-2)(\N{\eta}-3)}\nonumber\\
&&-\frac{4|\Q{n}{}{\eta}|^2}{\N{\eta}(\N{\eta}-1)(\N{\eta}-3)}+\frac{2}{(\N{\eta}-1)(\N{\eta}-2)}\,,\\
\eea
and for separate $\etaa$ and $\etab$ bins, they are given by~\cite{Bilandzic,Wang_formula}
\bea
\mn{2}_n(\etaa,\etab)&=&\frac{\Q{n}{}{\etaa}\Q{n}{*}{\etab}}{N_{\etaa}N_{\etab}}\,,\\
\mn{4}_n(\etaa,\etab)&=&\frac{1}{\N{\etaa}(\N{\etaa}-1)\N{\etab}(\N{\etab}-1)}\times\nonumber\\
&&[\Q{n}{2}{\etaa}\Q{n}{*2}{\etab}+\Q{2n}{}{\etaa}\Q{2n}{*}{\etab}\nonumber\\
&&-\Q{n}{2}{\etaa}\Q{2n}{*}{\etab}-\Q{2n}{}{\etaa}\Q{n}{*2}{\etab}]\,.
\eea
}

The two-particle cumulant is simply the event-average of $\mn{2}$. It is defined to be the square of two-particle cumulant anisotropy~\cite{v2method}. It is composed of three parts, the average flow, flow fluctuation, and nonflow. We may write
\bea
\Vn{}{2}\equiv\mnmn{2}&\equiv&v(\etaa)v(\etab)+\sigma(\etaa)\sigma(\etab)+\fl(\deta)\nonumber\\
&&+\nf(\etaa)\nf(\etab)+\delta(\deta)\,.\label{eq:V2}
\eea
where $v(\eta)$ is average flow and $\sigma(\eta)$ indicates effect of flow fluctuation. The term $\sigma(\etaa)\sigma(\etab)$ should be taken simply as the definition of flow fluctuation analogous to the fluctuation definition in a single variable quantity $x$ (i.e., $\mn{x^2}=\mn{x}^2+\sigma_x^2$); It could contain product terms of flow and flow fluctuations.
In general flow fluctuation can also depend on the $\eta$-gap between the particle pair; we have added this part of flow fluctuation by $\fl(\deta)$. 
It is important to note here that nonflow, generally depending on $\deta$, can also have a $\deta$-independent component. For example, back-to-back inter-jet correlations at low $\pt$ may not have a $\deta$ dependence due to the sampling of the underlying partonic kinematics, but may generally be $\eta$-dependent. When $\pt$ increases, the inter-jet correlation could develop $\deta$ dependence. 
In Eq.~(\ref{eq:V2}) we have therefore included two nonflow terms, $\delta(\deta)$ for the $\deta$-dependent nonflow and $\nf(\etaa)\nf(\etab)$ for the remaining $\deta$-independent component.
Generally $\fl(\deta)$ and $\delta(\deta)$ may also depend on $\etaa$ and $\etab$ which we have suppressed in Eq.~(\ref{eq:V2}) for clarity. We have also suppressed the harmonic order $n$.

The four-particle cumulant is given by $2\mnmn{2}^2-\mnmn{4}$ where $\mnmn{4}$ is the event-averaged four-particle moment. The four-particle cumulant anisotropy squared is defined to be~\cite{v2method} 
\bea
\Vn{}{4}\equiv\sqrt{2\mnmn{2}^2-\mnmn{4}}&\approx&v(\etaa)v(\etab)-\sigma(\etaa)\sigma(\etab)\nonumber\\
&&-\fl(\deta).\label{eq:V4}
\eea
It is composed of the average flow and the flow fluctuation effect. 
Nonflow is negligible in four-particle cumulant~\cite{Borghini_multipart,Borghini_multipart2}.
Note again that Eq.~(\ref{eq:V4}) is rather general, with the $\sigma$ terms simply indicating effects of flow fluctuations. However, flow fluctuations need to be small in order for the approximation in Eq.~(\ref{eq:V4}) to be valid.
Generally speaking the fluctuation terms in Eqs.~(\ref{eq:V2}) and (\ref{eq:V4}) are not necessarily the same because they come from cumulants of different orders; they are the same when fluctuations are Gaussian~\cite{v2method,LiYi}. However, for simplicity we have written them as if they are the same; this does not affect our general argument.

We will use short-hand notations, two-particle ``cumulant flow'' $\vt{}{2}$ and four-particle ``cumulant flow'' $\vt{}{4}$, respectively, to stand for the $\deta$-independent part of the correlation:
\bea
\Vt{}{2}&\equiv&\vt{}{2}(\etaa)\vt{}{2}(\etab)\nonumber\\
&\equiv&v(\etaa)v(\etab)+\sigma(\etaa)\sigma(\etab)+\nf(\etaa)\nf(\etab)\,,\nonumber\\
\Vt{}{4}&\equiv&\vt{}{4}(\etaa)\vt{}{4}(\etab)\nonumber\\
&\equiv&v(\etaa)v(\etab)-\sigma(\etaa)\sigma(\etab)\,.\label{eq:vt}
\eea

The average flow $v(\eta)$ is only a function of $\eta$ and is symmetric about mid-rapidity for symmetric heavy-ion collisions, $v(\eta)=v(-\eta)$; In this case we can write $v(|\eta|)$. The same holds for $\sigma(|\eta|)$ and $\nf(|\eta|)$.
%
Take the difference of the two- (and four-)particle cumulants for two pairs of $\eta$ bins, $(\etaa,\etab)$ and $(\etaa,-\etab)$, where $\etaa<\etab<0$. For symmetric collision systems, the $\deta$-independent terms in Eqs.~(\ref{eq:V2}) and (\ref{eq:V4}) cancel. Namely
\bea
\Delta\Vn{}{2}&\equiv&\Vn{}{2}(\etaa,\etab)-\Vn{}{2}(\etaa,-\etab)\nonumber\\
&\equiv&\Vn{}{2}(\detaa)-\Vn{}{2}(\detab)=\Delta\fl+\Delta\delta\,,\label{eq:dV2}\\
\Delta\Vn{}{4}&\equiv&\Vn{}{4}(\etaa,\etab)-\Vn{}{4}(\etaa,-\etab)\nonumber\\
&\equiv&\Vn{}{4}(\detaa)-\Vn{}{4}(\detab)\approx-\Delta\fl\,,\label{eq:dV4}
\eea
where
\bea
\Delta\fl&=&\fl(\detaa)-\fl(\detab)\,,\label{eq:dF}\\
\Delta\delta&=&\delta(\detaa)-\delta(\detab)\,.\label{eq:dd}
\eea
and 
\bea
\detaa&\equiv&\etab-\etaa\,,\\
\detab&\equiv&-\etab-\etaa\,.
\eea
Again the $\deta$-dependent fluctuation and nonflow terms $\fl(\deta)$ and $\delta(\deta)$ may also depend on $(|\etaa|,|\etab|)$ in case of symmetric collisions systems; these variables are suppressed in Eqs.~(\ref{eq:dF}) and (\ref{eq:dd}).

What remains in $\Delta\Vn{}{4}$ is $\Delta\fl$ which should yield information on $\fl(\eta)$. What remain in $\Delta\Vn{}{2}$ are $\Delta\fl$ and $\Delta\delta$. Given information on $\fl$ from $\Delta\Vn{}{4}$ and assuming the $\fl$ in $\Delta\Vn{}{2}$ is similar, one should be able to deduce valuable information on nonflow $\delta(\deta)$.

Note there is no assumption about the $\eta$-dependence of flow or flow fluctuation. On the other hand, our method relies on the symmetry of the average flow (and the other $\deta$-independent terms) about mid-rapidity, and thereby it applies only to symmetric collision systems. 

It is important to point out that our method essentially separates $\deta$-dependent and $\deta$-independent correlations, not particularly flow and nonflow. Because flow is a single particle property, the $\deta$-dependent correlation may be identified as nonflow. However, since flow is measured by particle correlations, there may be a $\deta$-dependent flow fluctuation component. The $\deta$-independent correlation should be dominated by flow. However, nonflow correlations are possibly present in back-to-back pairs of particles, such as away-side intra-jet correlations. This part of the nonflow is likely independent of $\deta$. Therefore the decomposed $\deta$-independent correlation is still contaminated by away-side nonflow. Consequently, the $\deta$-dependent correlation contains only part of the nonflow, possibly only from near-side correlations such as intra-jet correlations and resonance decays. 

\section{Illustration with event generators\label{AMPT}}

We illustrate our method using the \ampt\ (a multiphase transport) event generator~\cite{AMPT}. The \ampt\ model uses Monte Carlo Glauber initial geometry, liberates partons from the nucleons, follows a cascade evolution of parton-parton scatterings, and finally freezes out into hadrons through a coalescence mechanism. The scatterings generate collective flow among the partons and the final state hadrons~\cite{AMPTv2,Molnar}. We use all default settings of \ampt\ with string melting turned on. It was shown that string melting was needed to reproduce the $v_2$ measurements~\cite{AMPTv2}. We use \ampt\ for the purpose that it generates events which contain collective flow, flow fluctuations due to initial geometry fluctuations, and nonflow such as from resonance decays and jet-correlations. It is not critical for our purpose whether the physics in \ampt\ is all correct or not. 

We generate total $1.1\times10^7$ minimum-bias \ampt\ events. For illustration of our method, we focus on events with impact parameter \bperi, corresponding to approximately 10-60\% collision centrality. This is described in Sec.~\ref{sec:fluc}-\ref{sec:sanity}. In Sec.~\ref{sec:central} we check our method against central \ampt\ events with \bcent\ (approximately top 10\% centrality). We also use the \hijing\ event generator~\cite{hijing} to check our method. \hijing\ does not have flow but contain significant nonflow, and thus provides a critical test to our method. We generate total $4.7\times10^7$ minimum-bias \hijing\ events. This is described in Sec.~\ref{sec:hijing}.

In both models, strong and electromagnetic decay resonances are made to decay. For our study we use all particles with $\pt<2$~GeV/$c$ in our calculation of azimuthal cumulants. We include neutral particles for the \ampt\ study which significantly helps to increase statistics for the four-particle cumulant. We include only charged hadrons for the \hijing\ study because it is relatively cpu-cheap to generate more \hijing\ events.

\begin{figure*}[hbt]
\begin{center}
\includegraphics[width=0.3\textwidth]{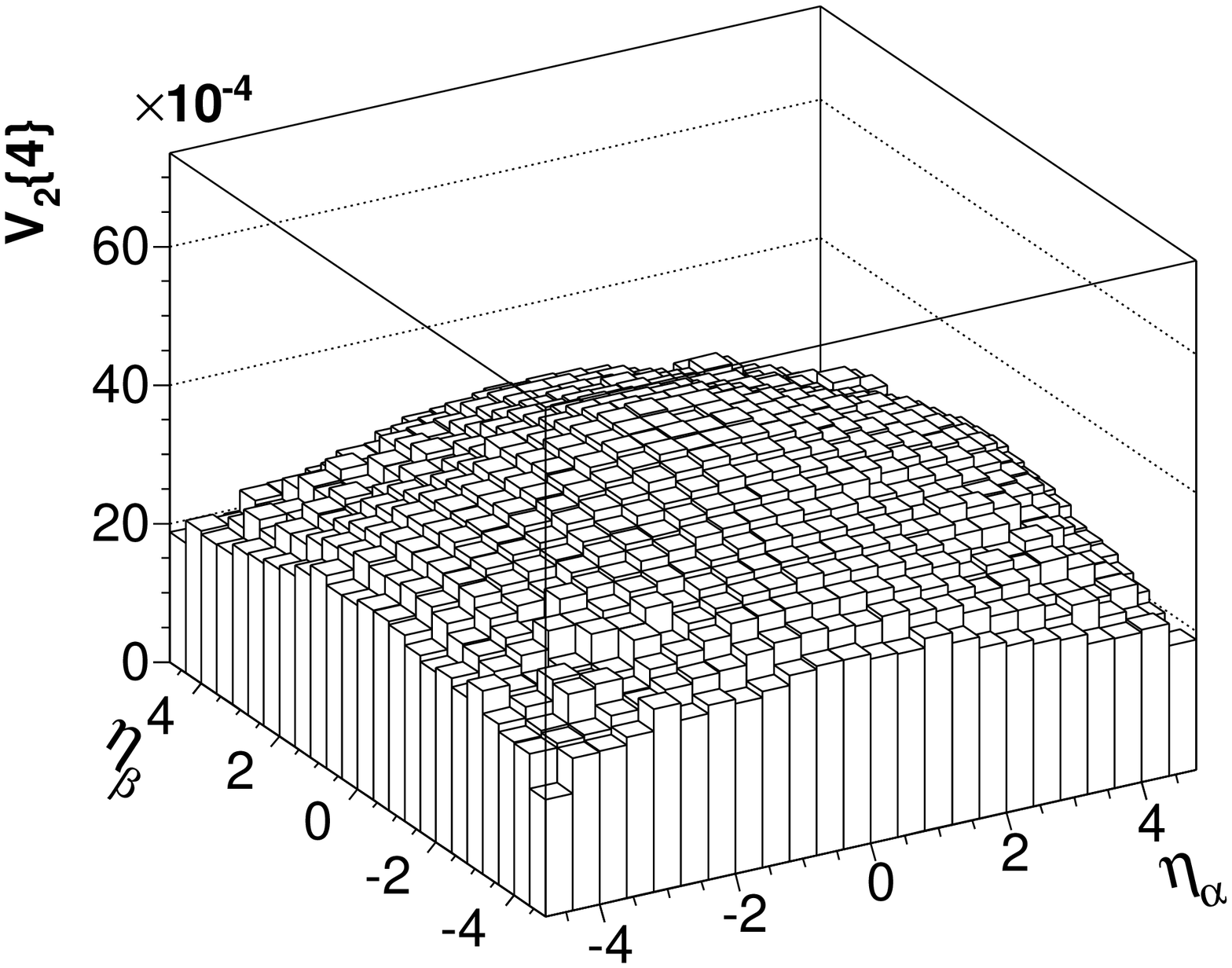}
\includegraphics[width=0.3\textwidth]{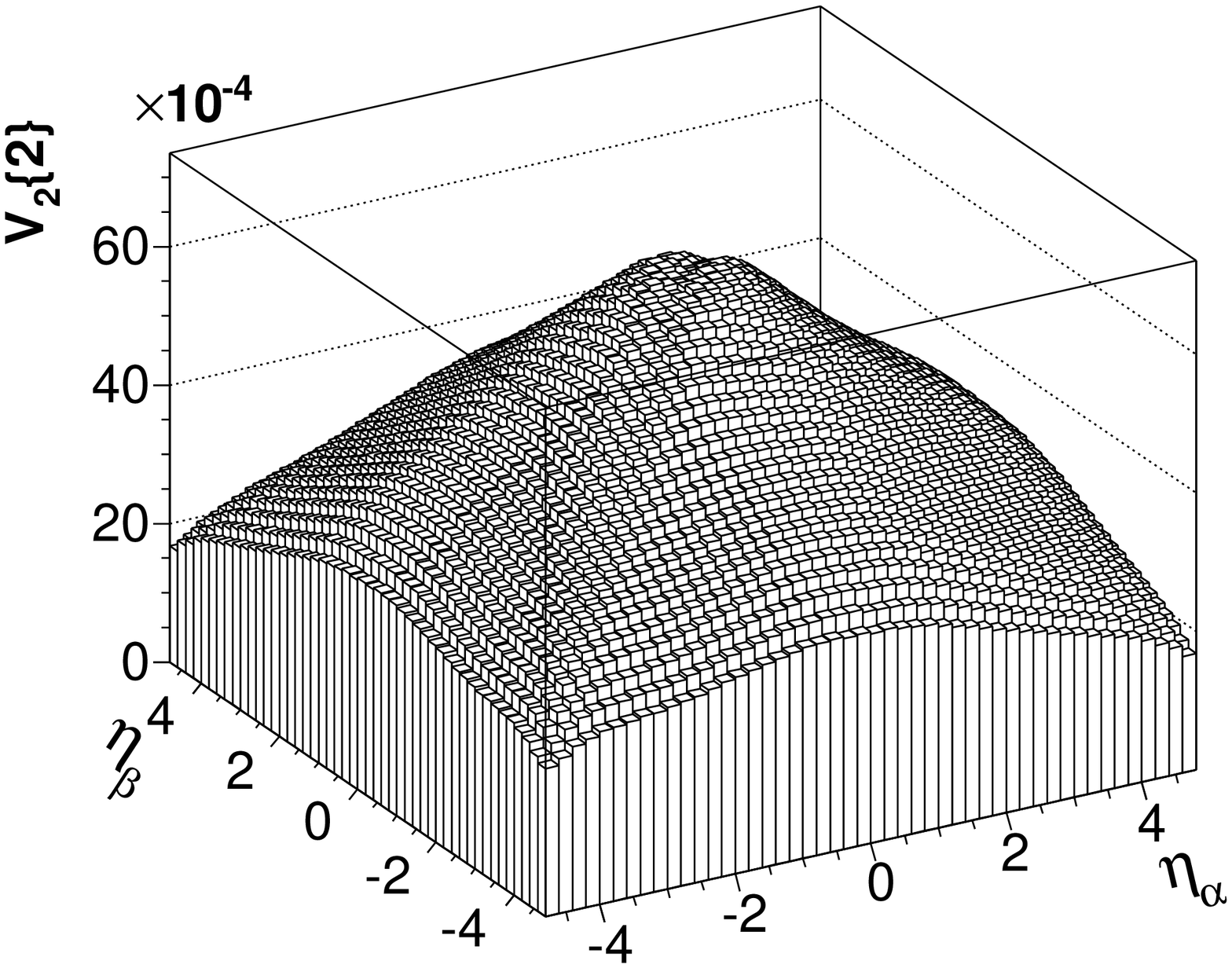}
\includegraphics[width=0.3\textwidth]{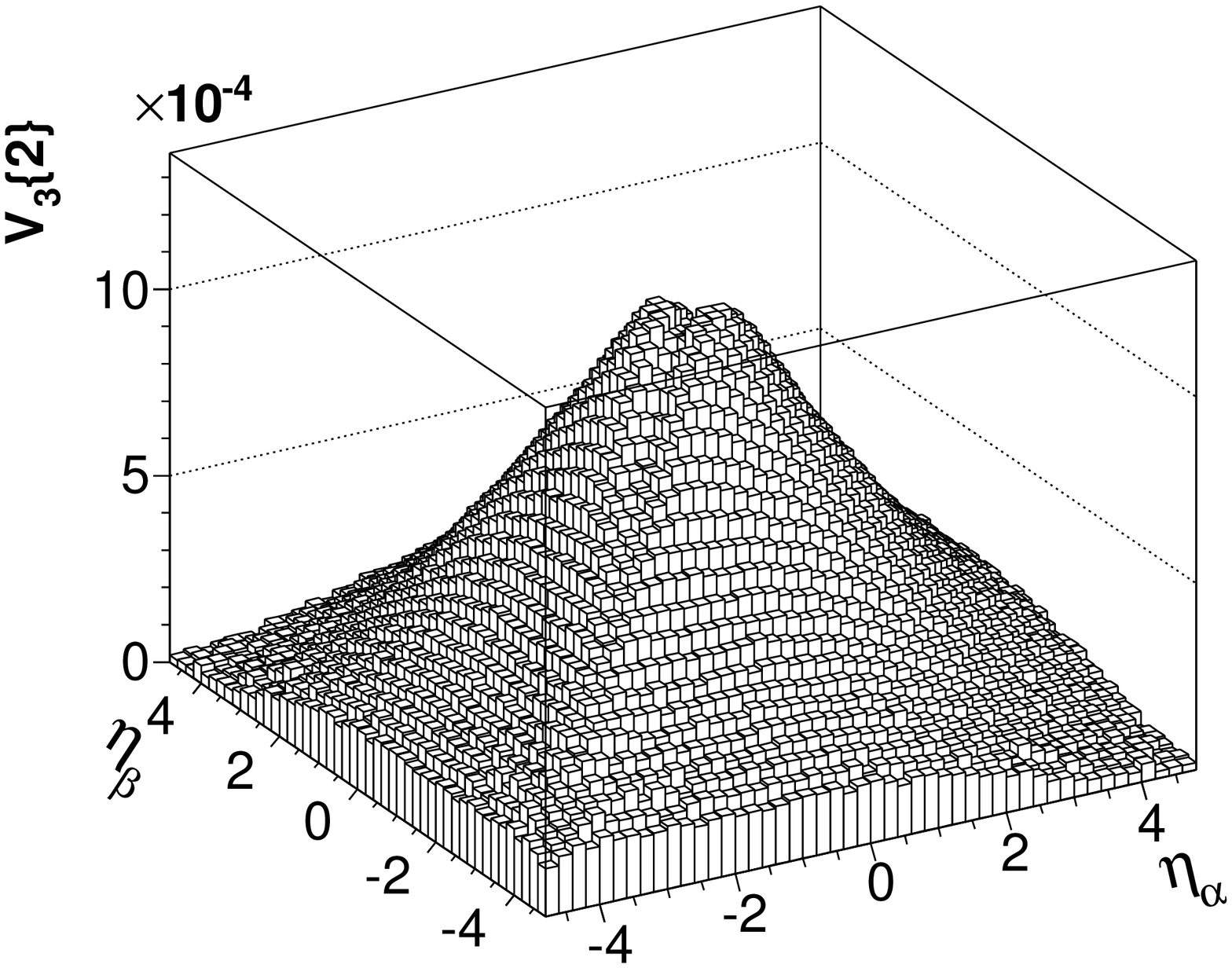}
\end{center}
\caption{The square root of the second-harmonic four-particle cumulant, $\Vn{2}{4}$ (left), the second-harmonic two-particle cumulant, $\Vn{2}{2}$ (middle), and the third-harmonic two-particle cumulant, $\Vn{3}{2}$ (right) in 200 GeV Au+Au collisions with \bperi\ (approximately 10-60\% centrality) generated by \ampt.}
\label{fig:V}
\end{figure*}

Figure~\ref{fig:V} shows the second-harmonic four- and two-particle cumulant results from \ampt\ events of \bperi, $\Vn{2}{4}$ (left panel) and $\Vn{2}{2}$ (middle panel), and the third-harmonic two-particle cumulant, $\Vn{3}{2}$ (right panel), as a function of $\etaa$ and $\etab$ of the two $\eta$ bins. The dip along the diagonal $\deta\sim0$ in the two-particle cumulants is not understood but has been also observed in two-particle correlations in Ref.~\cite{GLMa}. The $\deta\sim0$ dip is also observed in \hijing\ (see Fig.~\ref{fig:hijing} in Sec.~\ref{sec:hijing}). This suggests that its origin may reside in \hijing\ which is used as the initial input to \ampt\ parton cascading, not from final state rescattering processes. 

\subsection{$\deta$-dependent part of flow fluctuation~\label{sec:fluc}}

\begin{figure*}[hbt]
\begin{center}
\includegraphics[width=0.3\textwidth]{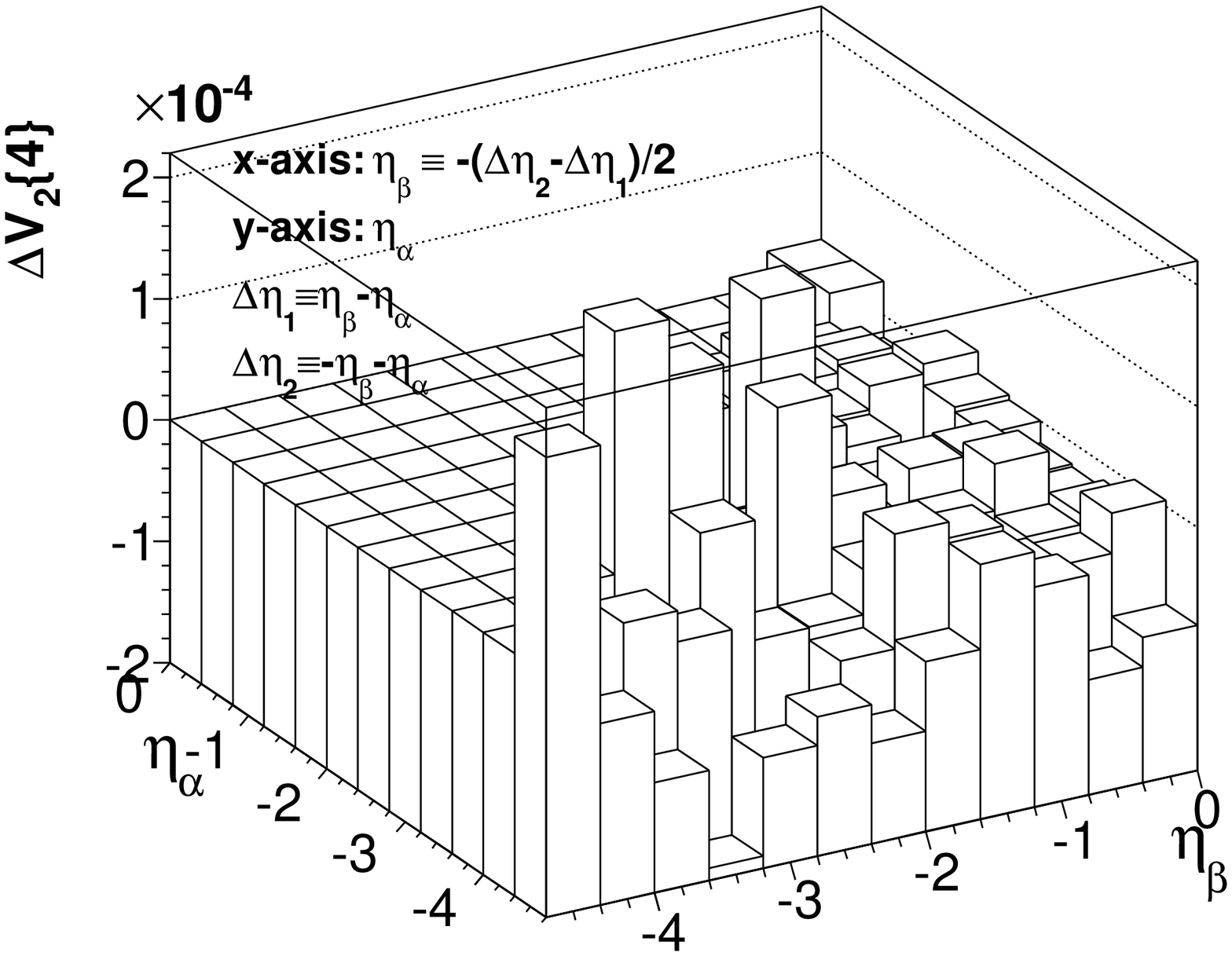}
\includegraphics[width=0.3\textwidth]{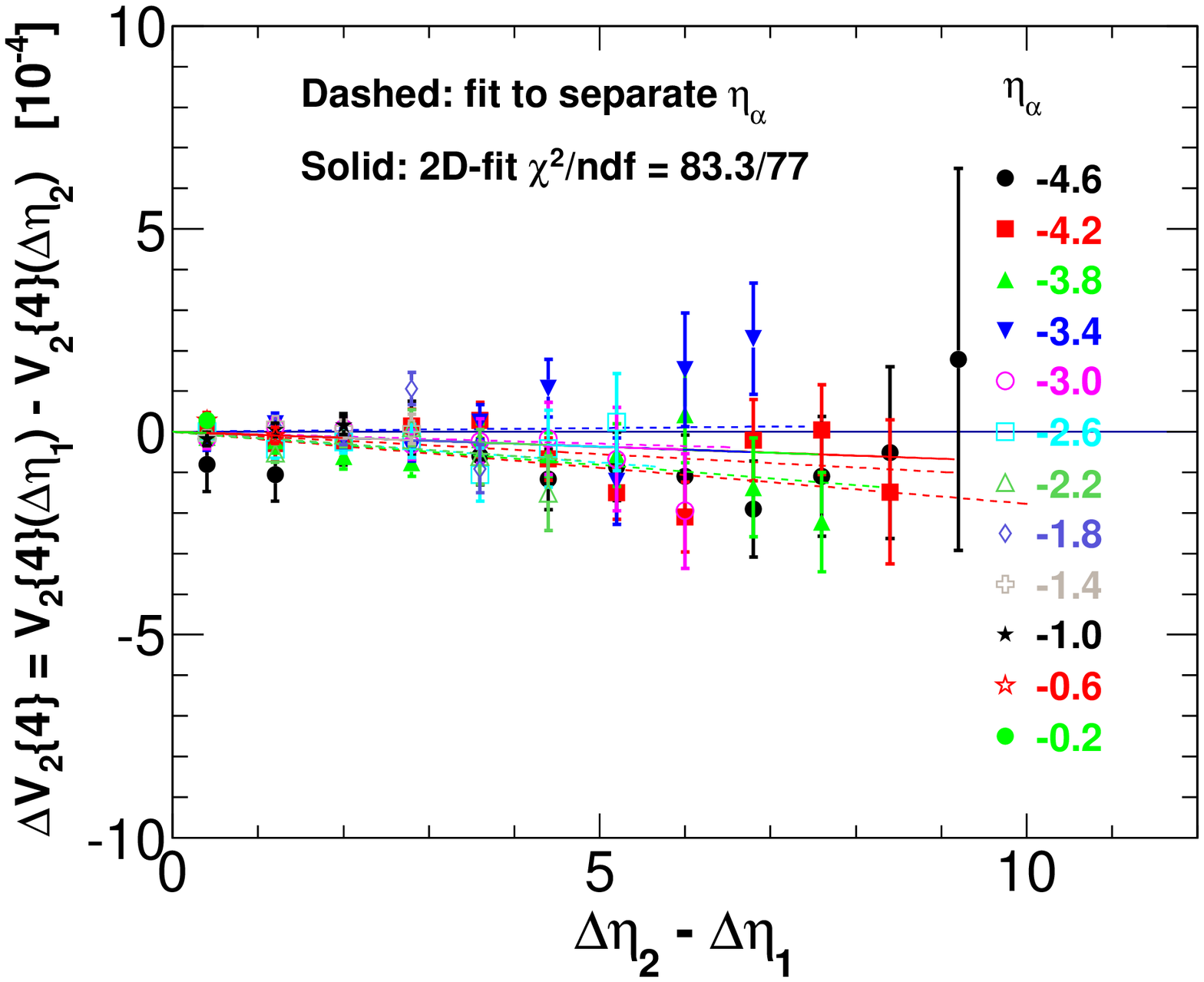}
\includegraphics[width=0.3\textwidth]{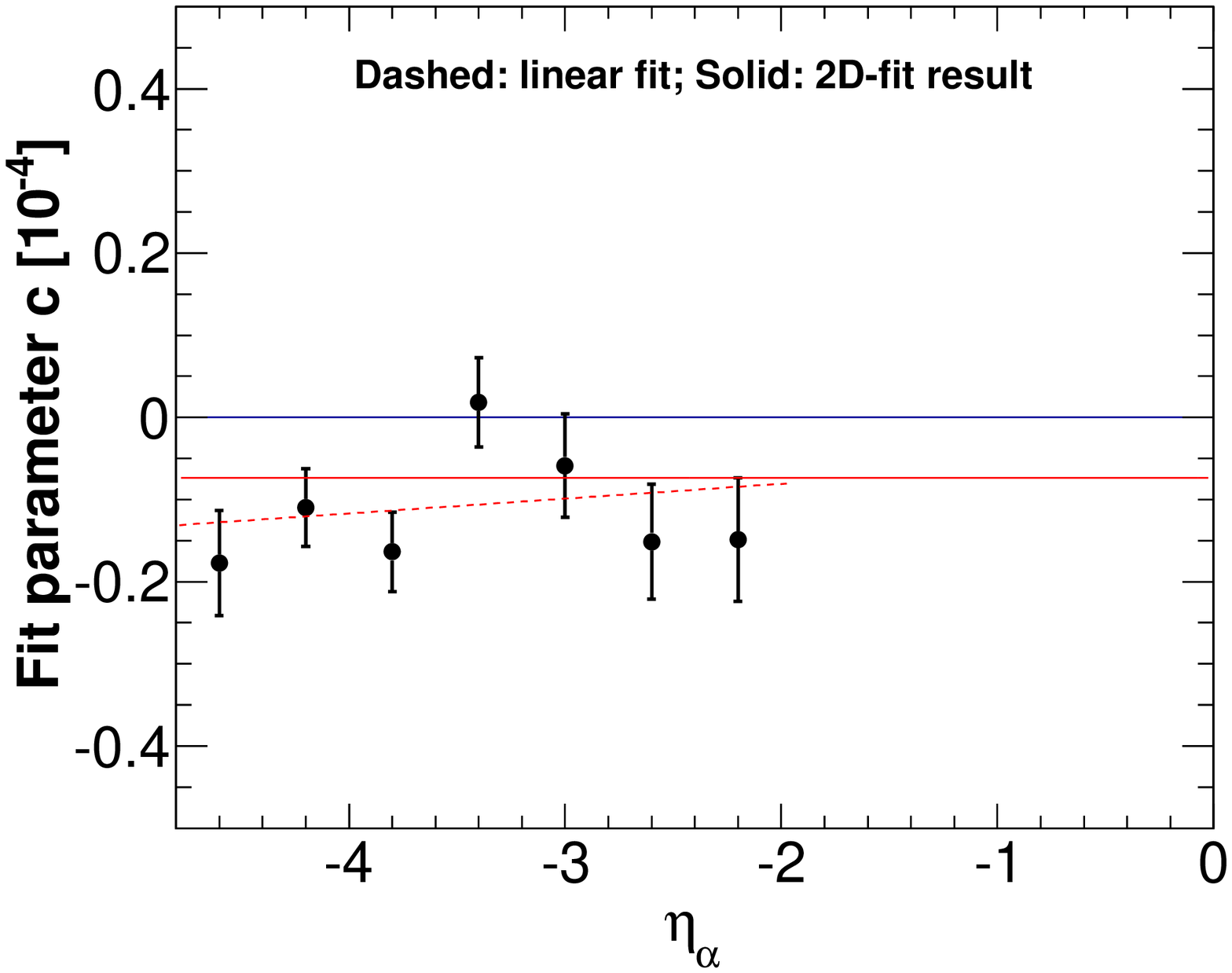}
\end{center}
\caption{(Color online) Left: Difference $\Delta\Vn{2}{4}$ between two pairs at $\eta$ bins of $(\etaa,\etab)$ and $(\etaa,-\etab)$. Middle: $\Delta\Vn{2}{4}$ as a function of $\detab-\detaa\equiv-2\etab$ for various values of $\etaa$. The dashed lines are linear fit to $c(\detab-\detaa)$ for each fixed value of $\etaa$. Right: The fit parameter $c$ as a function of $\etaa$. The dashed curve is a straight-line fit. A global 2D-fit of $c(\detab-\detaa)$ is applied to all data points in the middle panel resulting in the solid line. The fit parameter is $c=-0.07\pm0.02$ (indicated by the solid line in the right panel) with $\chisq=83/77$. Data are from \ampt\ simulation of 200 GeV Au+Au collisions with \bperi\ (approximately 10-60\% centrality).}
\label{fig:dV4}
\end{figure*}

First we take the difference of the four-particle cumulants $\Vn{2}{4}$ for two pairs of $\eta$ bins at $(\etaa,\etab)$ and $(\etaa,-\etab)$, see Eq.~(\ref{eq:dV4}). The contributions of average flow cancel and what remains is the difference $\Delta\fl$ between the fluctuation effects for $\eta$-gaps of $\detaa$ and $\detab$. Figure~\ref{fig:dV4} left panel shows the difference $\Delta\Vn{2}{4}$ plotted versus $(\etab,\etaa)$. 

In the middle panel of Fig.~\ref{fig:dV4} the difference $\Delta\Vn{2}{4}$ is plotted as a function of $\detab-\detaa\equiv-2\etab$ for various values of $\etaa$. The difference is not very different from zero, suggesting that second-harmonic flow fluctuations are nearly independent of $\deta$. Closer inspection seems to suggest a tendency of a decreasing $\Delta\Vn{2}{4}=\Vn{2}{4}(\detaa)-\Vn{2}{4}(\detab)$ with increasing $\detab-\detaa$. This implies that the flow correlation between two $\eta$ bins is larger when the $\eta$ bins are further apart. This is counterintuitive.

Nevertheless, we fit $\Delta\Vn{2}{4}$ with a linear function 
\be
\Delta\Vn{2}{4}=c(\detab-\detaa)\,,\label{eq:fluc_fit}
\ee
for each $\etaa$ where $c$ is a fit parameter. This functional form is driven by the observation in Fig.~\ref{fig:dV4} and the fact that $\Delta\Vn{}{4}$ must be zero when $\detaa=\detab$. We plot the single fit parameter $c$ in Fig.~\ref{fig:dV4} right panel as a function of $\etaa$ and find it is consistent with a constant. We therefore fit all the data points in Fig.~\ref{fig:dV4} middle panel with a single function of Eq.~(\ref{eq:fluc_fit}). The fit result is $c=-0.07\pm0.02$ with a $\chisq=83/77$.

The $\deta$-dependent part of flow fluctuation can be inferred as
\be\fl(\deta)=c(\deta-\detamax)\,,\label{eq:fluc}\ee
where $\detamax$ is chosen to be the maximum $\eta$-gap of 9.6 in our study. This is shown in the left panel of Fig.~\ref{fig:d} as function of $(\etaa,\etab)$. Note that the parameterization of Eq.~(\ref{eq:fluc}) is obtained from the restricted region of $\etaa<0$ and $\etaa<\etab$; The full region of the $(\etaa,\etab)$ space is extended by symmetry. 

We find $\Delta\Vn{3}{4}$ is consistent with zero but with large errors. 

\subsection{Extracting $\deta$-dependent nonflow correlation\label{sec:nonflow}}

\begin{figure*}[hbt]
\begin{center}
\includegraphics[width=0.3\textwidth]{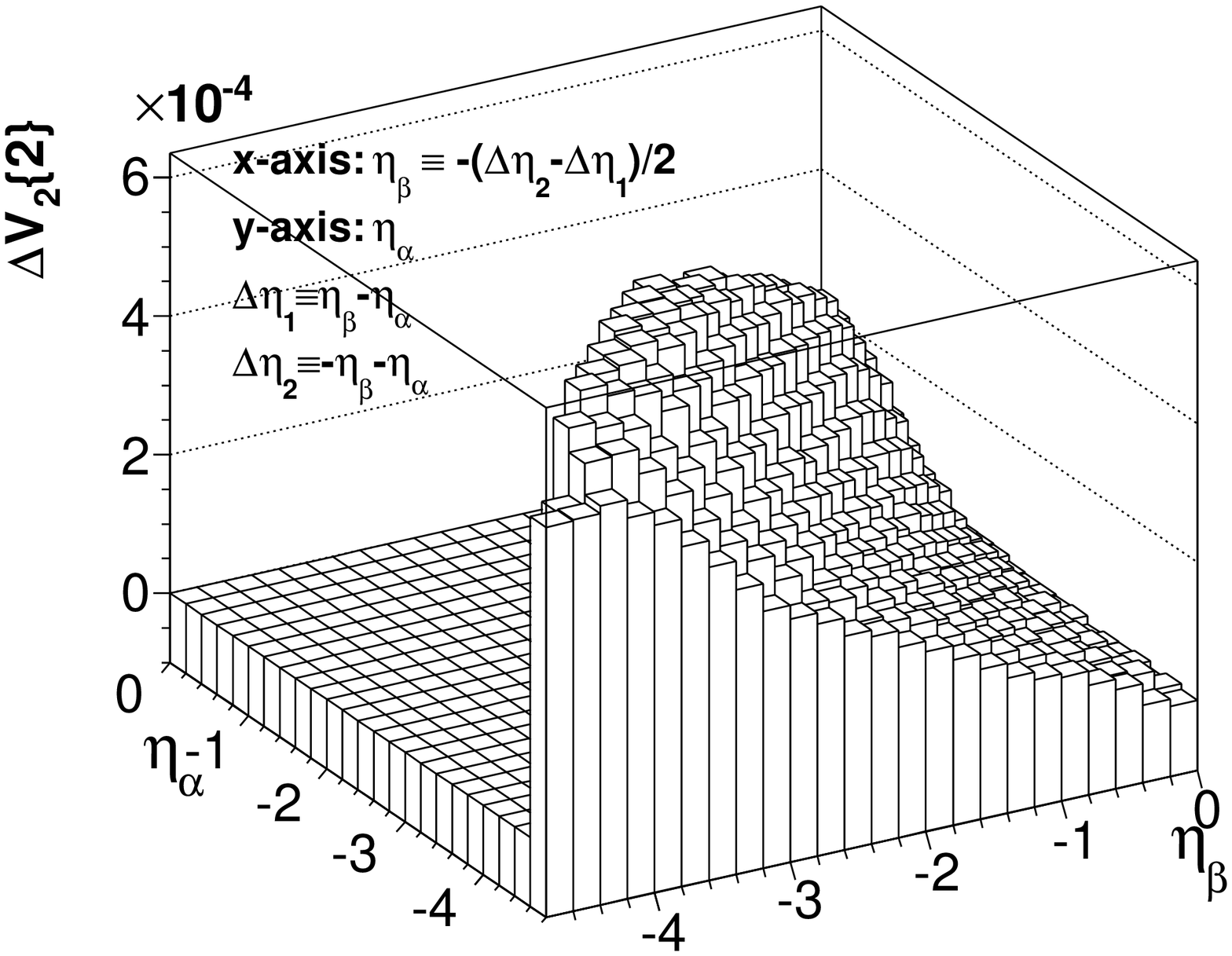}
\includegraphics[width=0.3\textwidth]{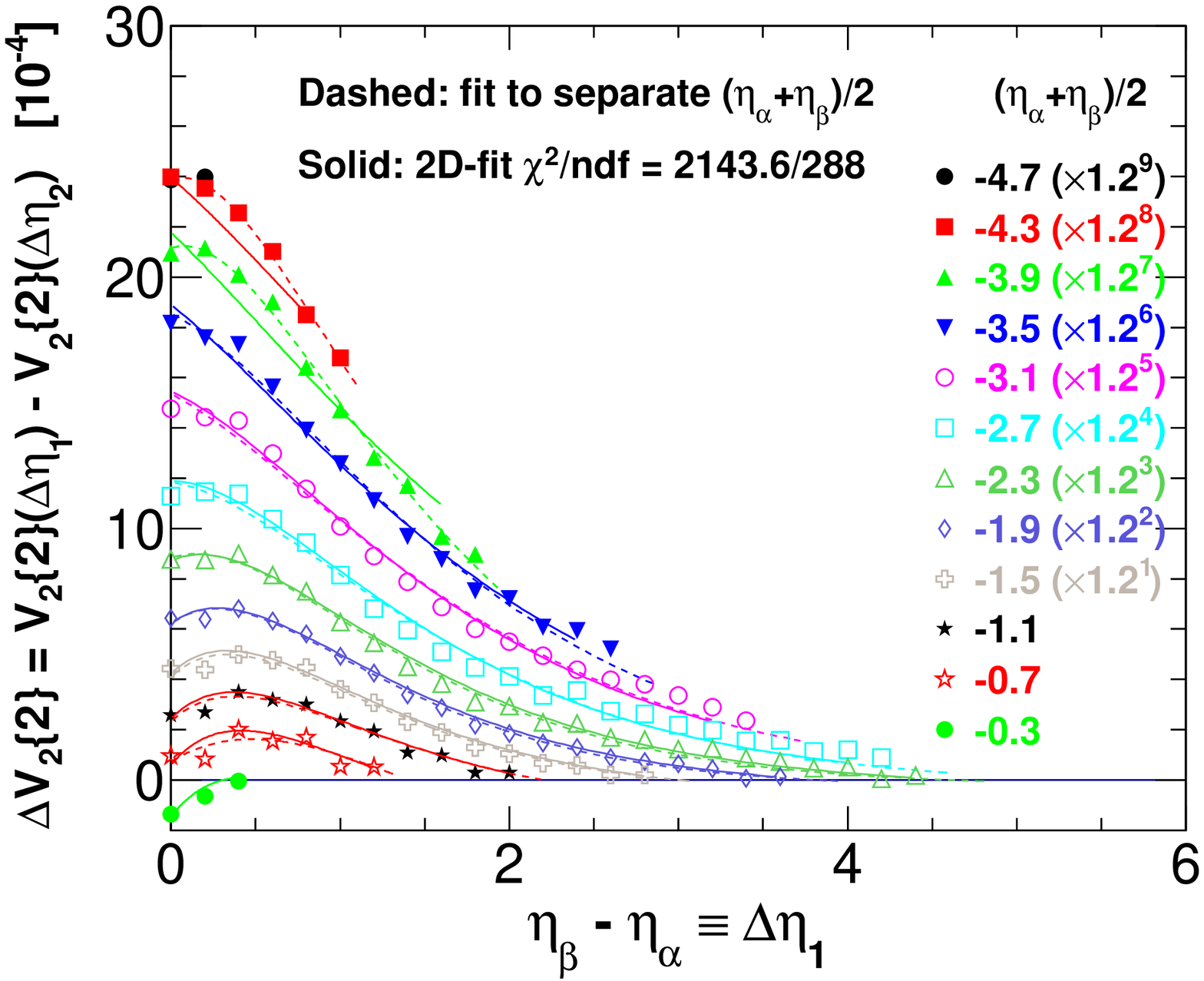}
\includegraphics[width=0.3\textwidth]{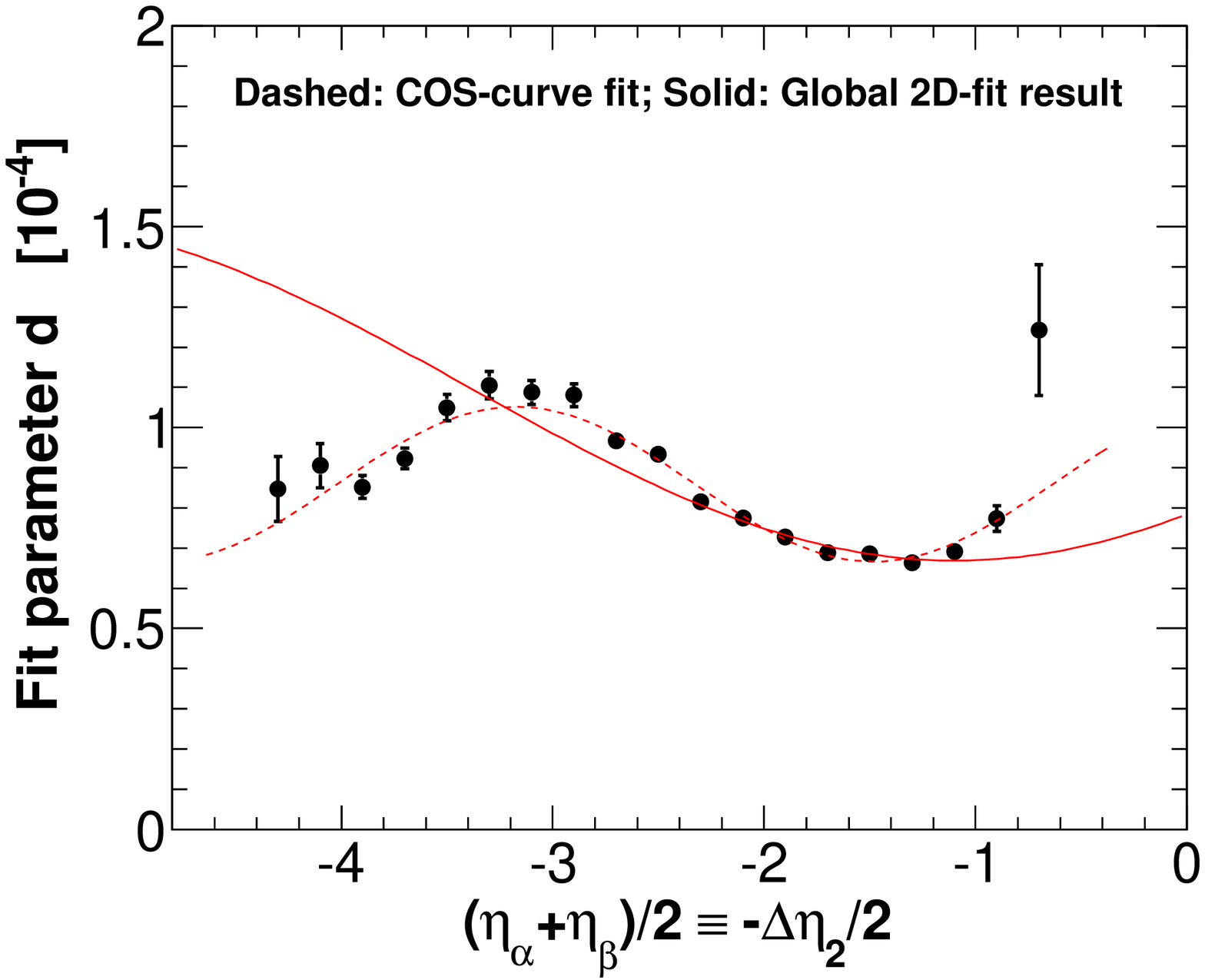}
\includegraphics[width=0.3\textwidth]{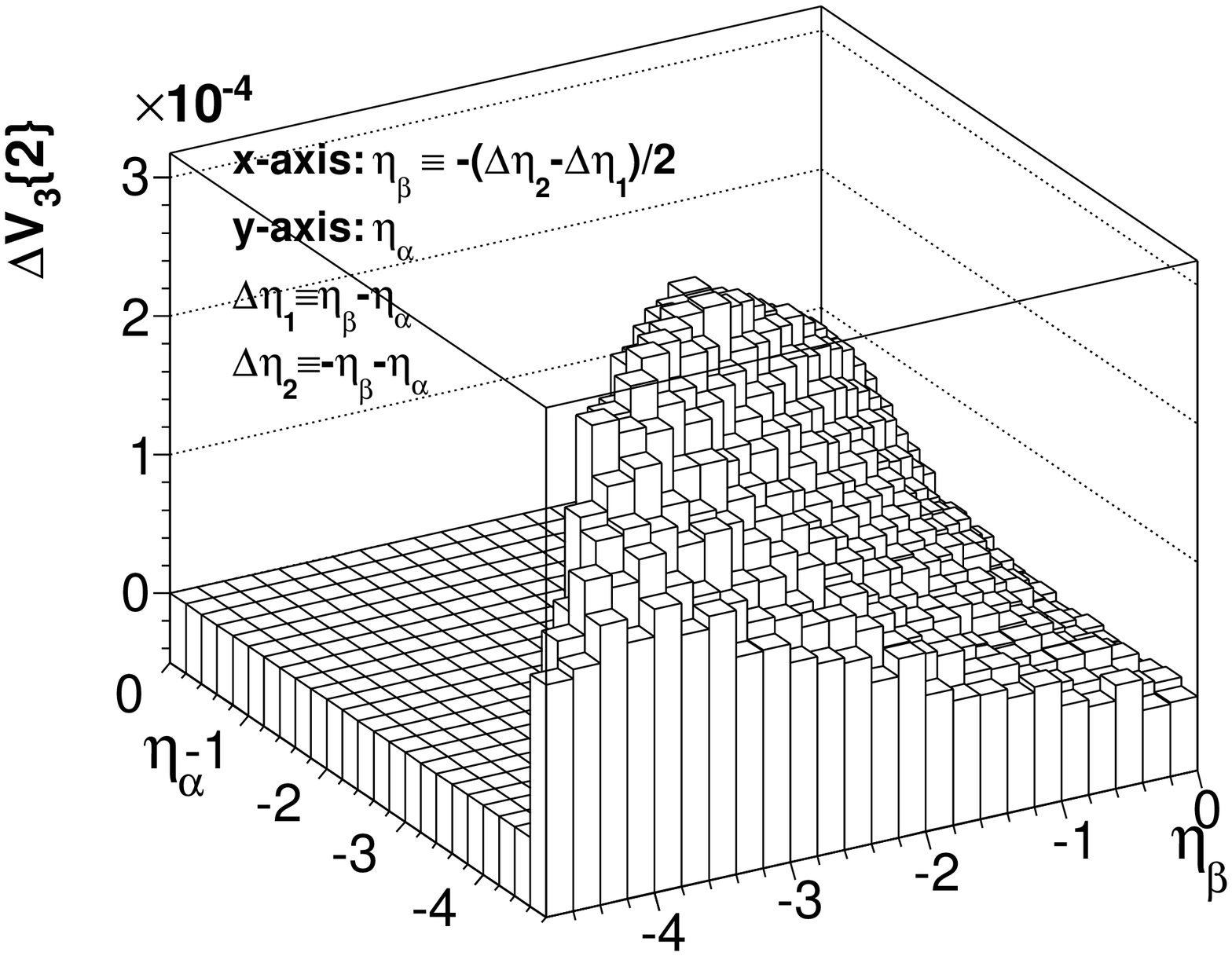}
\includegraphics[width=0.3\textwidth]{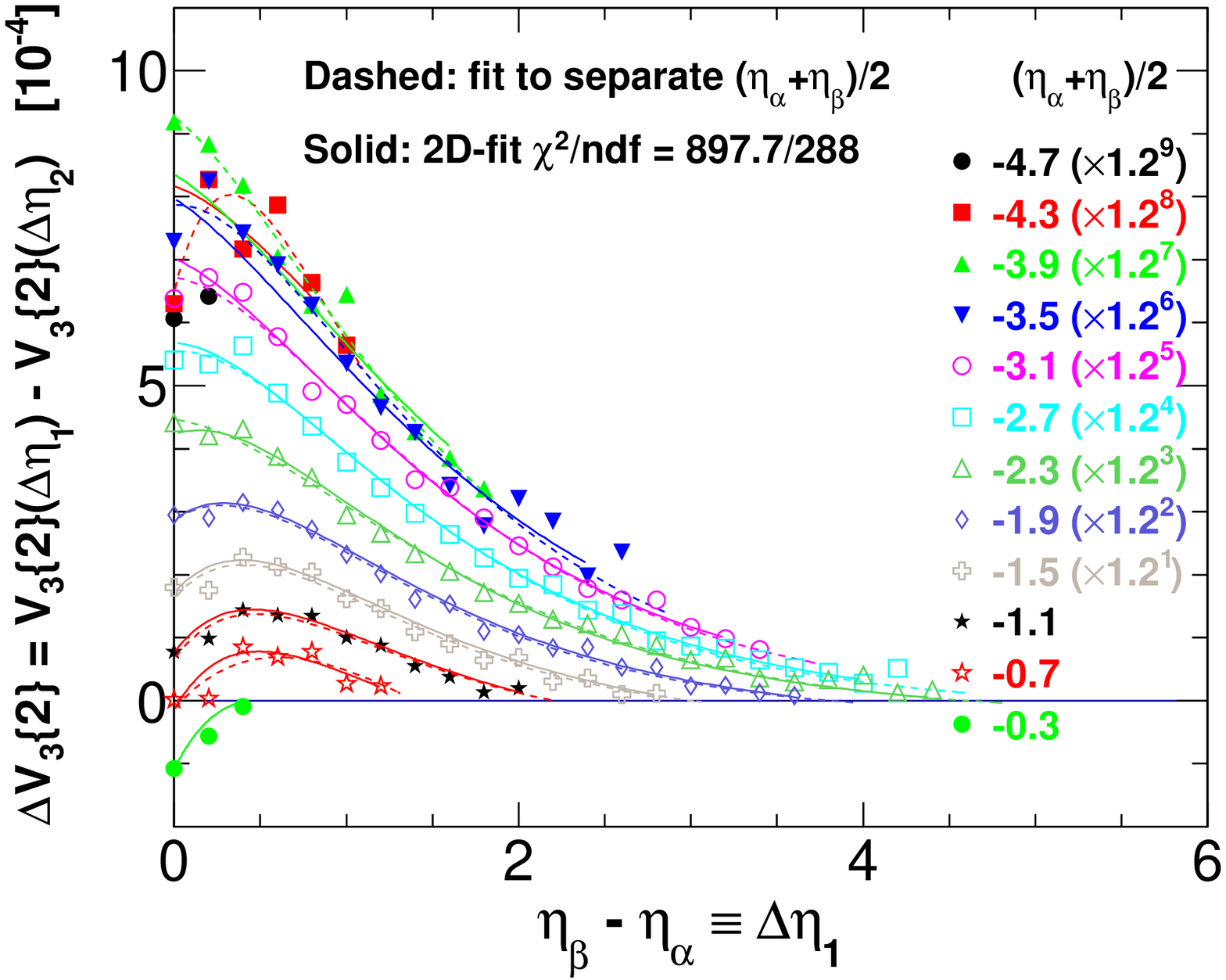}
\includegraphics[width=0.3\textwidth]{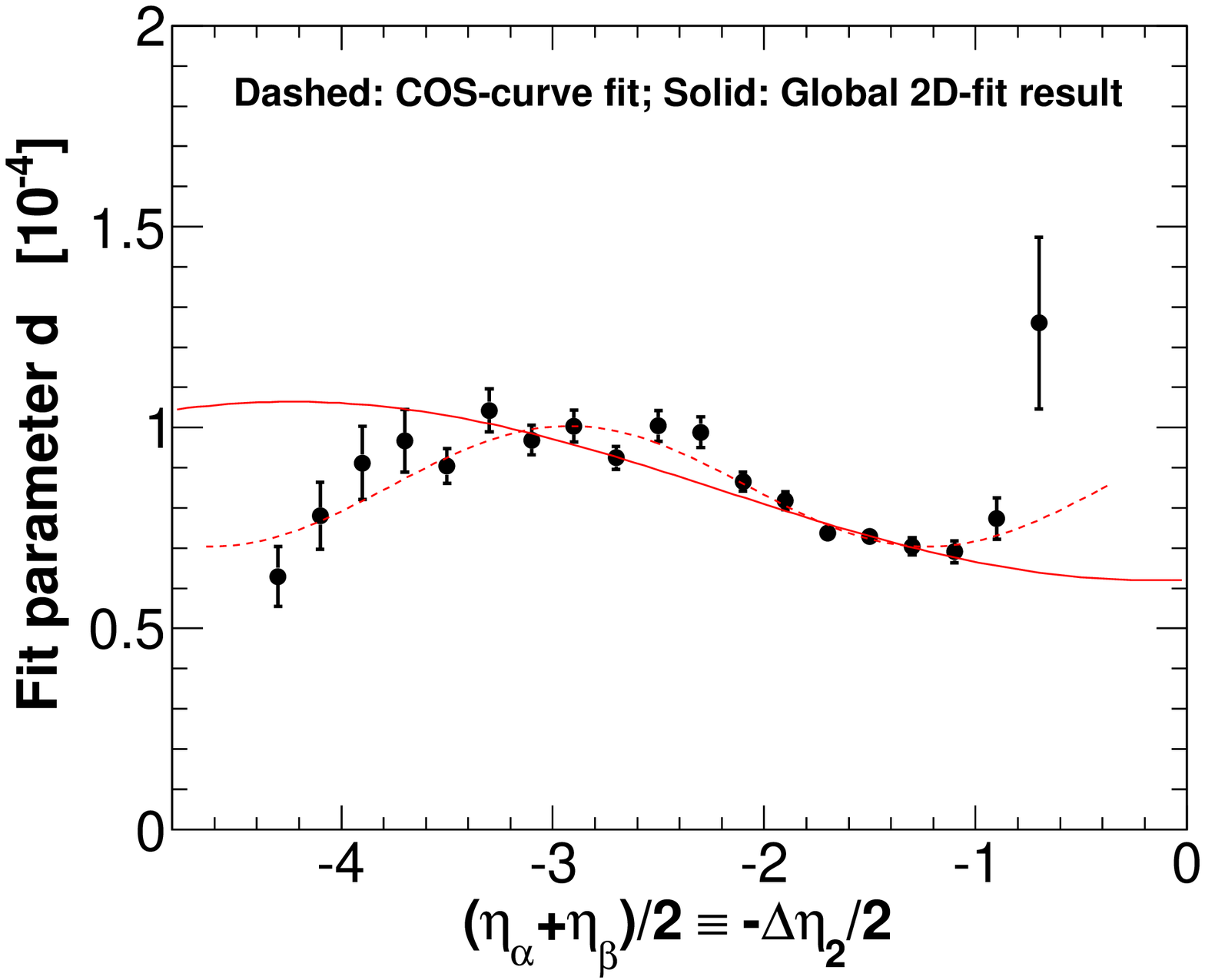}
\end{center}
\caption{(Color online) Left panels: difference $\Delta\Vn{2}{2}$ (upper) and $\Delta\Vn{3}{2}$ (lower) between $(\etaa,\etab)$ and $(\etaa,-\etab)$ as a function of $(\etab,\etaa)$. Middle panels: difference plotted as a function of $\etab-\etaa\equiv\detaa$ for different values of $(\etaa+\etab)/2\equiv-\detab/2$. Some of the data points are scaled by succussive factors of 1.2 (indicated in the legends) for clarity. The dashed curves are empirical fit by Eq.~(\ref{eq:nf_fit1D}) for each fixed $(\etaa+\etab)/2$, while the solid curves are results of a global 2D-fit to $\Delta\Vn{}{2}$ vs $(\detaa,\detab)$ by Eq.~(\ref{eq:nf_fit2D}). Right panels: parameter $d$ from Eq.~(\ref{eq:nf_fit1D}) fit separately for different $(\etaa+\etab)/2$ values plotted versus $(\etaa+\etab)/2$. The results motivate the Eq.~(\ref{eq:nf_par_fit}) fit in the dashed curve. The $d$ parameter from the global 2D-fit is shown in the solid curve. Data are from \ampt\ simulation of 200 GeV Au+Au collisions with \bperi\ (approximately 10-60\% centrality).}
\label{fig:dV2}
\end{figure*}

We now turn our attention to the two-particle cumulants. Take the difference of two-particle cumulants $\Vn{}{2}$ between two pairs of $\eta$ bins at $(\etaa,\etab)$ and $(\etaa,-\etab)$, see Eq.~(\ref{eq:dV2}). The contributions of average flow and the other $\deta$-independent terms cancel and what remains should be the difference in nonflow between $\detaa$ and $\detab$ (and any difference in the $\deta$-dependent part of flow fluctuation). The differences $\Delta\Vn{2}{2}$ and $\Delta\Vn{3}{2}$ are plotted in Fig.~\ref{fig:dV2} (left panels) as a function of $(\etab,\etaa)$. Similar to the procedure in Sec.~\ref{sec:fluc}, the rest of this section describes empirical ways to parameterize this difference.

As seen from the figure, $\Delta\Vn{}{2}$ appears to decrease exponentially along the off-diagonal $\etab-\etaa$ and vary modestly along the diagonal $\etaa+\etab$. We thus slice $\Delta\Vn{}{2}$ in bins of $(\etaa+\etab)/2$, and plot it as a function of $\etab-\etaa$ in Fig.~\ref{fig:dV2} (middle panels). Indeed the main feature is an exponential in $\etab-\etaa\equiv\detaa$ (with a downturn at small $\detaa$ which is related to the diagonal dip observed in Fig.~\ref{fig:V}). However, this is not the case for all values of $(\etaa+\etab)/2\equiv-\detab/2$. This is because $\Delta\Vn{}{2}$ is the {\em difference} of two-particle cumulant, must vanish at $\detaa=\detab$, and therefore ought to have another term corresponding to that at $\detab$. This term may be negligible for $\detab\gg\detab$ (as is the case when $\etaa$ and $\etab$ are both far removed from mid-rapidity; note $\etaa<\etab<0$), but comparable to the $\detaa$ term when $\detab\sim\detaa$ (as is the case when $\etaa$ and $\etab$ are both near mid-rapidity). This explains why the data points for $(\etaa+\etab)/2\sim0$ are quite different from others where $(\etaa+\etab)/2$ is far from zero. Even for those other data points, the large $\detaa$ tails (i.e.~large negative $\etaa$ and $\etab\sim0$ and thus $\detaa\sim\detab$) of $\Delta V_{2}$ seem to (and should) deviate from the simple exponential drop. 

In order to gain more insights, we fit each slice of data points in Fig.~\ref{fig:dV2} middle panels by function
\be
\Delta\Vn{}{2}=(a+b\detaa)e^{-\detaa/d} - (a+b\detab)e^{-\detab/d}\,,\label{eq:nf_fit1D} 
\ee
where $\detaa\equiv\etab-\etaa$ is treated as variable and $\detab\equiv-(\etab+\etaa)$ is fixed for each slice. 
The linear component is to mimic the down-turn at small $\detaa$. We have kept the two, presumably nonflow, terms of identical functional form, but evaluated at $\detaa$ and $\detab$, respectively.
The individual fit result is superimposed as the dashed curve in Fig.~\ref{fig:dV2} middle panels for each slice of data points. 

We find that the fit parameters $a$, $b$, and $d$ are not constant but vary modestly with $\etaa+\etab$. This is not surprising because, although $\delta(\deta)$ may be only a function of $\deta$, its magnitude likely depends on the average $\mn{\eta}$ of the particle pair. As an example we show the fit parameter $d$ as a function of $\mnetaa\equiv(\etaa+\etab)/2$ in Fig.~\ref{fig:dV2} right panels. We find the parameters $a$, $b$, and $d$ seem to all follow a cosine curve. We thus fit them to functions of 
\bea
a(\mnetaa)&=&a_0+a_1\cos[a_2(\mnetaa-a_3)]\,,\nonumber\\
b(\mnetaa)&=&b_0+b_1\cos[b_2(\mnetaa-b_3)]\,,\nonumber\\
d(\mnetaa)&=&d_0+d_1\cos[d_2(\mnetaa-d_3)]\,.\label{eq:nf_par_fit}
\eea
The fit result to $d$ is superimposed in Fig.~\ref{fig:dV2} right panels as the dashed curve.

Given the above observations, we now fit the entire set of data points in each middle panel of Fig.~\ref{fig:dV2} simultaneously by a single two-dimensional function
\bea
\Delta\Vn{}{2}&=&[a(\mnetaa)+b(\mnetaa)\detaa]e^{-\detaa/d(\mnetaa)}-\nonumber\\
&&[a(\mnetab)+b(\mnetab)\detab]e^{-\detab/d(\mnetab)}\,,\label{eq:nf_fit2D}
\eea
treating $\etab$ and $\etaa$ as the two variables. We call this ``global 2D-fit.''
Note we have kept the two terms of identical functional form, corresponding to the $\deta$-dependent correlations of the two pairs. The parameters are given by the function form of Eq.~(\ref{eq:nf_par_fit}), but we have written them as a function of the average $\mneta$ of the respective pair. Namely, $\mnetaa$ is the mean $\eta$ of the first pair $(\etaa,\etab)$ with $\eta$-gap of $\detaa$ and $\mnetab$ is the mean $\eta$ of the second pair $(\etaa,-\etab)$ with $\eta$-gap of $\detab$,
\bea
\mnetaa&=&(\etaa+\etab)/2\,,\nonumber\\
\mnetab&=&(\etaa-\etab)/2\,.
\eea

We perform the global 2D-fit for both $\Delta\Vn{2}{2}$ and $\Delta\Vn{3}{2}$. The fit parameters and $\chisq$ are tabulated in Table~\ref{tab}. The fit results are superimposed as the solid curves in Fig.~\ref{fig:dV2} middle panels. As can be seen, the global 2D-fit does a fairly good job to describe all the data points. The 2D-fit result for $d$ is superimposed in the right panel and there the 2D-fit does not seem to perform well which is not obvious from the middle panels of Fig.~\ref{fig:dV2}. However, we note that $d$ is plotted as a function of $\etaa+\etaa\equiv2\mnetaa$, but the $d$'s in the two r.h.s. terms of Eq.~(\ref{eq:nf_fit2D}) are functions of $\mnetaa$ and $\mnetab$, respectively. The 2D-fit is expected to be worse than the stand-alone fit to $d$ in the dashed curve because of extra constraints in the 2D-fit.

\begin{table*}
\caption{Fit parameters for $\deta$-dependent nonflow correlation (plus any $\deta$-dependent flow fluctuation effect), $\delta(\deta)+\fl(\deta)$, and fit $\chisq$ by Eq.~(\ref{eq:nf_fit2D}) to 10-60\% (\bperi) and 0-10\% (\bcent) \ampt\ events and to 10-60\% (\bperi) \hijing\ events.}
\label{tab}
\begin{tabular}{c|cc|cc|cc}
\hline
& \multicolumn{2}{|c}{\ampt\ \bperi} & \multicolumn{2}{|c}{\ampt\ \bcent} & \multicolumn{2}{|c}{\hijing\ \bperi} \\
Parameter & $\delta_2$ & $\delta_3$ & $\delta_2$ & $\delta_3$ & $\delta_2$ & $\delta_3$ \\
\hline
$a_0$ & $ 5.54 \pm0.02 $ & $ 1.99 \pm0.02 $ & $ 2.1 \pm0.2 $ & $ -5.1  \pm0.2   $ & $8.44 \pm0.09 $ & $-10.5 \pm0.1   $ \\
$a_1$ & $-0.82 \pm0.02 $ & $ 0.84 \pm0.03 $ & $-0.4 \pm0.2 $ & $  6.0  \pm0.2   $ & $7.64 \pm0.09 $ & $-10.9 \pm0.1   $ \\
$a_2$ & $ 1.71 \pm0.02 $ & $ 1.32 \pm0.02 $ & $ 0.85\pm0.08$ & $  0.181\pm0.003 $ & $0.118\pm0.002$ & $0.0651\pm0.0007$ \\
$a_3$ & $ 2.10 \pm0.06 $ & $ 1.72 \pm0.07 $ & $ 7.6 \pm1.0 $ & $ 31.3  \pm0.6   $ & $22.3 \pm0.3  $ & $50.2  \pm0.6   $ \\
$b_0$ & $ 9.7  \pm0.1  $ & $ 8.3  \pm0.1  $ & $ 5.7 \pm0.2 $ & $-44.0  \pm0.6   $ & $3.6  \pm0.1  $ & $1.0   \pm0.2   $ \\
$b_1$ & $ 6.9  \pm0.1  $ & $-6.7  \pm0.1  $ & $ 2.1 \pm0.2 $ & $ 71.0  \pm0.8   $ & $2.1  \pm0.1  $ & $1.2   \pm0.2   $ \\
$b_2$ & $ 0.652\pm0.009$ & $-0.321\pm0.001$ & $ 1.2 \pm0.1 $ & $ 0.0178\pm0.0002$ & $0.76 \pm0.01 $ & $0.9   \pm0.1   $ \\
$b_3$ & $ 0.46 \pm0.04 $ & $34.5  \pm0.1  $ & $-0.8 \pm0.2 $ & $ 45.3  \pm0.6   $ & $8.5  \pm0.2  $ & $0.1   \pm0.4   $ \\
$d_0$ & $ 1.094\pm0.007$ & $ 0.84 \pm0.01 $ & $ 0.85\pm0.03$ & $  5.953\pm0.071 $ & $2.35 \pm0.02 $ & $6.2   \pm0.1   $ \\
$d_1$ & $ 0.426\pm0.008$ & $-0.22 \pm0.01 $ & $-0.07\pm0.01$ & $ -5.30 \pm0.07  $ & $1.93 \pm0.02 $ & $-5.9  \pm0.1   $ \\
$d_2$ & $ 0.689\pm0.004$ & $ 0.77 \pm0.03 $ & $ 2.5 \pm0.2 $ & $  0.12  \pm0.01 $ & $0.272\pm0.002$ & $0.11  \pm0.03  $ \\
$d_3$ & $ 3.46 \pm0.03 $ & $-0.15 \pm0.07 $ & $-1.98\pm0.05$ & $ -0.5  \pm0.3   $ & $11.0 \pm0.1  $ & $0.3   \pm0.7   $ \\
$\chisq$ & 2144 / 288 & 898 / 288 & 306 / 288 & 303 / 288 & 417 / 288 & 142 / 66 \\
\hline
\end{tabular}
\end{table*}

\begin{figure*}[hbt]
\begin{center}
\includegraphics[width=0.3\textwidth]{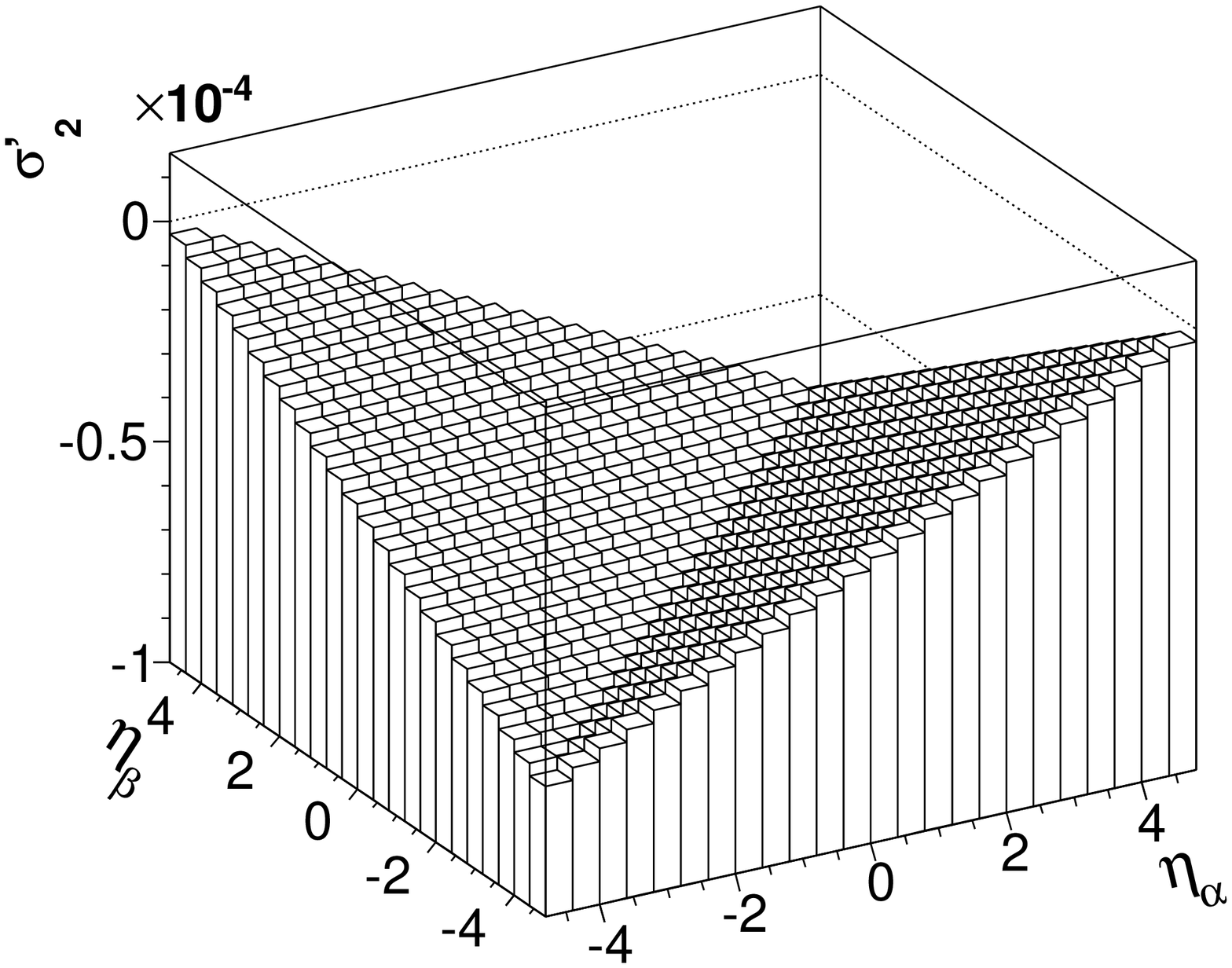}
\includegraphics[width=0.3\textwidth]{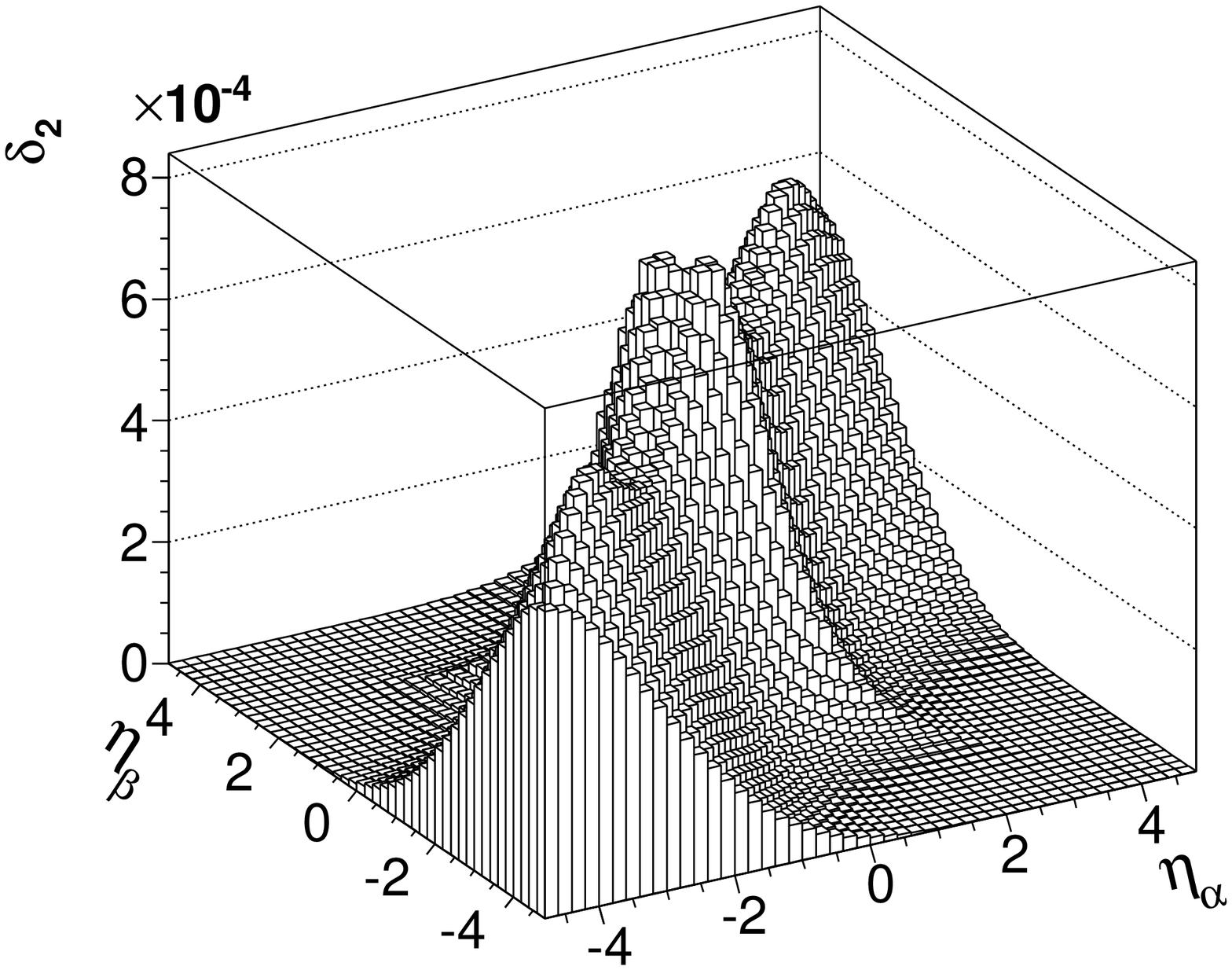}
\includegraphics[width=0.3\textwidth]{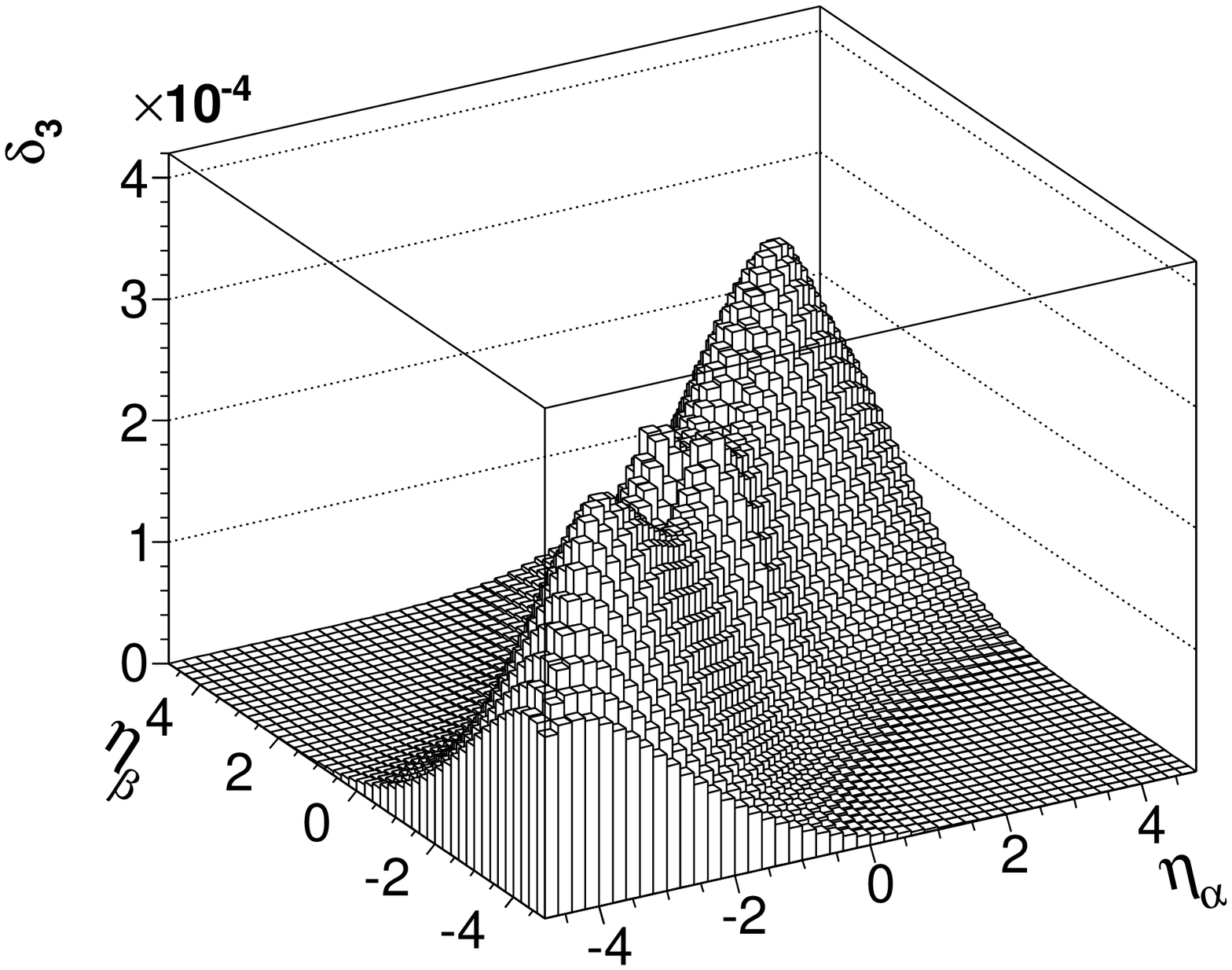}
\end{center}
\caption{Parameterized $\deta$-dependent part of the second-harmonic flow fluctuation $\fl_2(\deta)$ (left), the second-harmonic $\deta$-dependent nonflow correlation $\delta_2(\deta)$ (middle) and the third-harmonic $\deta$-dependent nonflow correlation $\delta_3(\deta)$ (right) in \ampt\ 200 GeV Au+Au collisions with \bperi\ (approximately 10-60\% centrality), plotted as a function of $(\etaa,\etab)$.}
\label{fig:d}
\end{figure*}

The relatively large $\chisq$ indicates that the fit function of Eq.~(\ref{eq:nf_fit2D}) is an imperfect description of the $\deta$-dependent correlations in \ampt. The imperfection of the fit function is amplified by the small statistical errors on $\Delta\Vn{}{2}$. While a fit function better describing the data is desirable, Eq.~(\ref{eq:nf_fit2D}) is adequate for the purpose of illustrating our method, because the deviations of the fit function from the data points are small compared to the magnitudes of the data points themselves, as shown in the middle panels of Fig.~\ref{fig:dV2}.

From the fitted function Eq.~(\ref{eq:nf_fit2D}) to the two-particle cumulant difference, we deduce the $\deta$-dependent correlation itself (plus any $\deta$ dependent fluctuation effect) as 
\be\delta(\deta)+\fl(\deta)=[a(\mneta)+b(\mneta)\deta]e^{-\deta/d(\mneta)}\,.\label{eq:nf}\ee
Mathematically there is an arbitrariness in this procedure--an arbitrary function in $(\etaa,\etab)$ is allowed in the deduced $\deta$-dependent correlations, which would be cancelled in the two-particle cumulant difference in Eq.~(\ref{eq:nf_fit2D}). Physically, however, $\delta(\deta)$ should depend on $\deta$ and any $\deta$-independent function only in $(\etaa,\etab)$ is included in $\Vt{}{2}$. Moreover, nonflow, corresponding to the $\deta$-dependent correlations, should vanish with very large $\eta$-gap-- In Eq.~(\ref{eq:nf}) we have defined zero nonflow at $\deta=\infty$, but nonflow is practically zero with $\eta$-gap between the forward and backward beam rapidities.

Figure~\ref{fig:d} shows the parameterized second- and third-harmonic $\deta$-dependent correlations, presumably the nonflow $\delta_2(\deta)$ and $\delta_3(\deta)$, as function of $(\etaa,\etab)$, in the middle and right panels, respectively. Again, the parameterization of Eq.~(\ref{eq:nf}) is obtained from the restricted region of $\etaa<0$ and $\etaa<\etab$, and is extended to the full $(\etaa,\etab)$ space by symmetry. The variation of the $\deta$-dependent correlation amplitude along the diagonal is the result of the cosine variation in the fit parameter $a(\mneta)$. The peaks of $\delta$ at $(\etaa,\etab)=(0,0)$ probably indicate excessive nonflow correlation pairs at mid-rapidity. 
The reason for the shoulders may be due to an interplay between the rapidity distributions of the same-side correlation strength and the particle $dN/d\eta$ of the underlying background event; the $\delta$ is a per-pair quantity, normalized by the total number of particle pairs between a given pair of $\eta$ bins.

The modest variation in the fit parameter $d(\mneta)$ shown in the right panels of Fig.~\ref{fig:dV2} does not show up clearly in the presentation of Fig.~\ref{fig:d}. The fit parameter $b$ is to take into account the apparent artifact of the $\deta\sim0$ dip in \ampt, and is not very important to our study.

\subsection{Factorization of the $\deta$-independent correlation\label{sec:sanity}}

\begin{figure*}[hbt]
\begin{center}
\includegraphics[width=0.3\textwidth]{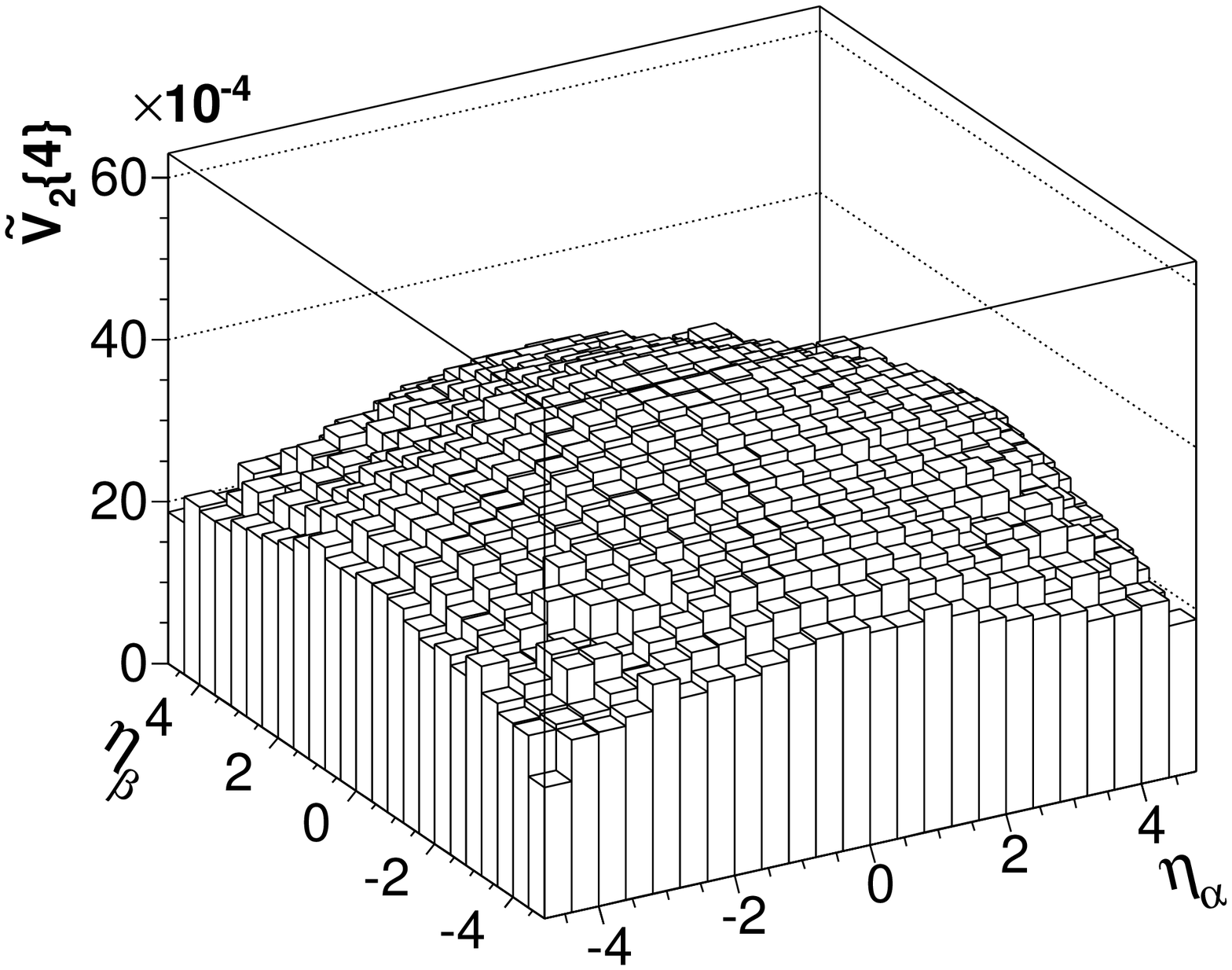}
\includegraphics[width=0.3\textwidth]{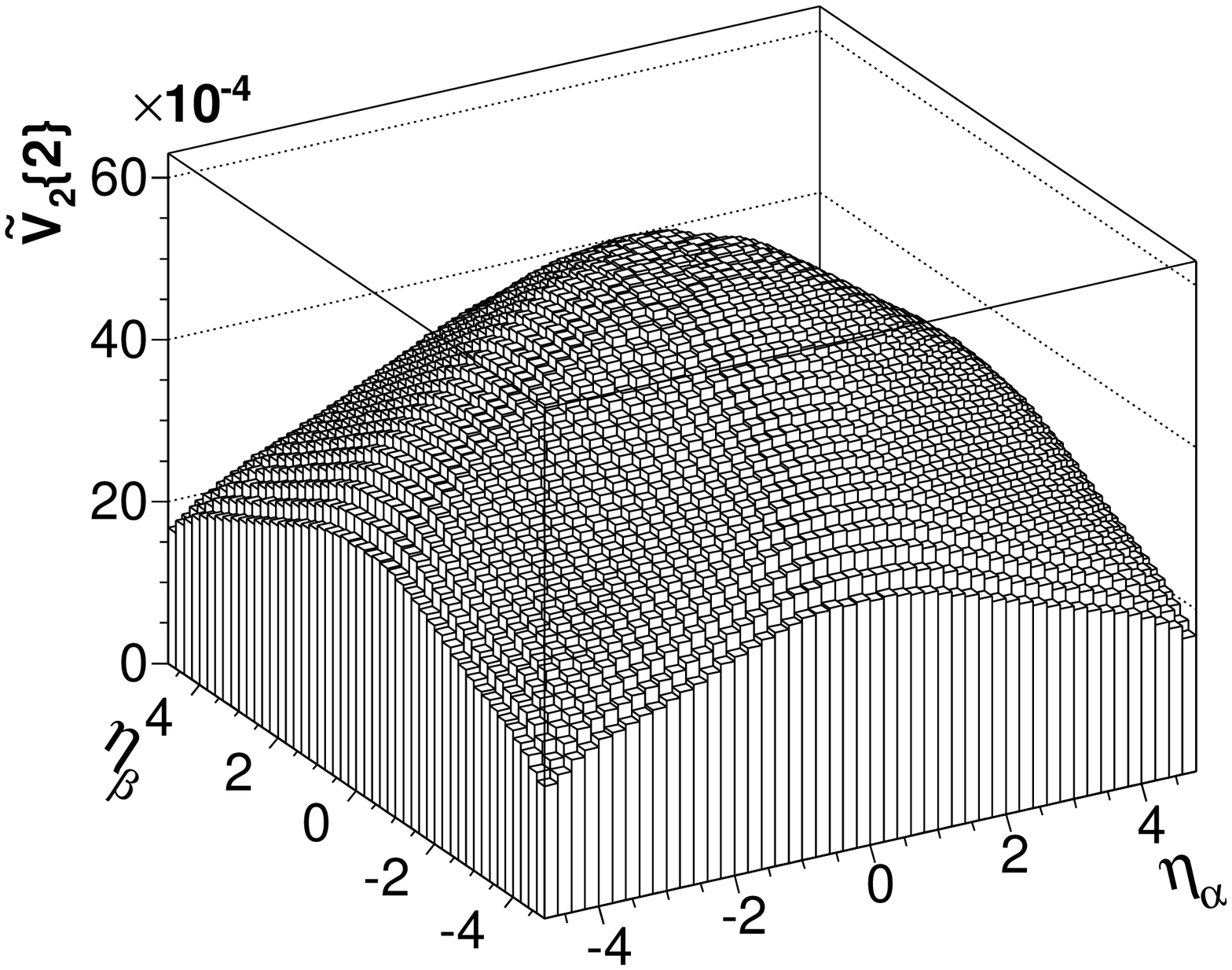}
\includegraphics[width=0.3\textwidth]{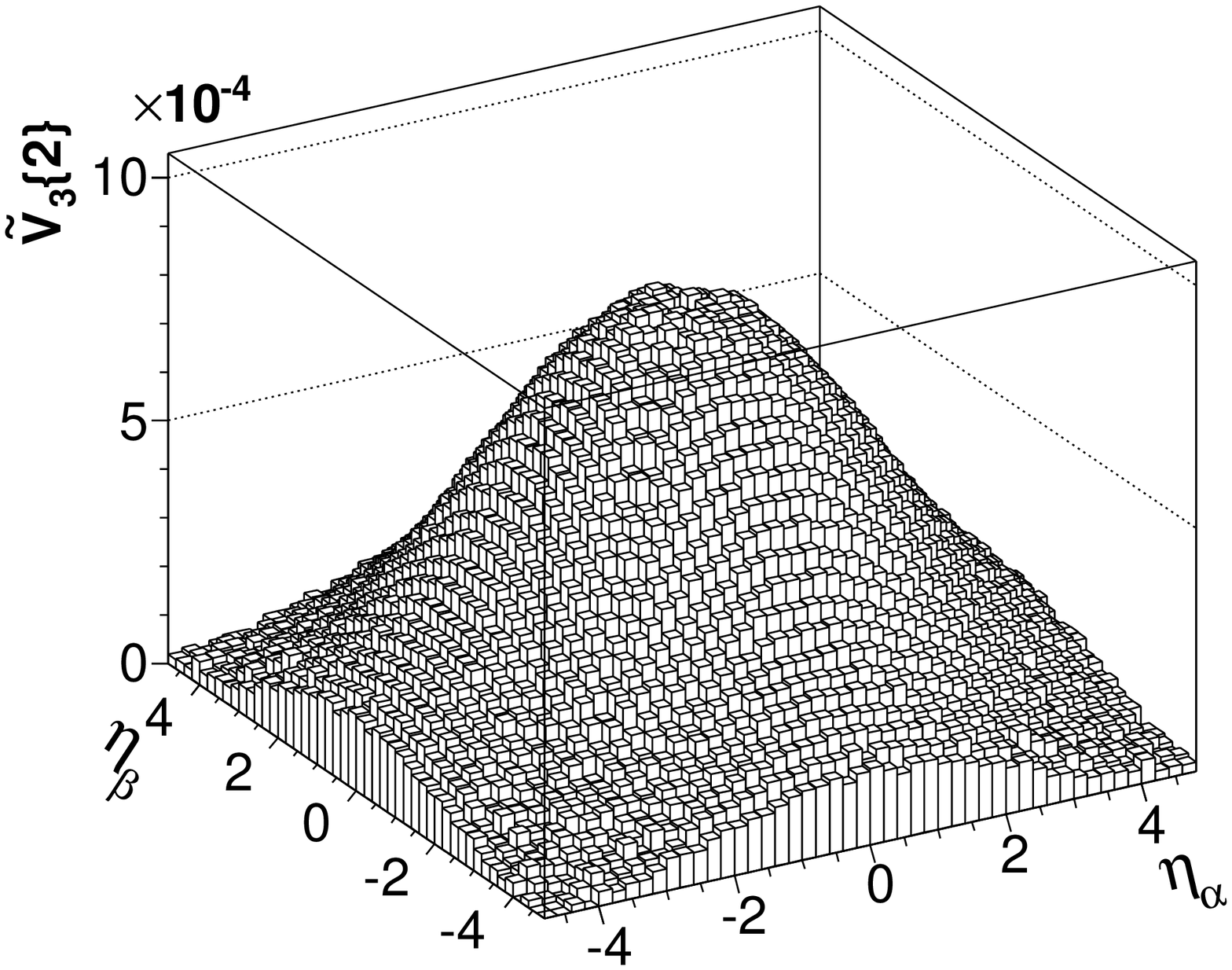}
\end{center}
\caption{The second-harmonic four-particle cumulant subtracted by the parameterized $\deta$-dependent flow fluctuation effect: $\Vt{2}{4}$ (left panel). The two-particle cumulants subtracted by the parameterized $\deta$-dependent nonflow correlation (and $\deta$-dependent flow fluctuation effect): second-harmonic $\Vt{2}{2}$ (middle panel) and third-harmonic $\Vt{3}{2}$ (right panel). The results are plotted versus $(\etaa,\etab)$. Data are from \ampt\ simulation of 200 GeV Au+Au collisions with \bperi\ (approximately 10-60\% centrality).}
\label{fig:vv}
\end{figure*}

\begin{figure*}[hbt]
\begin{center}
\includegraphics[width=0.3\textwidth]{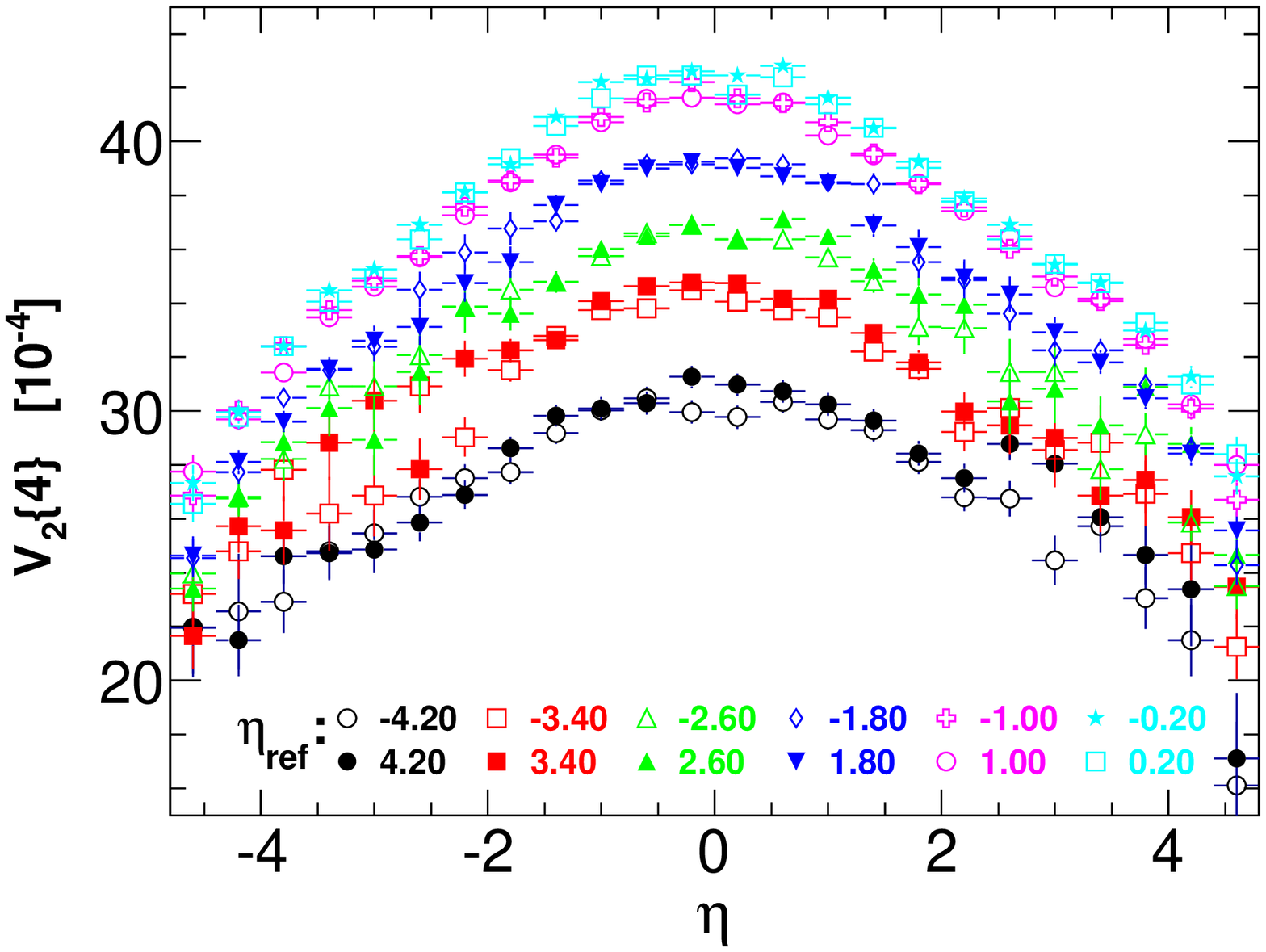}
\includegraphics[width=0.3\textwidth]{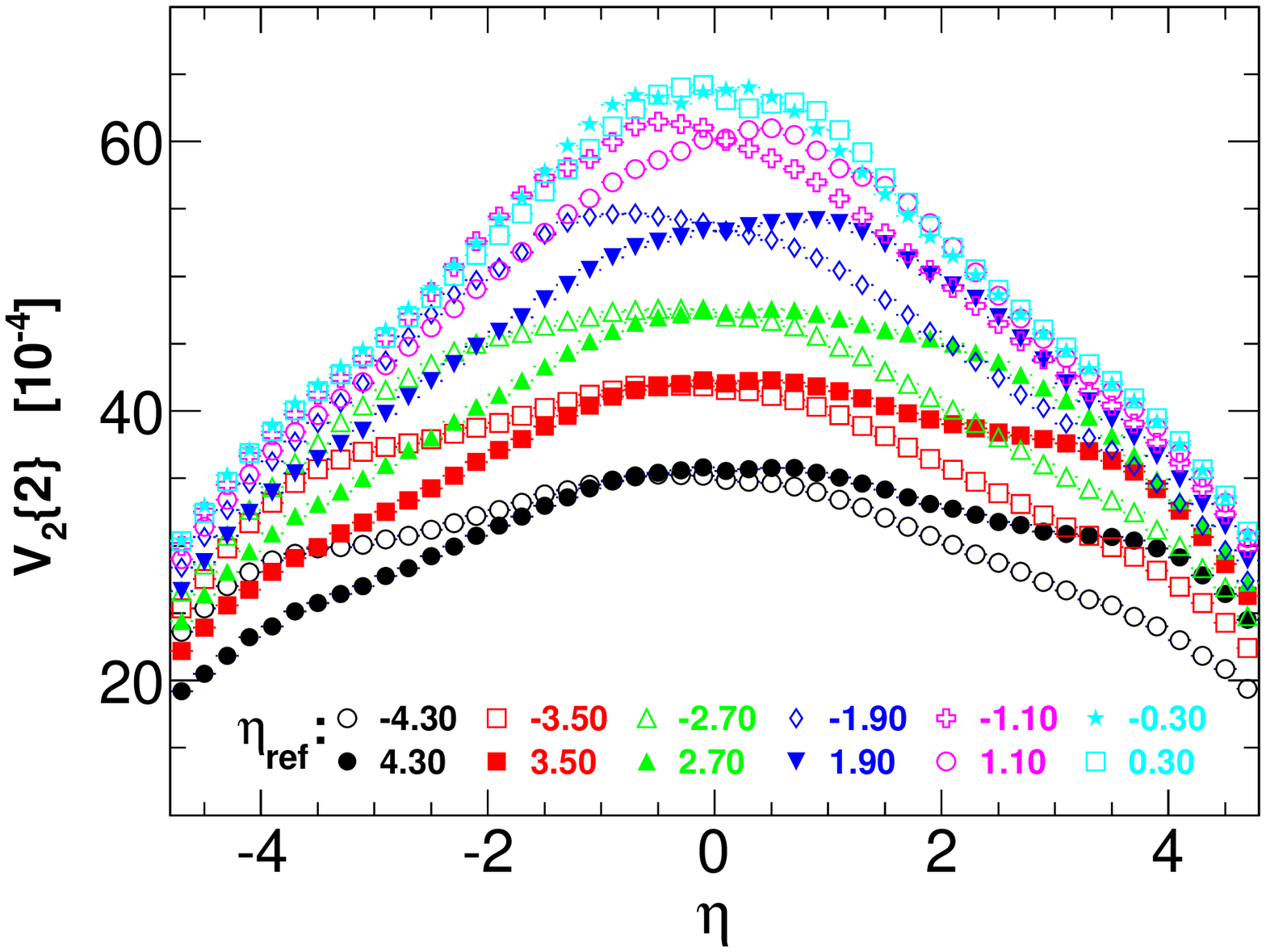}
\includegraphics[width=0.3\textwidth]{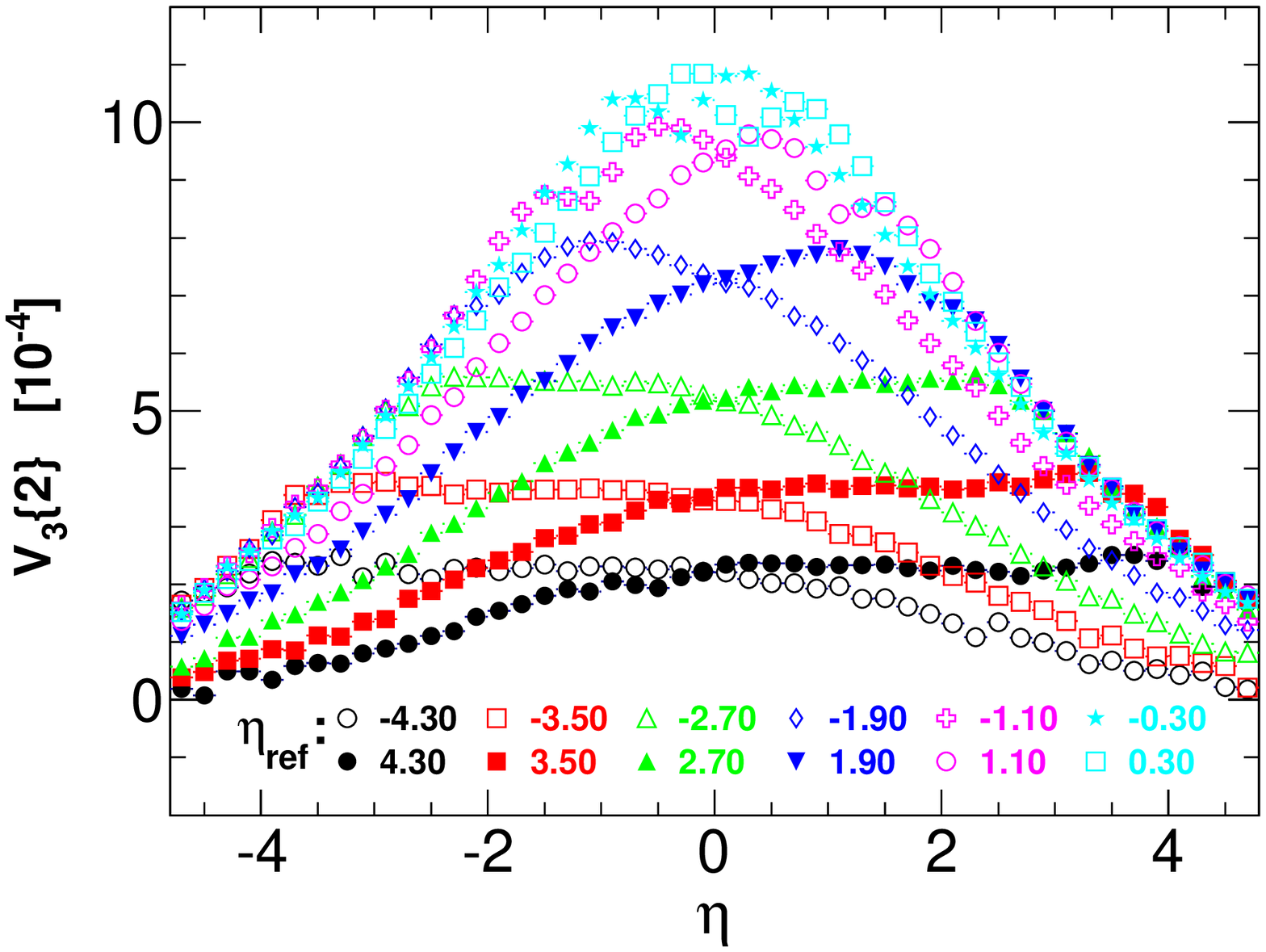}
\includegraphics[width=0.3\textwidth]{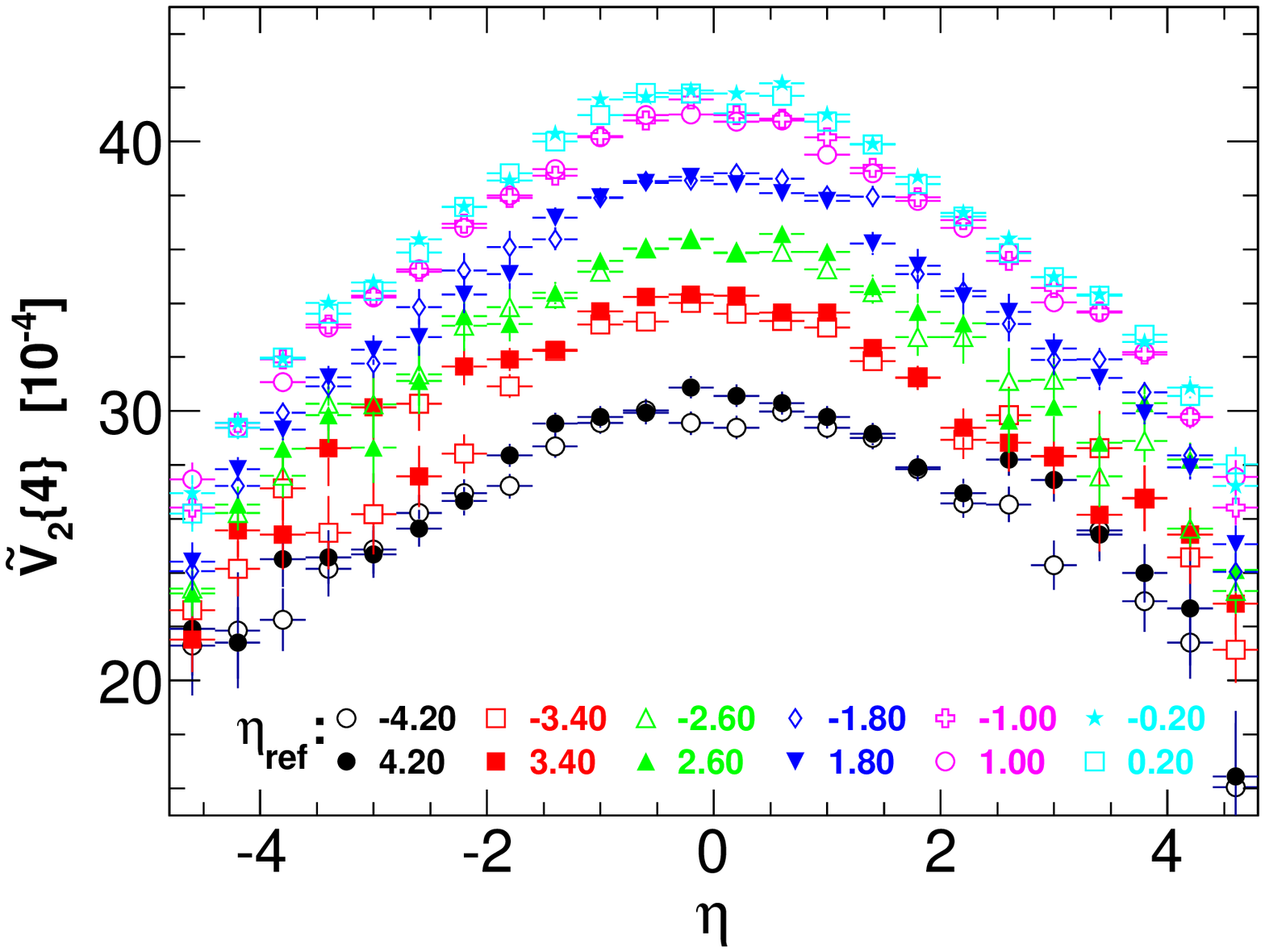}
\includegraphics[width=0.3\textwidth]{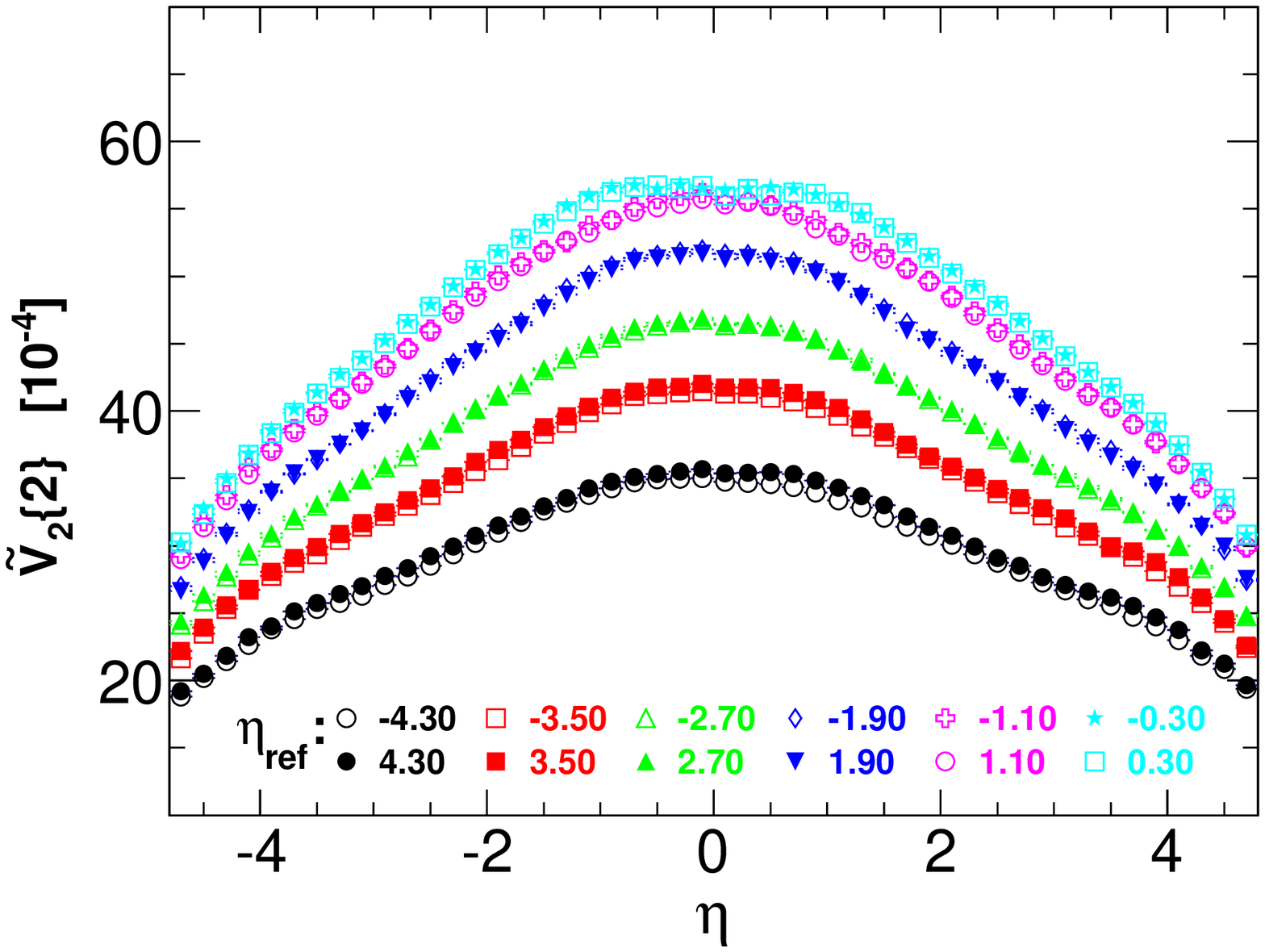}
\includegraphics[width=0.3\textwidth]{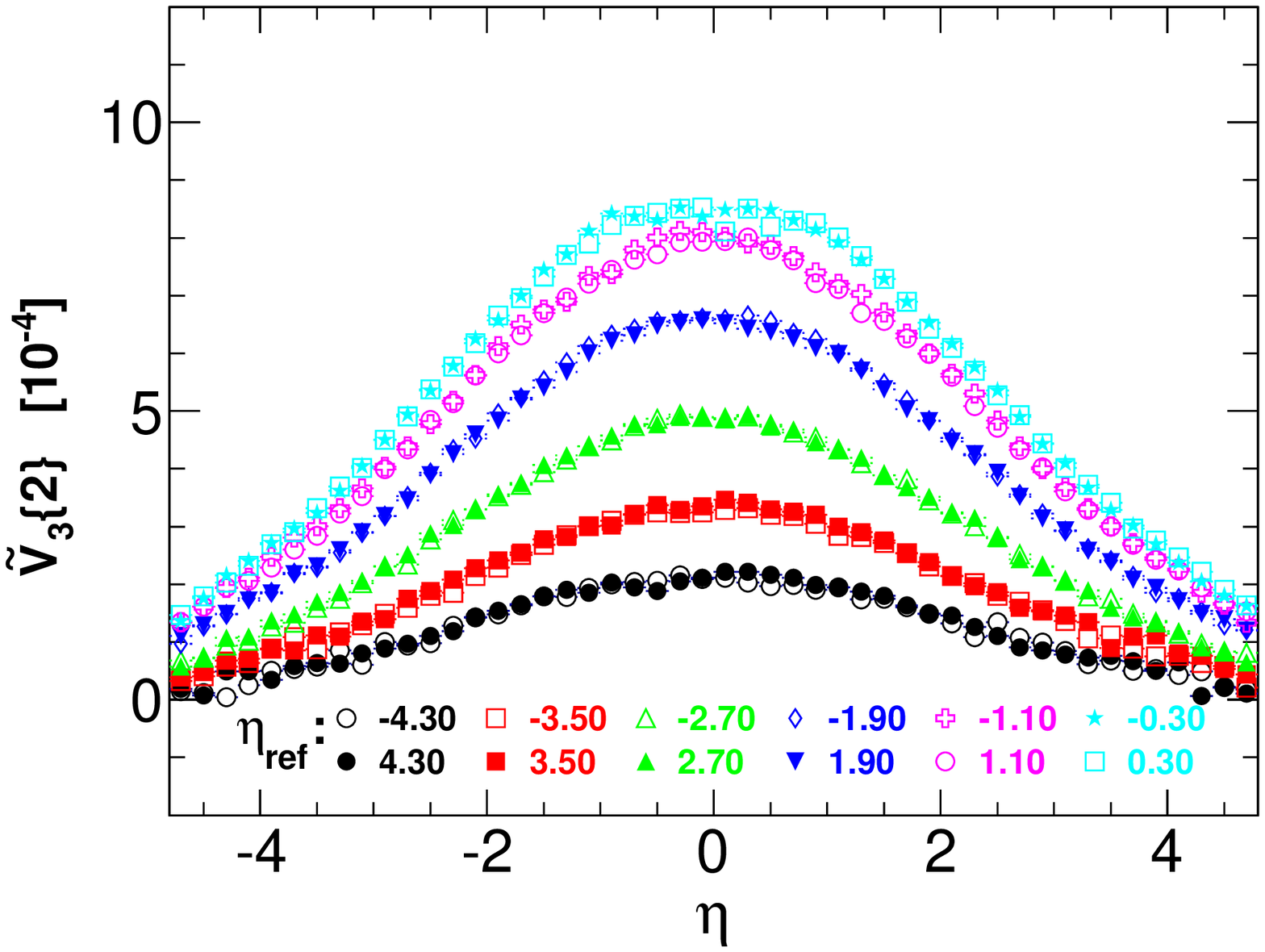}
\end{center}
\caption{(Color online) Slices of the 2D plots of $V_n$ in Fig.~\ref{fig:V} (upper panels) and those of $\tV_n$ in Fig.~\ref{fig:vv} (lower panels) at selected $\etaref$ values. Data are from \ampt\ simulation of 200 GeV Au+Au collisions with \bperi\ (approximately 10-60\% centrality).}
\label{fig:slice}
\end{figure*}

The $\deta$-independent cumulants, $\Vt{}{2}$ and $\Vt{}{4}$, are obtained by subtracting the parameterized $\deta$-dependent correlations from the raw two- and four-particle cumulants $\Vn{}{2}$ and $\Vn{}{4}$ in Fig.~\ref{fig:V}. As mentioned in Sec.~\ref{method}, because flow is a single particle property, the $\deta$-dependent component can be attributed to nonflow. Consequently the $\deta$-independent components $\Vt{}{2}$ and $\Vt{}{4}$ are dominated by flow.  We note that $\Vt{}{2}$ and $\Vt{}{4}$ include effects of the $\deta$-independent flow fluctuation. The $\Vt{}{2}$, in addition, may include any $\deta$-independent nonflow contributions. The $\deta$-independent nonflow may arise from back-to-back inter-jet correlations, however in the low $\pt$ region studied in this work it should be small. In the following discussion, for simplicity, we refer to $\Vt{}{2}$ and $\Vt{}{4}$ as ``flow'' cumulants or ``flow'' and the $\deta$-dependent $\delta$ as ``nonflow,'' unless otherwise noted. It is important to keep in mind that the former (that we called flow) may contain away-side nonflow while the latter (that we called nonflow) may not contain the away-side nonflow. Our nonflow may be primarily caused by near-side correlations that depends on $\deta$.

The results of $\Vt{}{2}$ and $\Vt{}{4}$ are shown in Fig.~\ref{fig:vv}. 
In this section we investigate the properties of the obtained $\Vt{}{2}$ and $\Vt{}{4}$ to check whether they resemble what we would expect for flow.
Figure~\ref{fig:slice} lower panels show slices of the obtained $\deta$-independent correlations in Fig.~\ref{fig:vv} at selected values for one of the $\eta$ bins (we call this $\etaref$). The data are plotted versus the value of the other $\eta$ bin (we simply call $\eta$). The data points follow a smooth pattern. The curves for symmetric $\etaref$ values are consistent with each other as flow should be. The magnitudes of the data points depend on the $\etaref$ value because the flow is not constant as a function of $\eta$. For comparison, the slices of the 2D plots of the raw cumulants from Fig.~\ref{fig:V} are shown in the upper panels of Fig.~\ref{fig:slice}. The data are not symmetric for symmetric $\etaref$ values. The peaks (bumps) are due to nonflow contributions that are strongest at small $\deta$. 

On the other hand, the slices of the four-particle cumulants, with and without subtracting the $\deta$-dependent fluctuation effect, are similar as shown in the left panels of Fig.~\ref{fig:slice}. This is because the $\deta$-dependent part of flow fluctuation is small and can be safely neglected (see Fig.~\ref{fig:d} left panel).

\begin{figure*}[hbt]
\begin{center}
\includegraphics[width=0.3\textwidth]{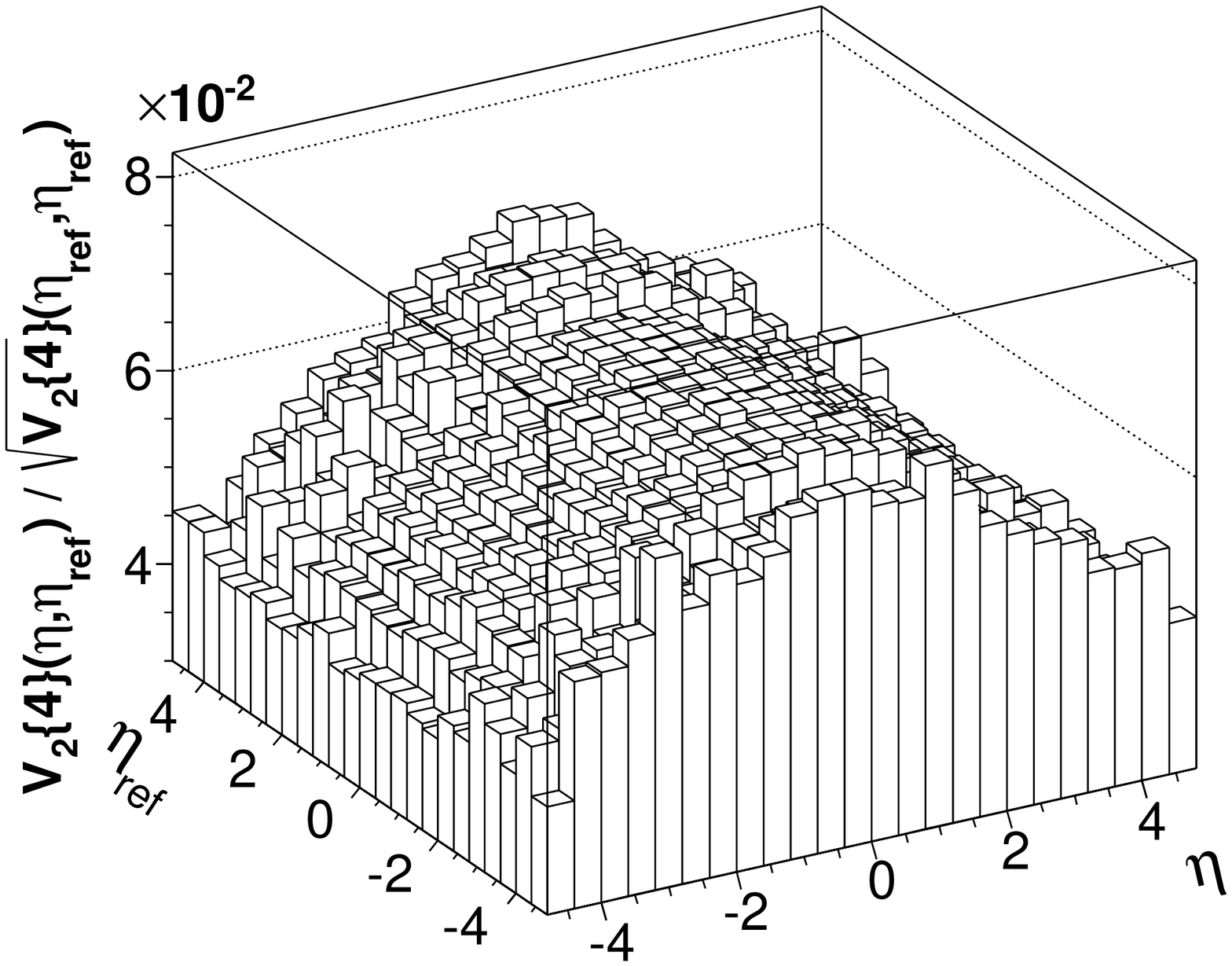}
\includegraphics[width=0.3\textwidth]{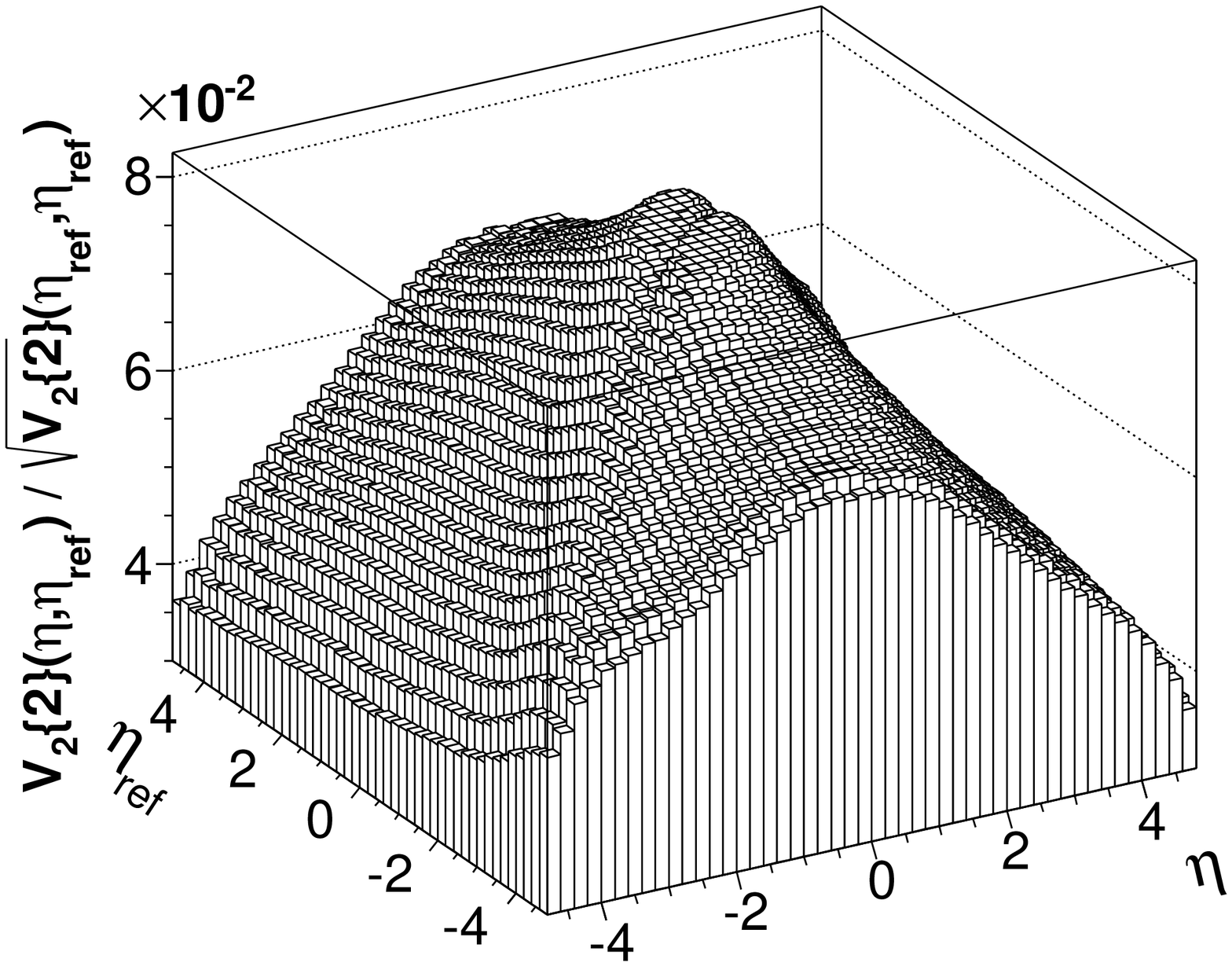}
\includegraphics[width=0.3\textwidth]{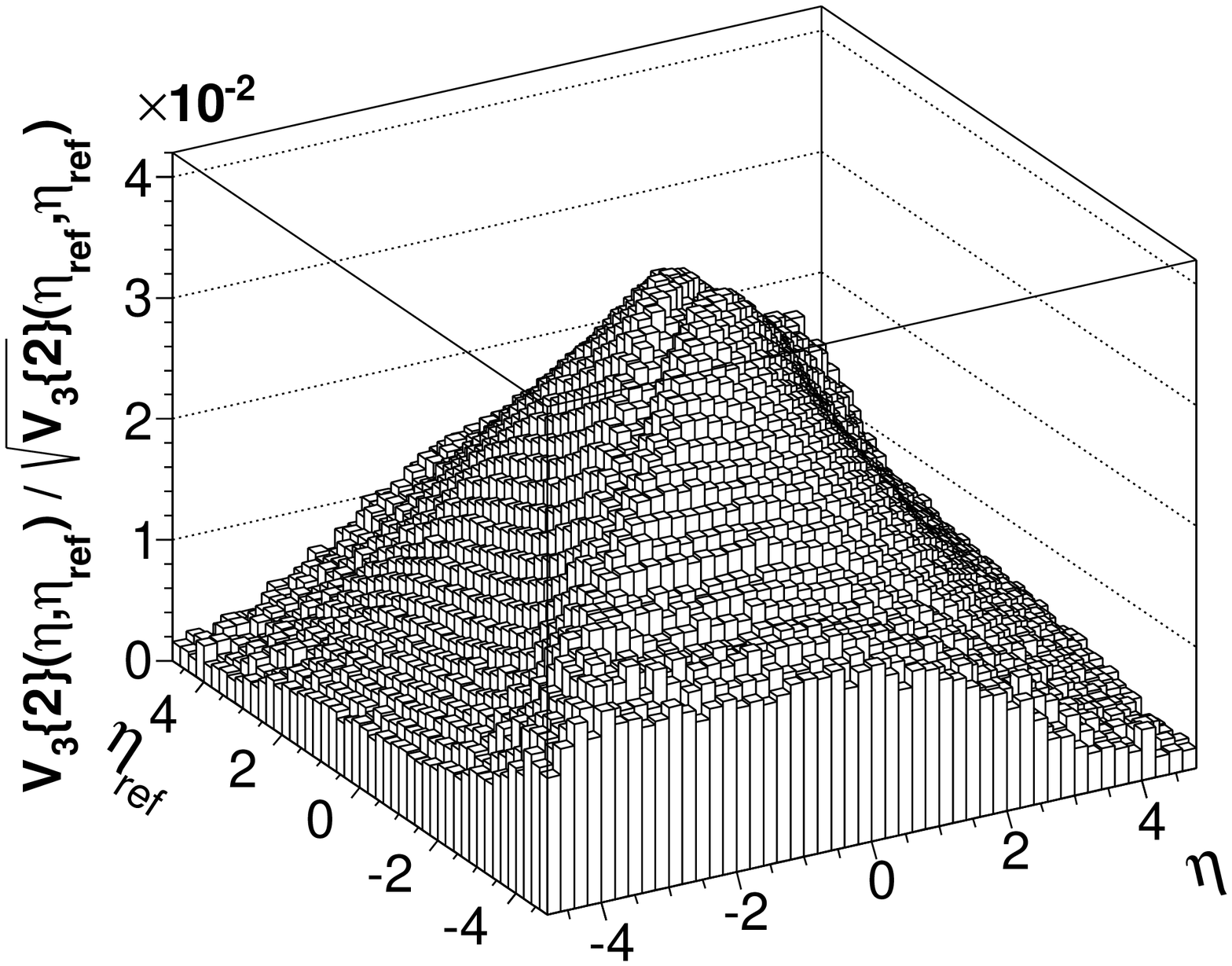}
\includegraphics[width=0.3\textwidth]{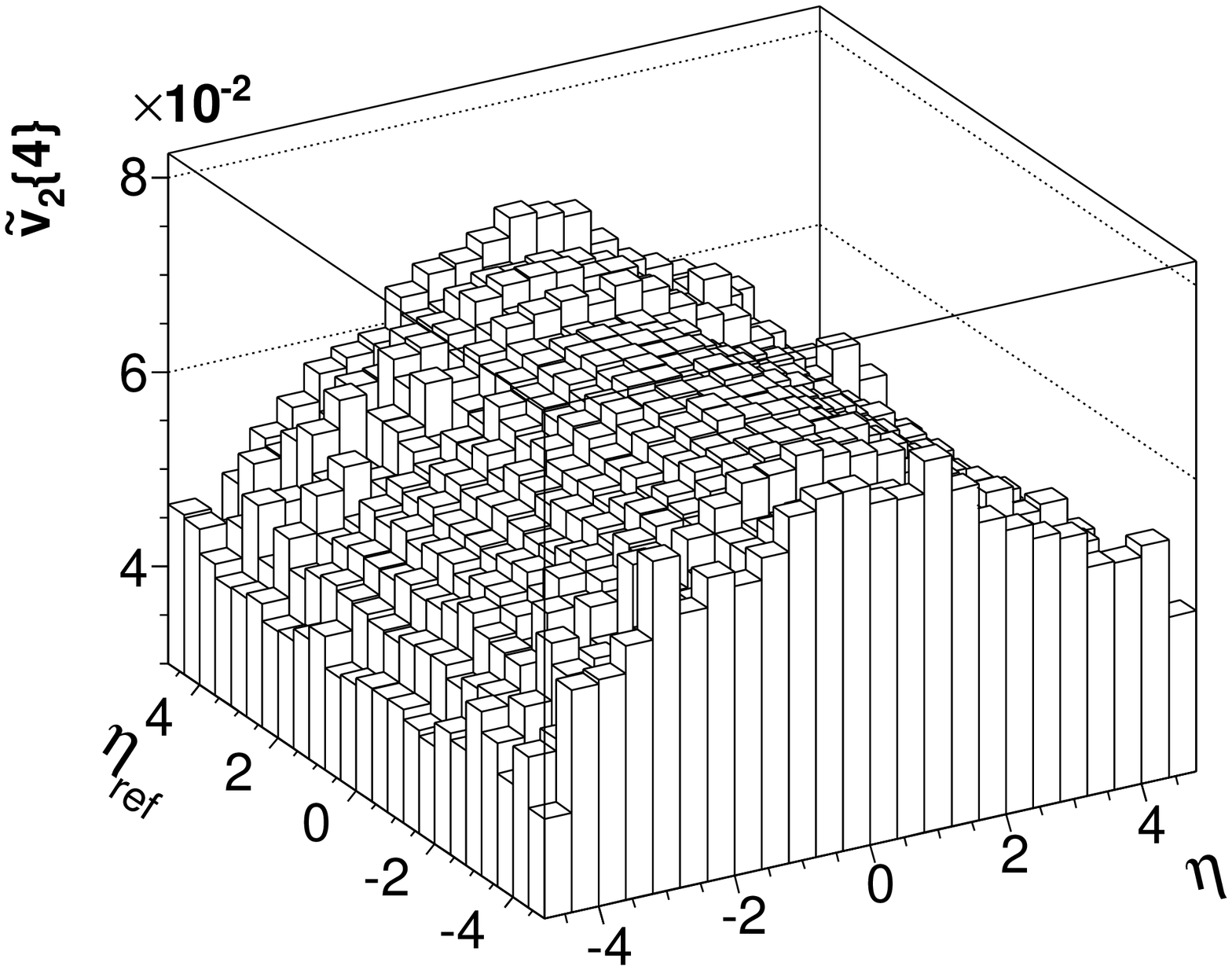}
\includegraphics[width=0.3\textwidth]{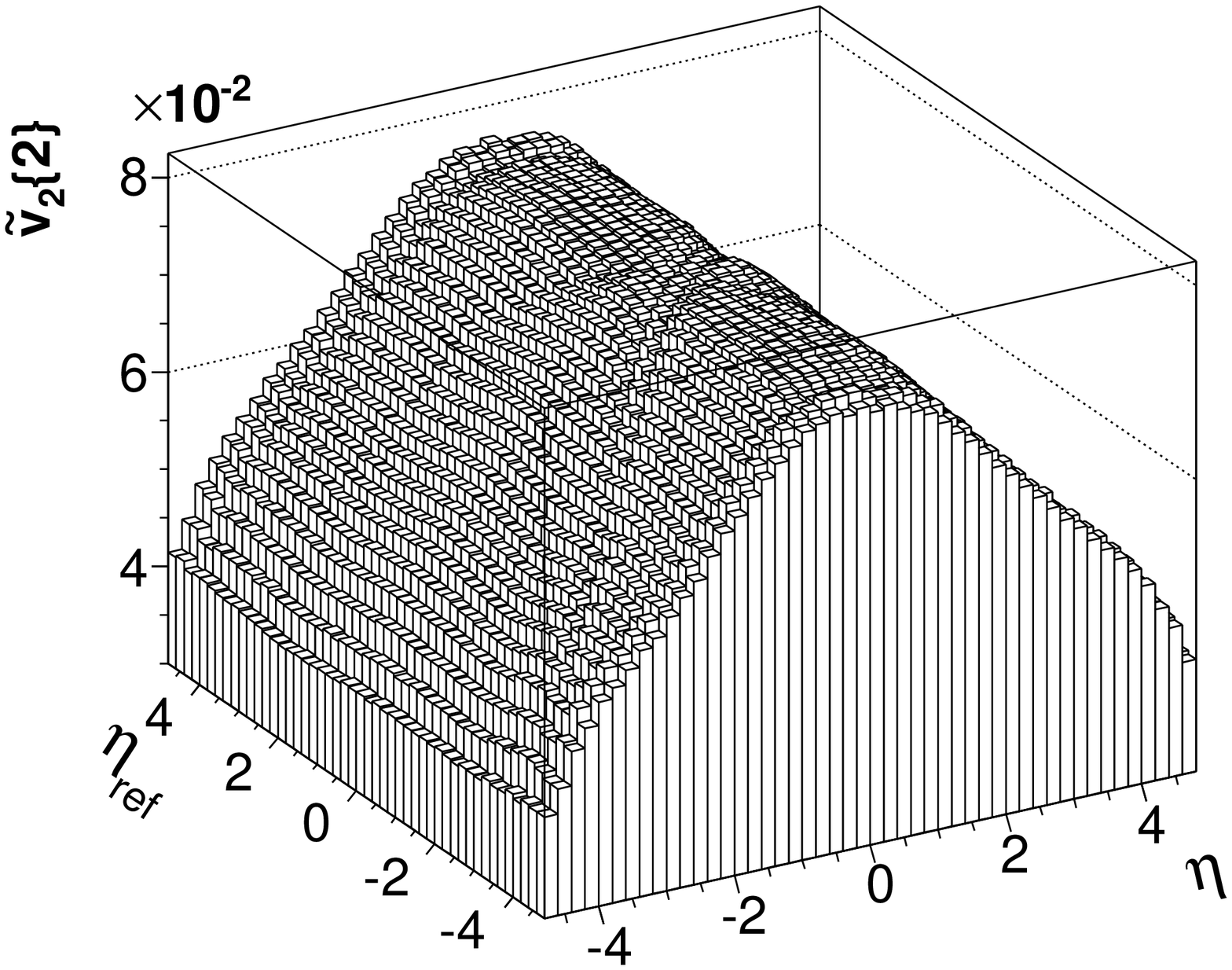}
\includegraphics[width=0.3\textwidth]{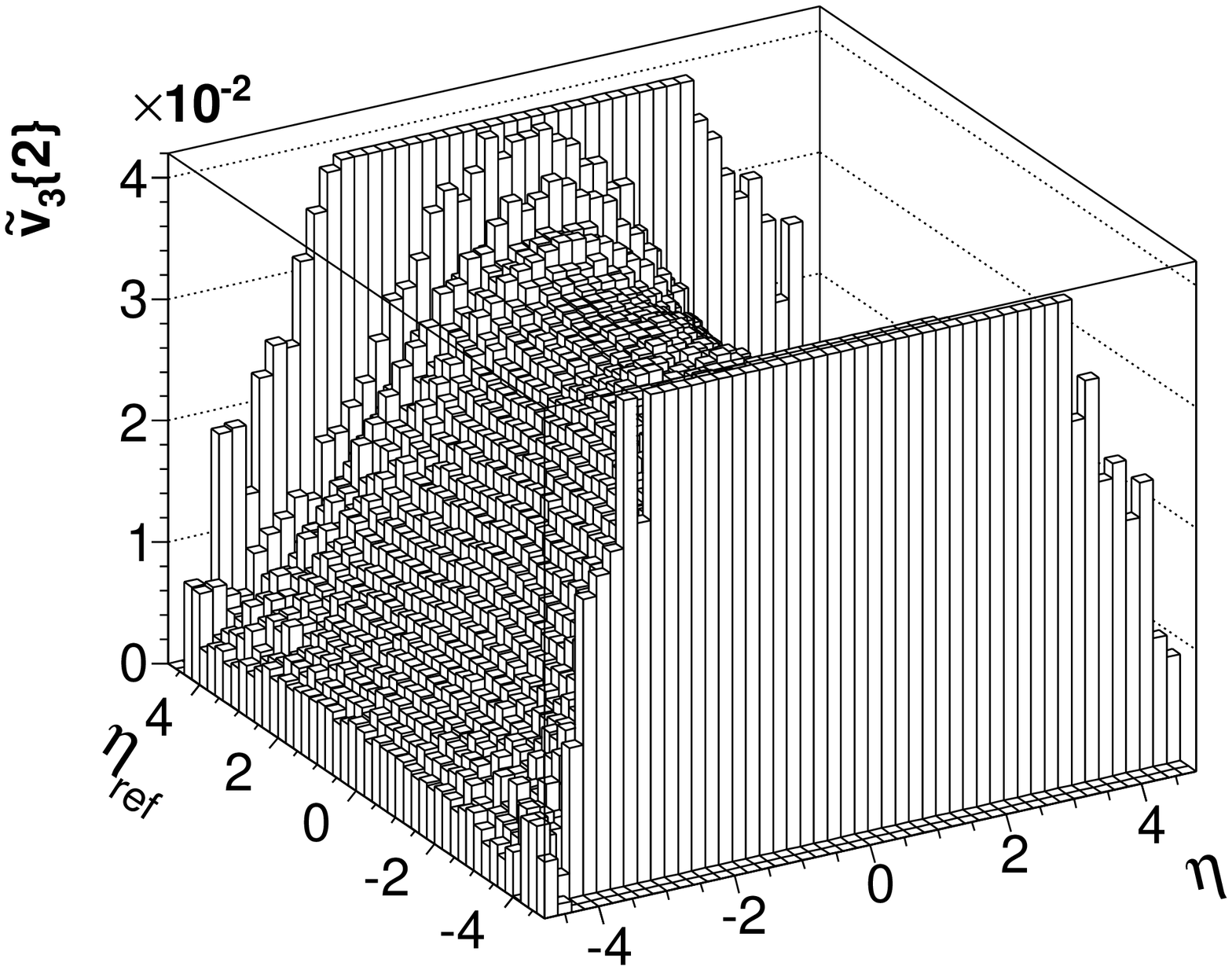}
\end{center}
\caption{Study of factorization. The upper panels show $V_n(\eta,\etaref)/\sqrt{V_n(\etaref,\etaref)}$ with the raw cumulants $V_n$ from Fig.~\ref{fig:V}. The lower panels show $\tv_n(\eta)=\tV_n(\eta,\etaref)/\tv_n(\etaref)$ with the $\tV_n$ results in Fig.~\ref{fig:vv}. The lower panel results are $\etaref$ independent, indicating factorization of the decomposed $\deta$-independent two-particle cumulant $\tV_n(\eta,\etaref)$. Data are from \ampt\ simulation of 200 GeV Au+Au collisions with \bperi\ (approximately 10-60\% centrality).}
\label{fig:fact}
\end{figure*}

\begin{figure*}[hbt]
\begin{center}
\includegraphics[width=0.3\textwidth]{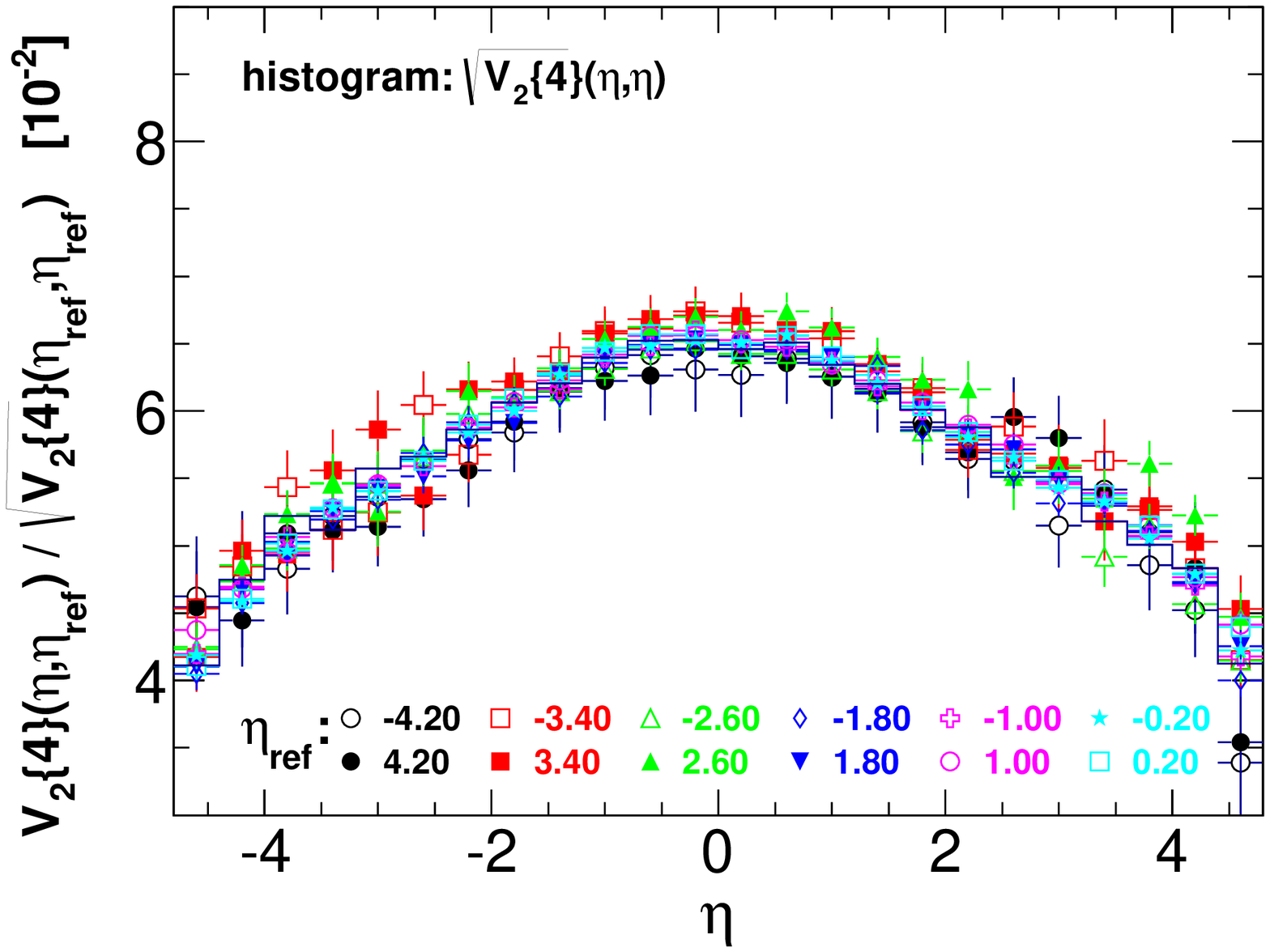}
\includegraphics[width=0.3\textwidth]{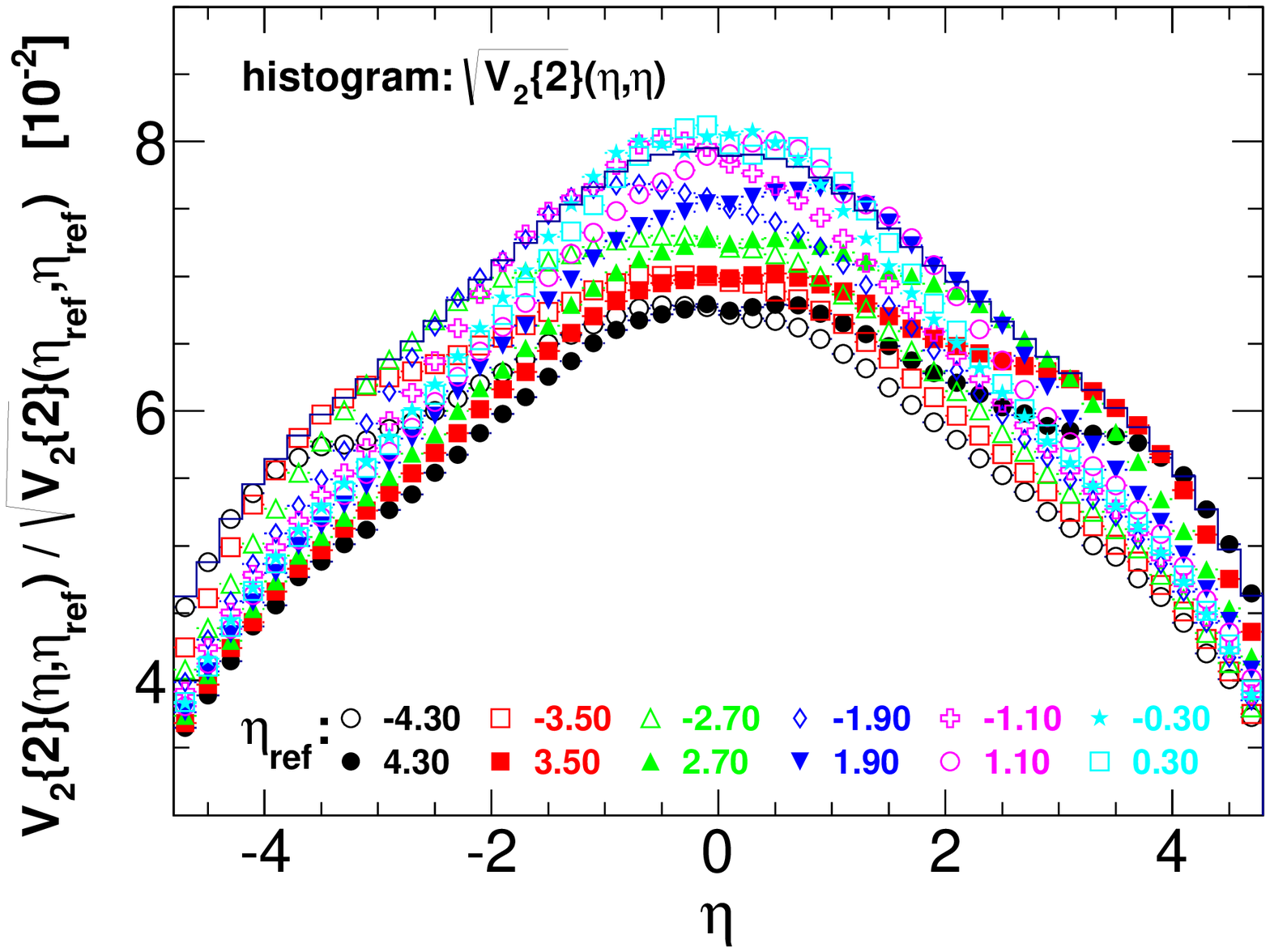}
\includegraphics[width=0.3\textwidth]{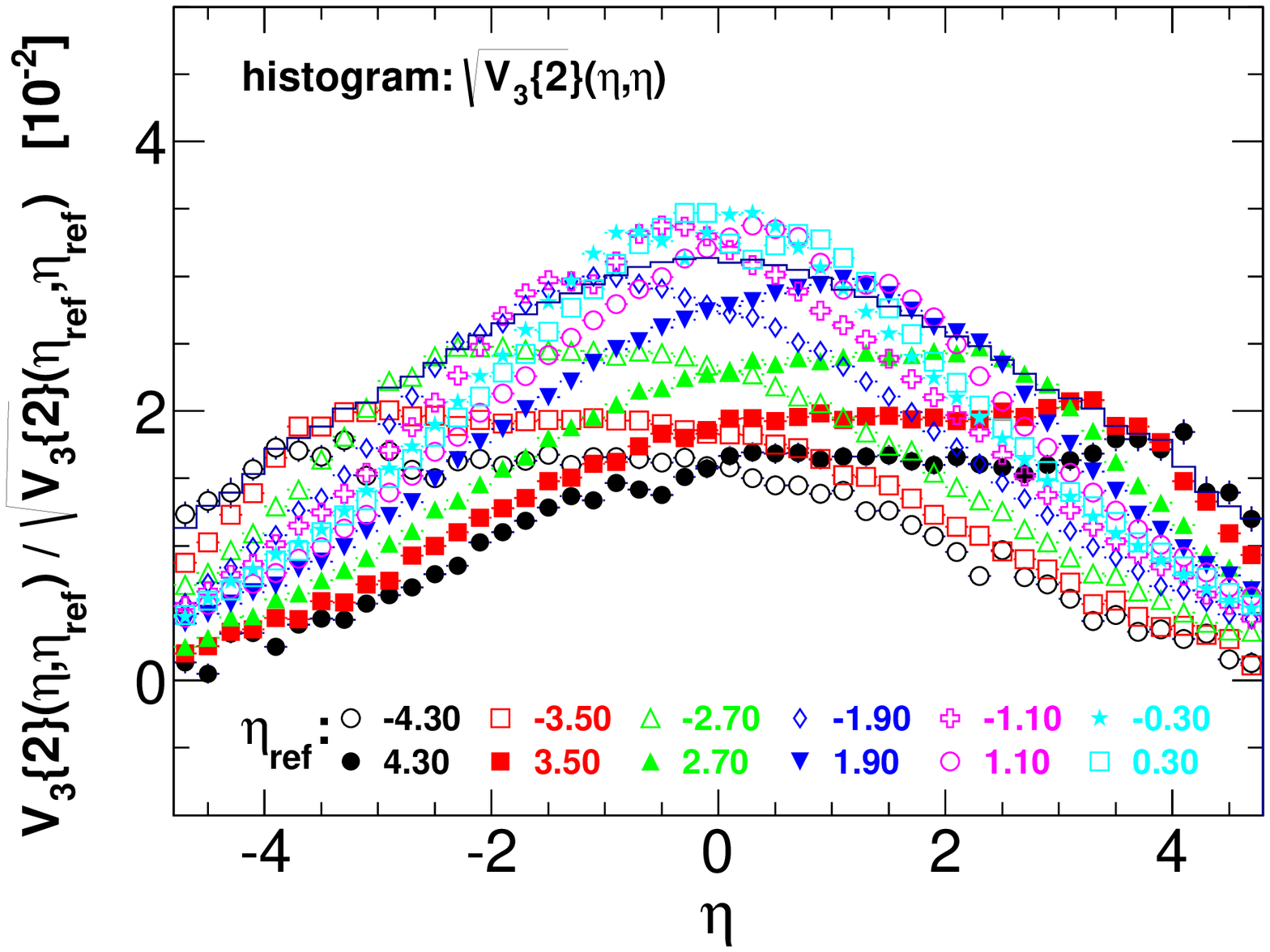}
\includegraphics[width=0.3\textwidth]{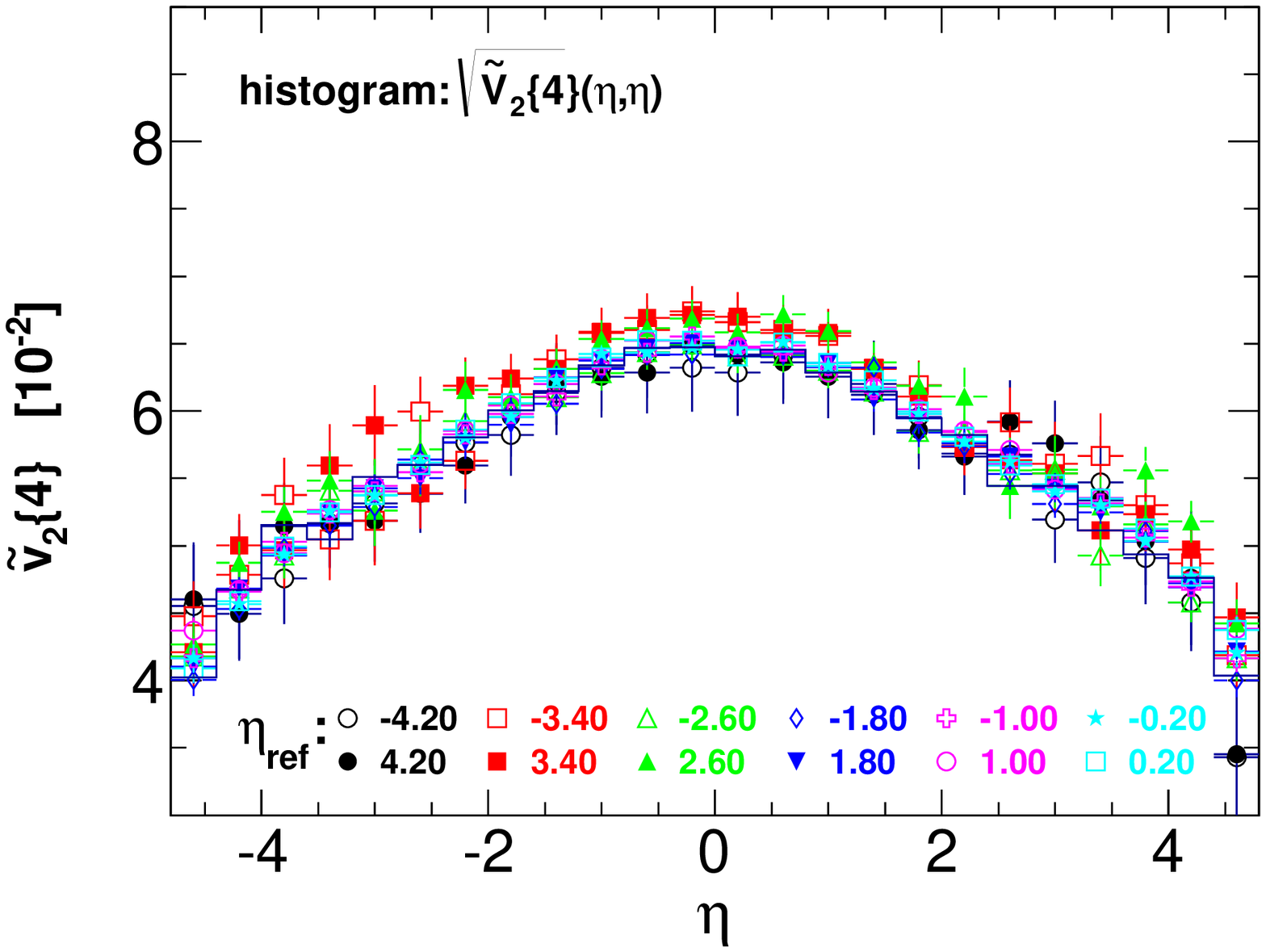}
\includegraphics[width=0.3\textwidth]{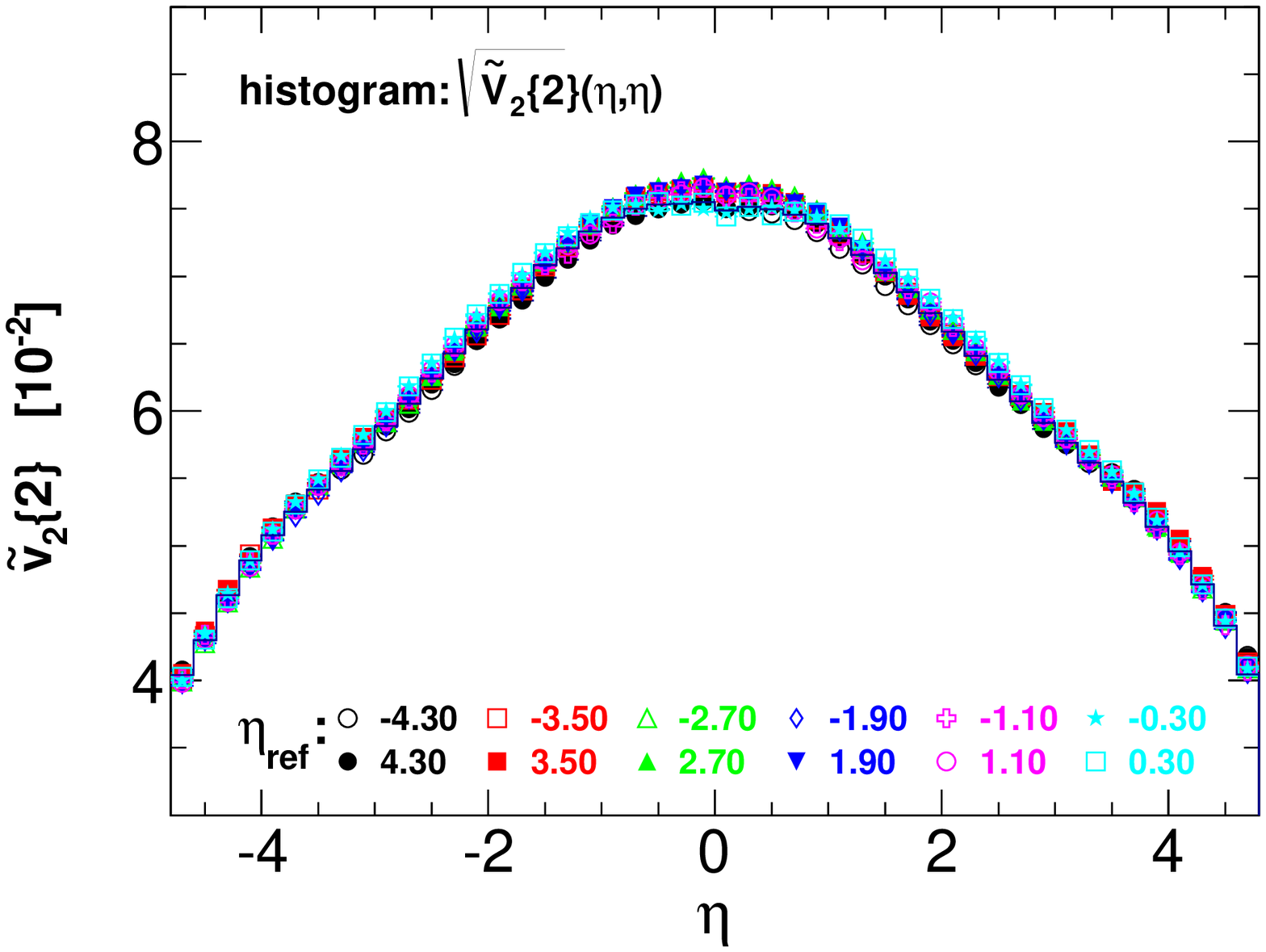}
\includegraphics[width=0.3\textwidth]{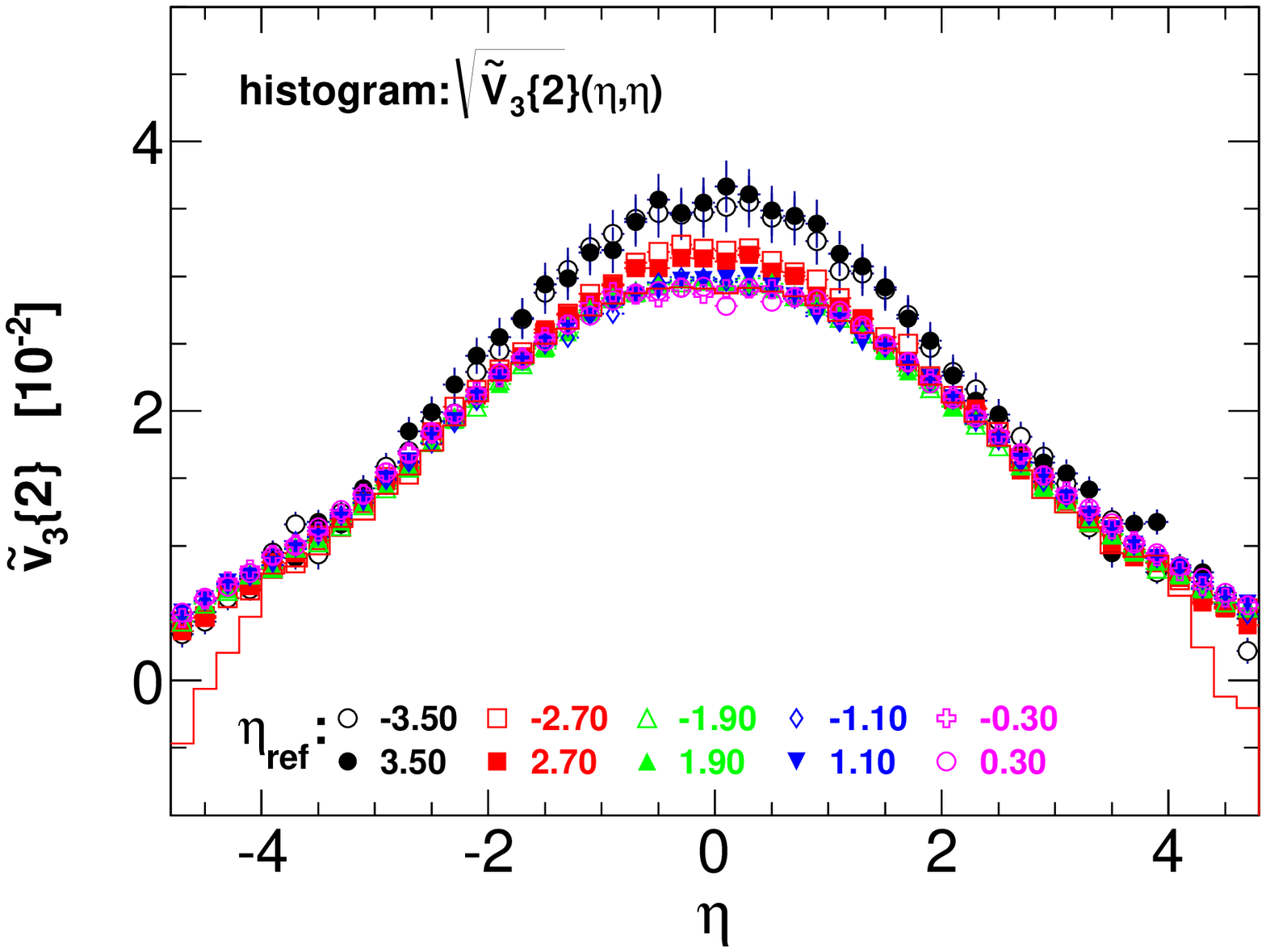}
\end{center}
\caption{(Color online) Slices of plots in Fig.~\ref{fig:fact} at selected $\etaref$, showing more quantitatively factorization in the decomposed $\deta$-independent two-particle cumulant: $\tv_n=\tV_n(\eta,\etaref)/\tv_n(\etaref)$ (lower panels) and non-factorization in the raw cumulant measurement $V_n(\eta,\etaref)/\sqrt{V_n(\etaref,\etaref)}$ (upper panels). Data are from \ampt\ simulation of 200 GeV Au+Au collisions with \bperi\ (approximately 10-60\% centrality).}
\label{fig:fact_slice}
\end{figure*}

Next we check factorization of the flow that might be obtained from the diagonal elements in Fig.~\ref{fig:vv} by 
\be\tv(\etaref)=\tV(\etaref,\etaref)\left/\sqrt{|\tV(\etaref,\etaref)|}\right.\,.\label{eq:diag}\ee
Because $\tV(\etaref,\etaref)$ can be negative due to fluctuations, we take the square root of its absolute value but retain its sign following a recipe from ATLAS~\cite{ATLAS}. Note in Eq.~(\ref{eq:diag}) we have suppressed the harmonic number subscript $n$ and the cumulant ranks, such as in $\vt{n}{2}$ and $\vt{n}{4}$. We divide the results in Fig.~\ref{fig:vv} by $\tv(\etaref)$ to obtain $\tv(\eta)$. The results are plotted in Fig.~\ref{fig:fact} (lower panels) as a function of $(\etaref,\eta)$. The results show that $\tv(\eta)$ is independent of $\etaref$, indicating that the nonflow subtracted cumulants are factorized as $\tv(\etaref)\tv(\eta)$. (The bed-headboard slices at large $|\etaref|$ are due to division by a small, close to zero, value of $\tv(\etaref)$, with large statistical errors.) Again for comparison the upper panel of Fig.~\ref{fig:fact} shows the results obtained by the same procedure but using the raw cumulants in Fig.~\ref{fig:V}. Clearly no factorization is observed for the raw two-particle cumulants. On the other hand, the four-particle raw cumulant does show factorization; this is because the $\deta$-dependent fluctuation and nonflow effects are negligible.

Figure~\ref{fig:fact_slice} shows $\etaref$ slices of the 2D plots in Fig.~\ref{fig:fact}. Note we have avoided the problematic bed-headboard regions at large $|\etaref|$ (some effect is still noticeable for $|\etaref|=0.35$). Figure~\ref{fig:fact_slice} demonstrates more quantitatively the factorization of the $\deta$-independent correlations in the lower panels (all data points fall on a common curve) and the non-factorization of the raw two-particle cumulants in the upper panels (``flow'' would depend on the reference particle used to measure it). 

It is important to note that flow factorization is not required by our method. The only requirement of our method is that the average flow (and other $\deta$-independent terms) be symmetric about mid-rapidity, which must be true in symmetric collision systems. For instance, the obtained nonflow subtracted cumulants could have the functional form of $u(\etaa)u(\etab)+w(\etaa)w(\etab)$ which does not generally factorize~\cite{Kikola}. 
Therefore the factorization of the obtained $\deta$-independent correlation is a result of the analysis using our method, which indicates that our method may have succeeded to separate the major parts of flow and nonflow in \ampt. 

To conclude this section, we have shown that the $\deta$-independent correlations in \ampt\ obtained by our method satisfies what we would naively expect for flow--The diagonal element of the $\deta$-independent correlation does not depend on the reference particle $\etaref$ used to measure it. This is so despite that the $\deta$-independent correlations include flow fluctuation effects and possibly $\deta$-independent nonflow contributions. The factorization may indicate that the flow fluctuation is proportional to the average flow and the nonflow effect from back-to-back inter-jet correlations is small compared to flow in the studied low-$\pt$ region.

\subsection{Test with \ampt\ central events\label{sec:central}}

We have been focusing on medium central events from \ampt\ where flow is significant. In this section we check the approximately 10\% most central events from \ampt\ cutting on impact parameter \bcent.

The statistical errors on the central data are noticeably larger. The statistics in $\Vn{}{4}$ are too poor to tell whether there is a significant $\deta$-dependent flow fluctuation effect. However, it seems reasonable to assume the effect is negligible based on the results from the 10-60\% centrality. 

\begin{figure*}[hbt]
\begin{center}
\includegraphics[width=0.3\textwidth]{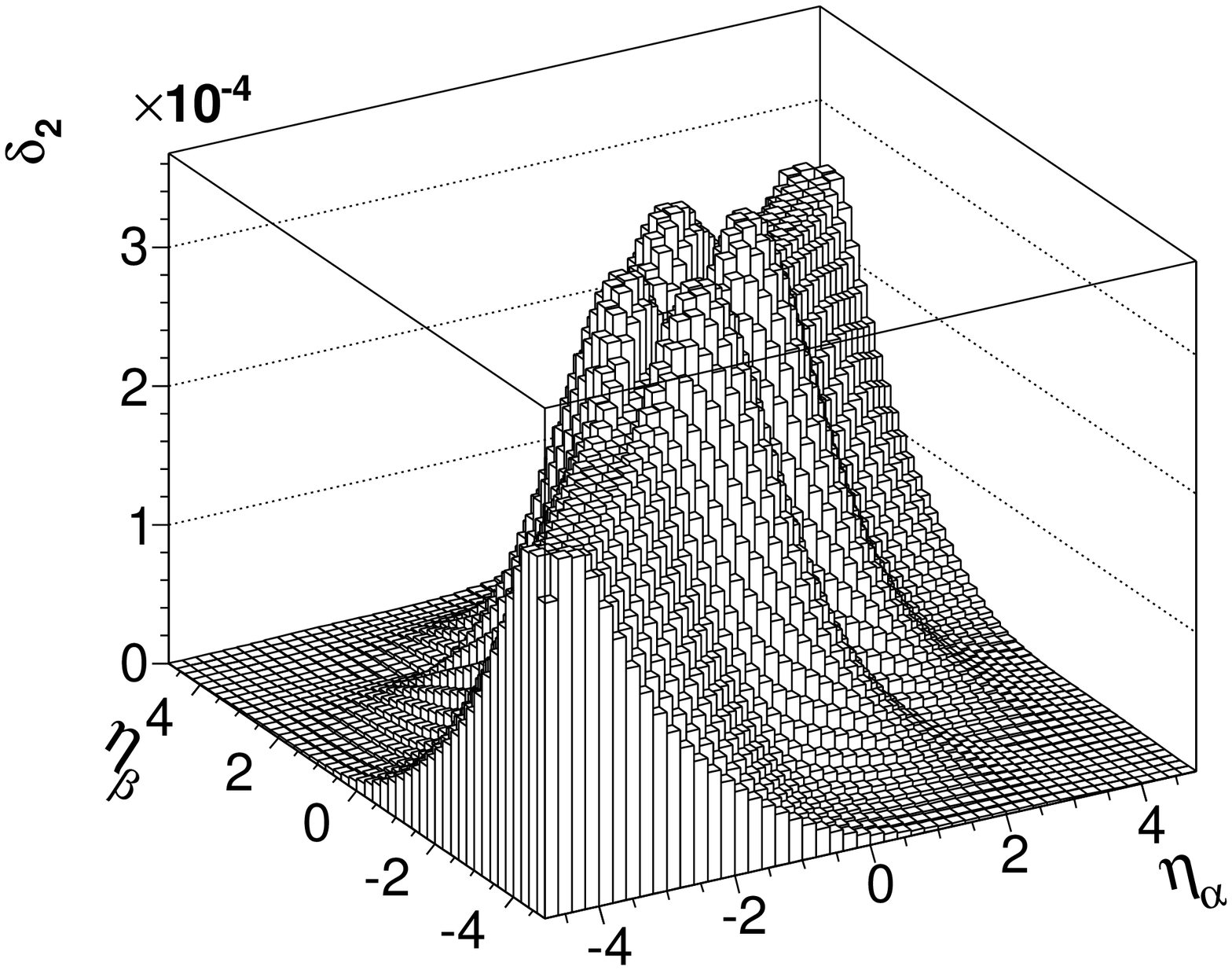}
\includegraphics[width=0.3\textwidth]{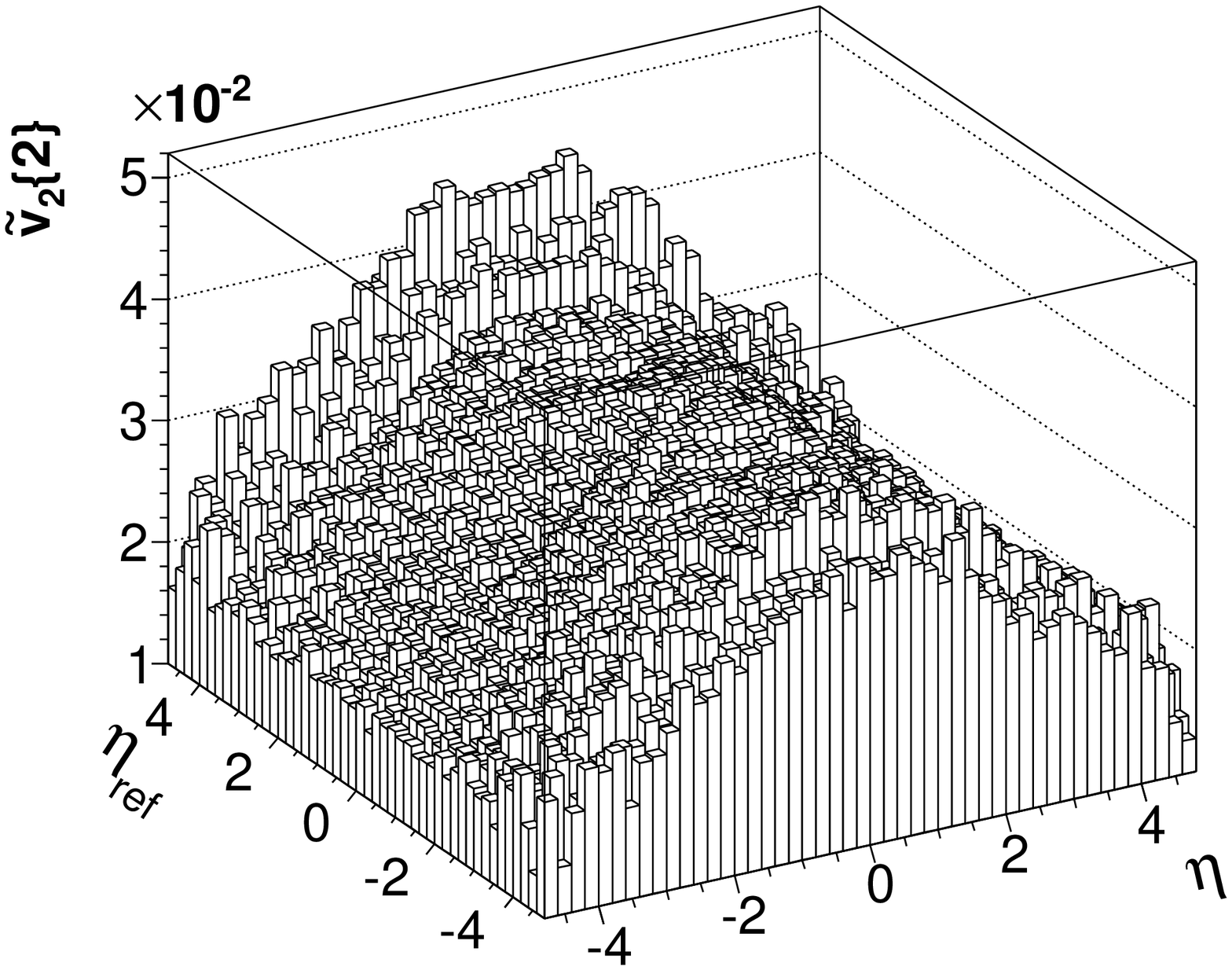}
\includegraphics[width=0.3\textwidth]{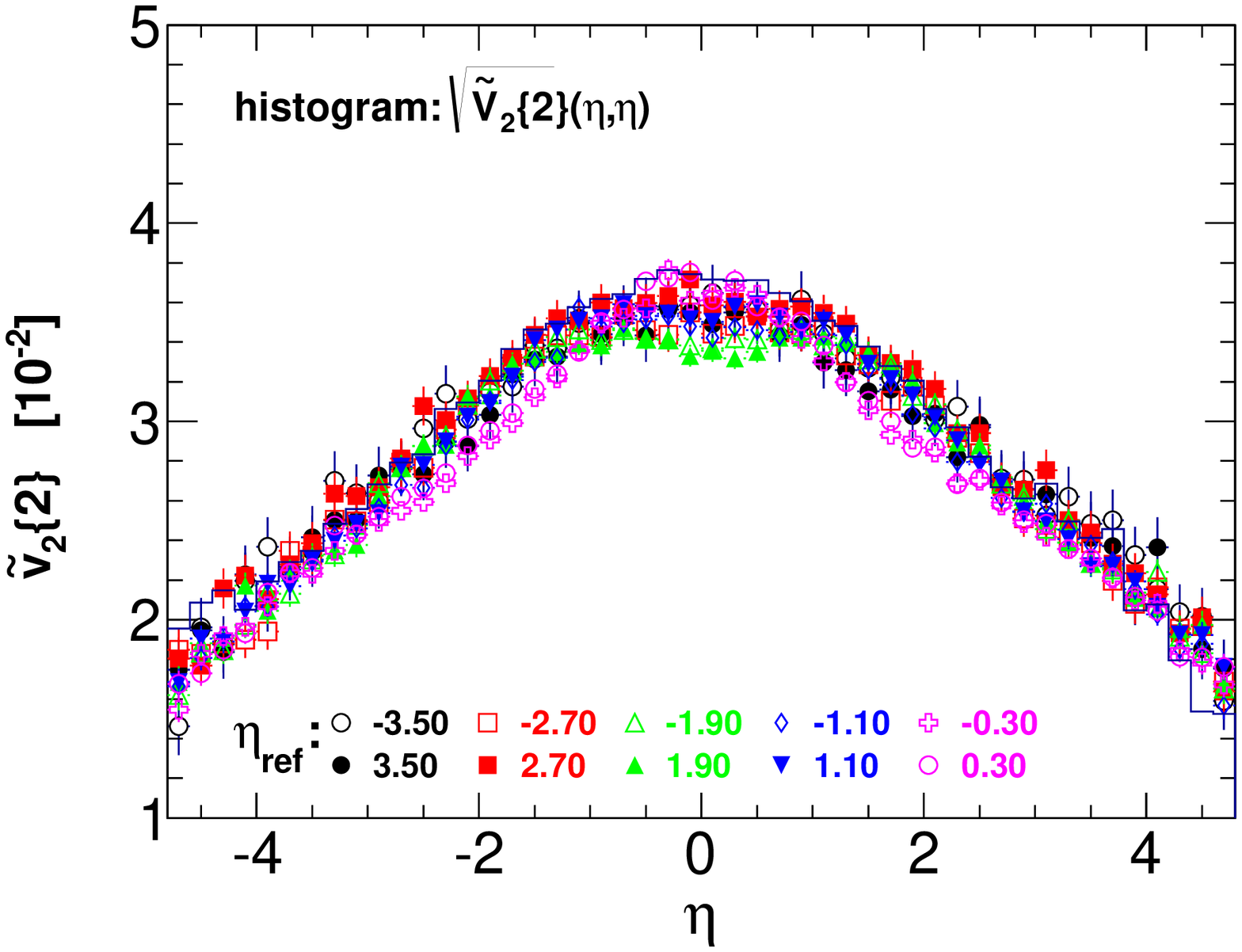}
\includegraphics[width=0.3\textwidth]{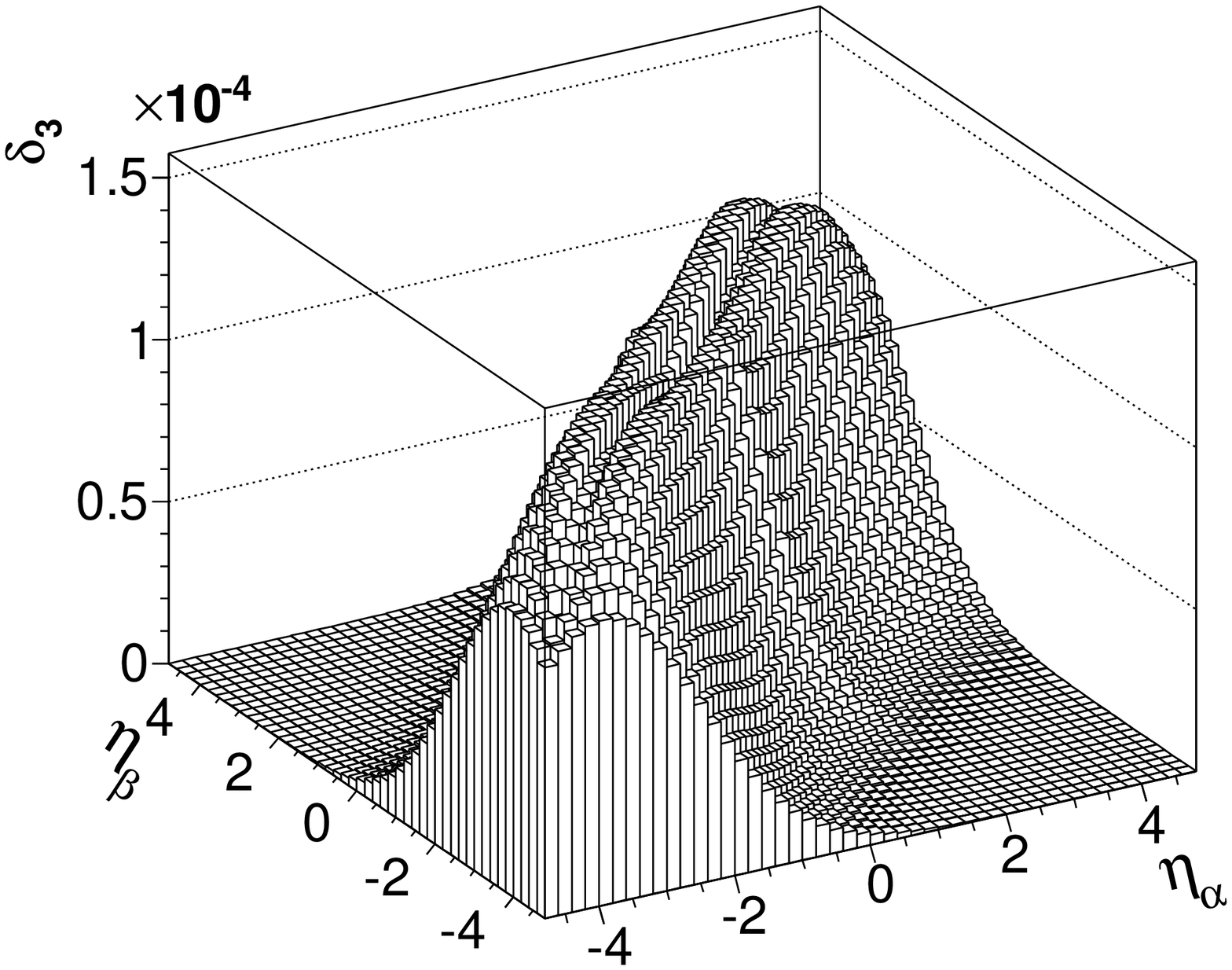}
\includegraphics[width=0.3\textwidth]{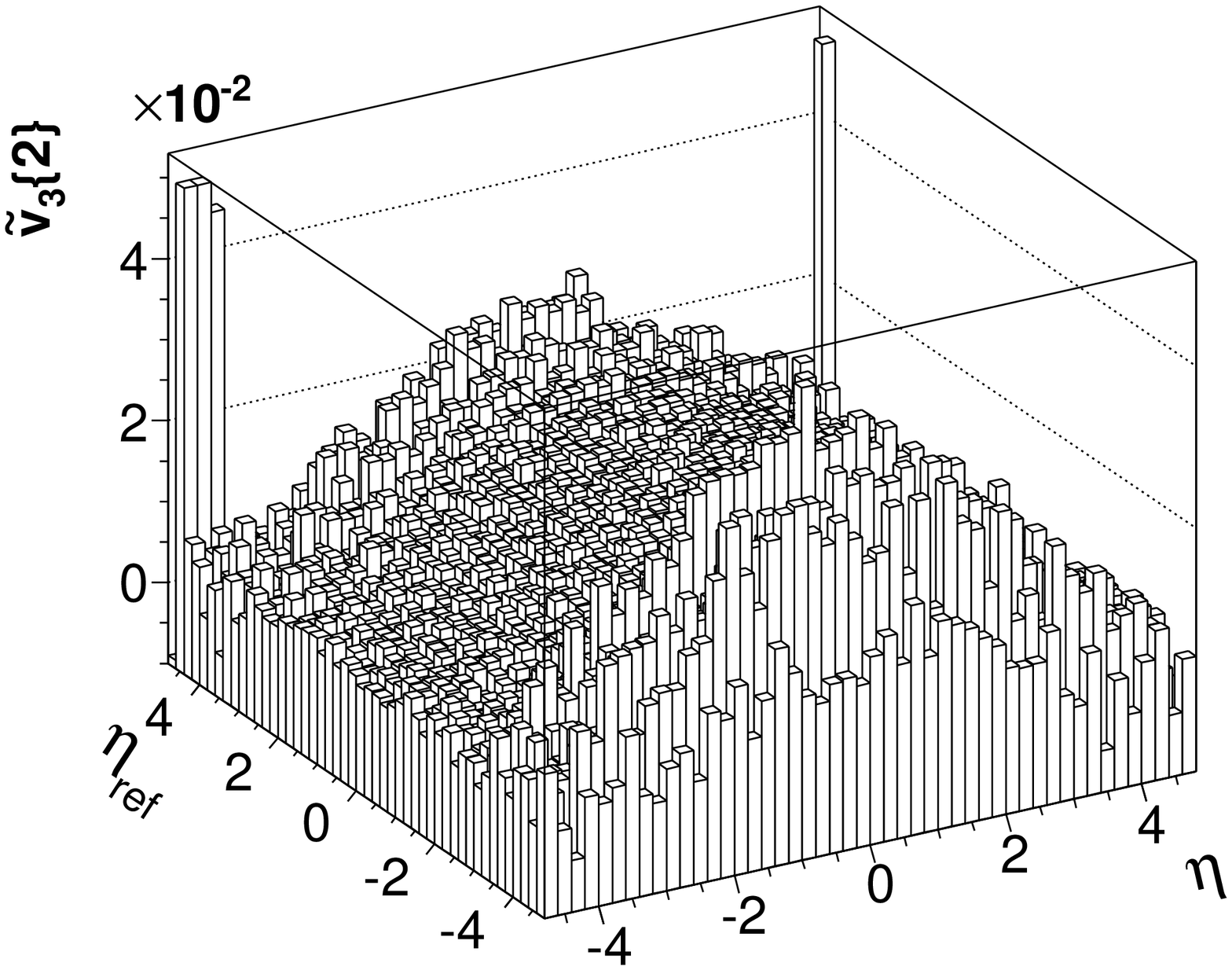}
\includegraphics[width=0.3\textwidth]{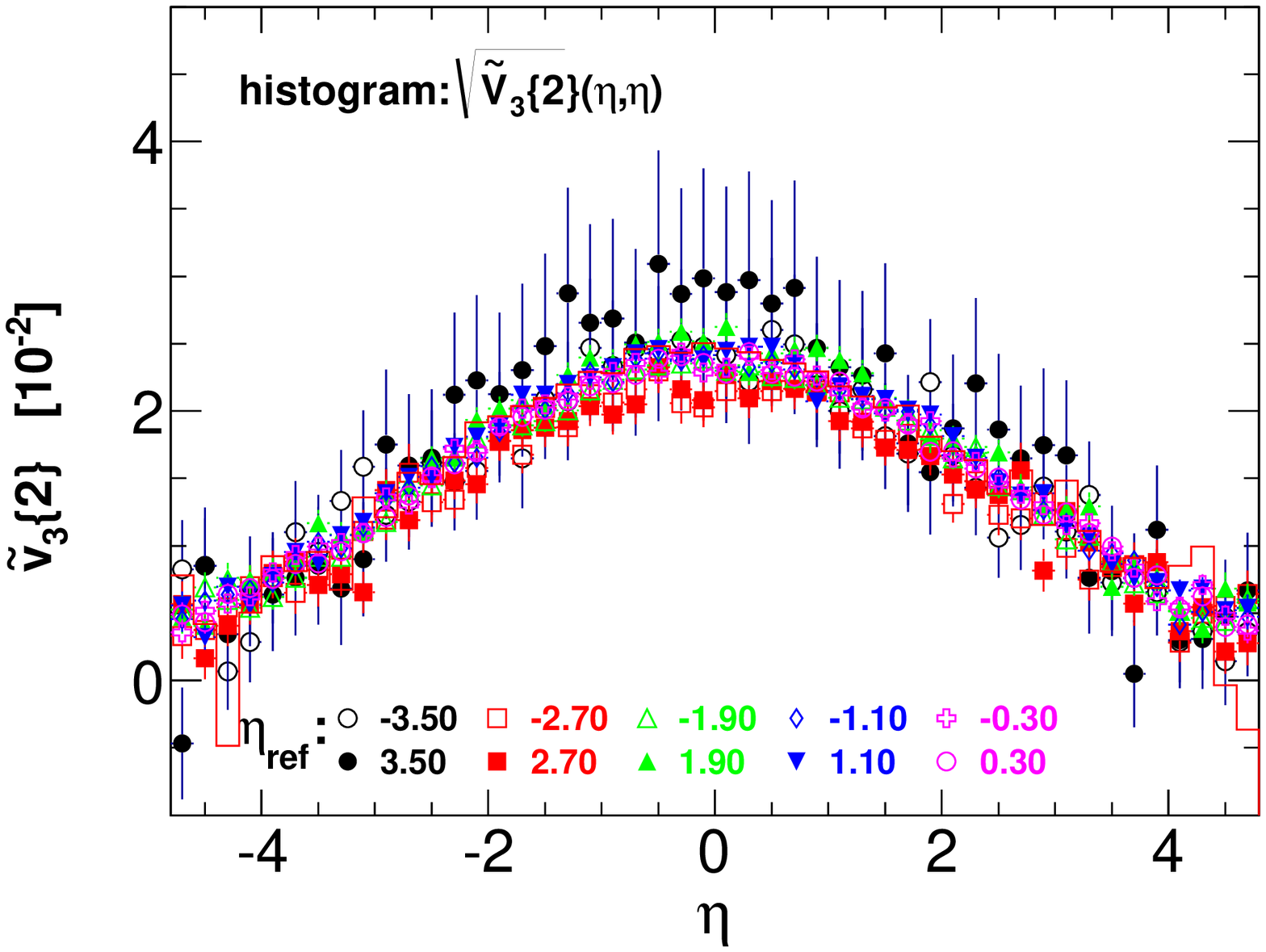}
\end{center}
\caption{(Color online) Results for \ampt\ 200 GeV central Au+Au collisions with \bcent\ (approximately top 10\% centrality). Left panels: Parameterized second-harmonic $\deta$-dependent nonflow correlation $\delta_2$ (upper panel) and third-harmonic $\deta$-dependent nonflow correlation $\delta_3$ (lower panel). Middle panels: decomposed $\deta$-independent single-particle-level ``flow'' correlations $\vt{2}{2}$ (upper panel) and $\vt{3}{2}$ (lower panel) as a function of $\eta$ (and $\etaref$ used to obtain it). The results are independent of $\etaref$. Right panels: Slices of plots in the middle panels at selected $\etaref$, showing more quantitatively factorization in the decomposed single-particle-level ``flow'' correlations $\vt{2}{2}$ (upper panel) and $\vt{3}{2}$ (lower panel).}
\label{fig:central}
\end{figure*}

Given the statistical errors, the nonflow fit model of Eq.~(\ref{eq:nf_fit2D}) seems adequate. The global 2D-fit gives a $\chisq\sim1$; see Table~\ref{tab}. Figure~\ref{fig:central} shows the central collision results from our method. The left panels show the fitted $\deta$-dependent correlation, $\delta(\deta)+\fl(\deta)$. 

Figure~\ref{fig:central} middle panels show the extracted $\vt{2}{2}$ and $\vt{3}{2}$ as a function of $\eta$ and $\etaref$, similar to results in Fig.~\ref{fig:fact}. The $\deta$-independent correlation results are again approximately independent of $\etaref$; the bed-headboard regions which show up in both $\vt{2}{2}$ and $\vt{3}{2}$ are due to the division by the small magnitude (with large errors) at large $|\etaref|$ in central collisions. The $\etaref$ independence indicates factorization of the $\deta$-independent correlation that is expected for flow. The $\etaref$ independence is more quantitatively demonstrated in the right panels of Fig.~\ref{fig:central} where the slices of $\vt{n}{2}$ in $\etaref$ are depicted; all results fall on top of each other.

\subsection{Test against the \hijing\ model\label{sec:hijing}}

In this section we test our method against the \hijing\ event generator~\cite{hijing}. \hijing\ is a quantum chromodynamics inspired phenomenological model focusing on hard parton scattering processes, with particle production from soft processes parameterized. Since it does not have final state interactions, it does not generate collective flow. However, it contains significant nonflow, mainly from jet-correlations. Thus \hijing\ provides a critical test to our method.

\begin{figure*}[hbt]
\begin{center}
\includegraphics[width=0.3\textwidth]{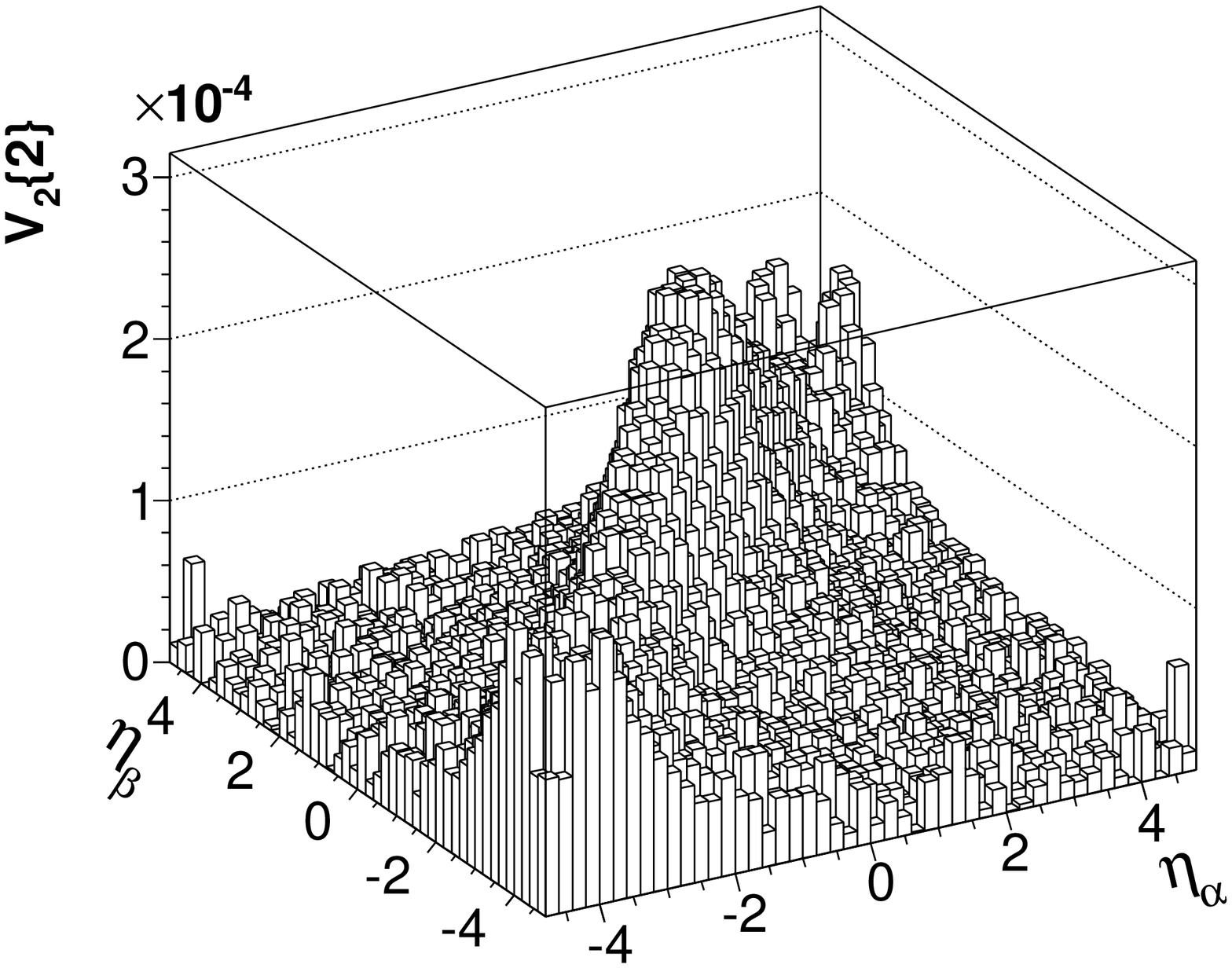}
\includegraphics[width=0.3\textwidth]{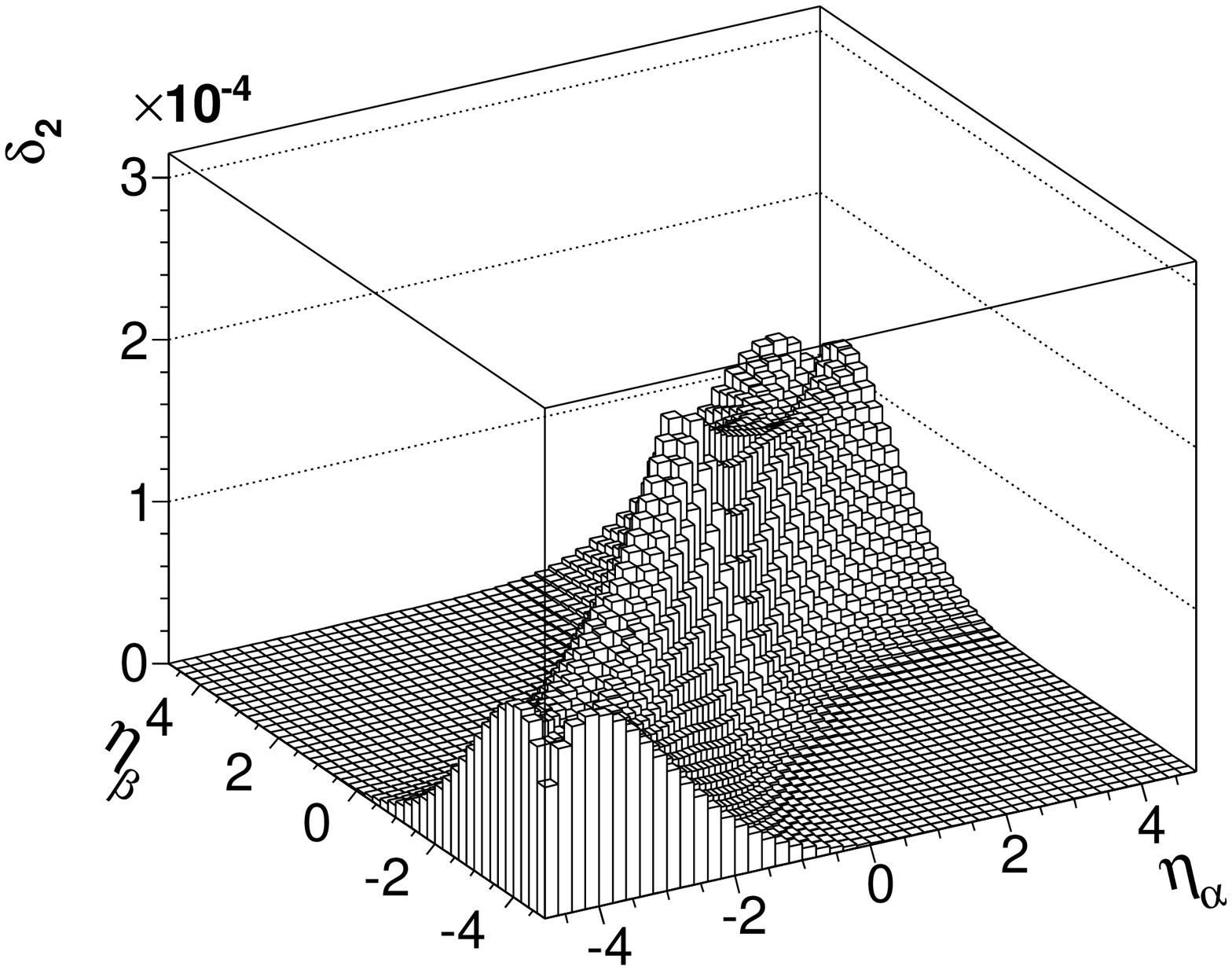}
\includegraphics[width=0.3\textwidth]{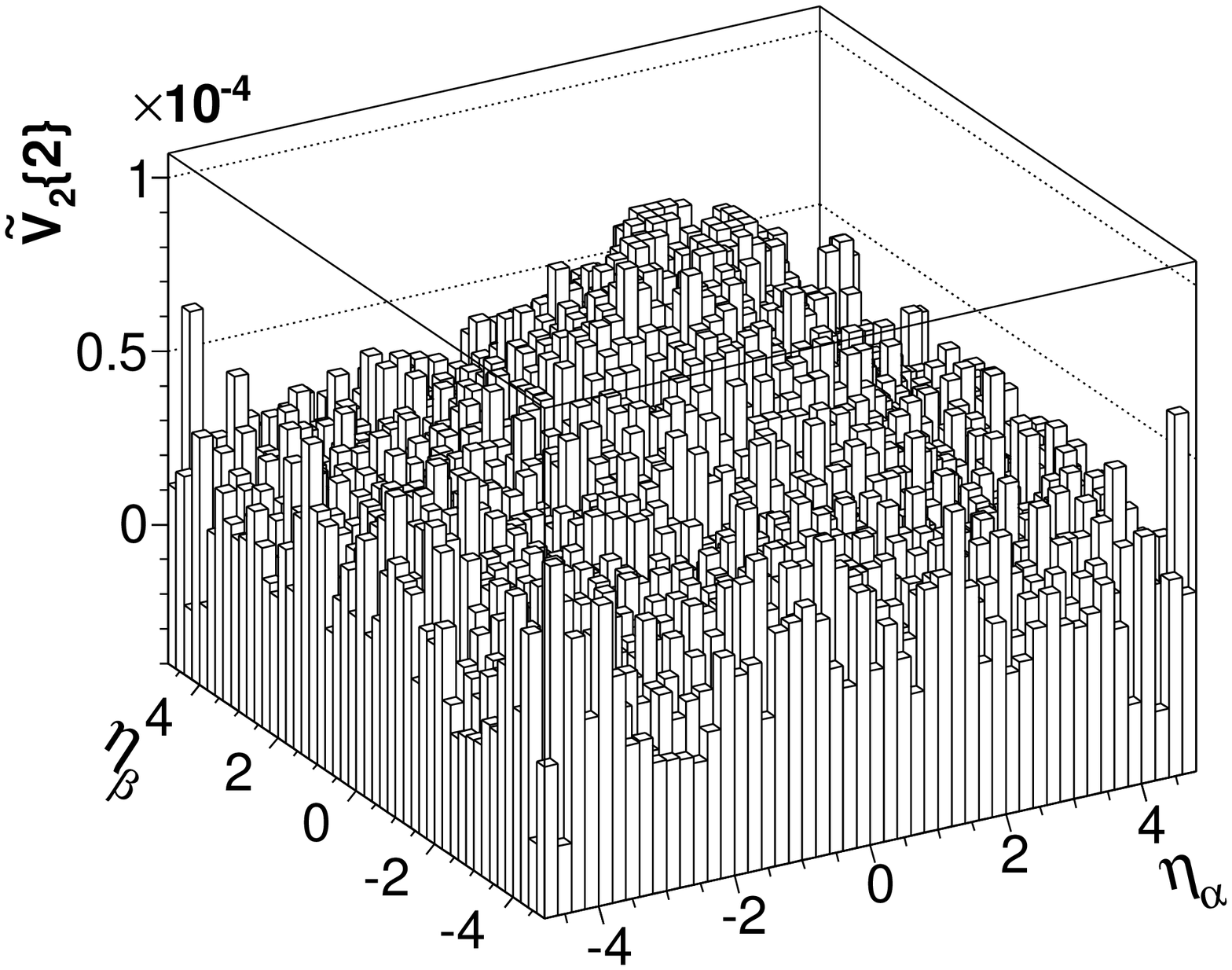}
\includegraphics[width=0.3\textwidth]{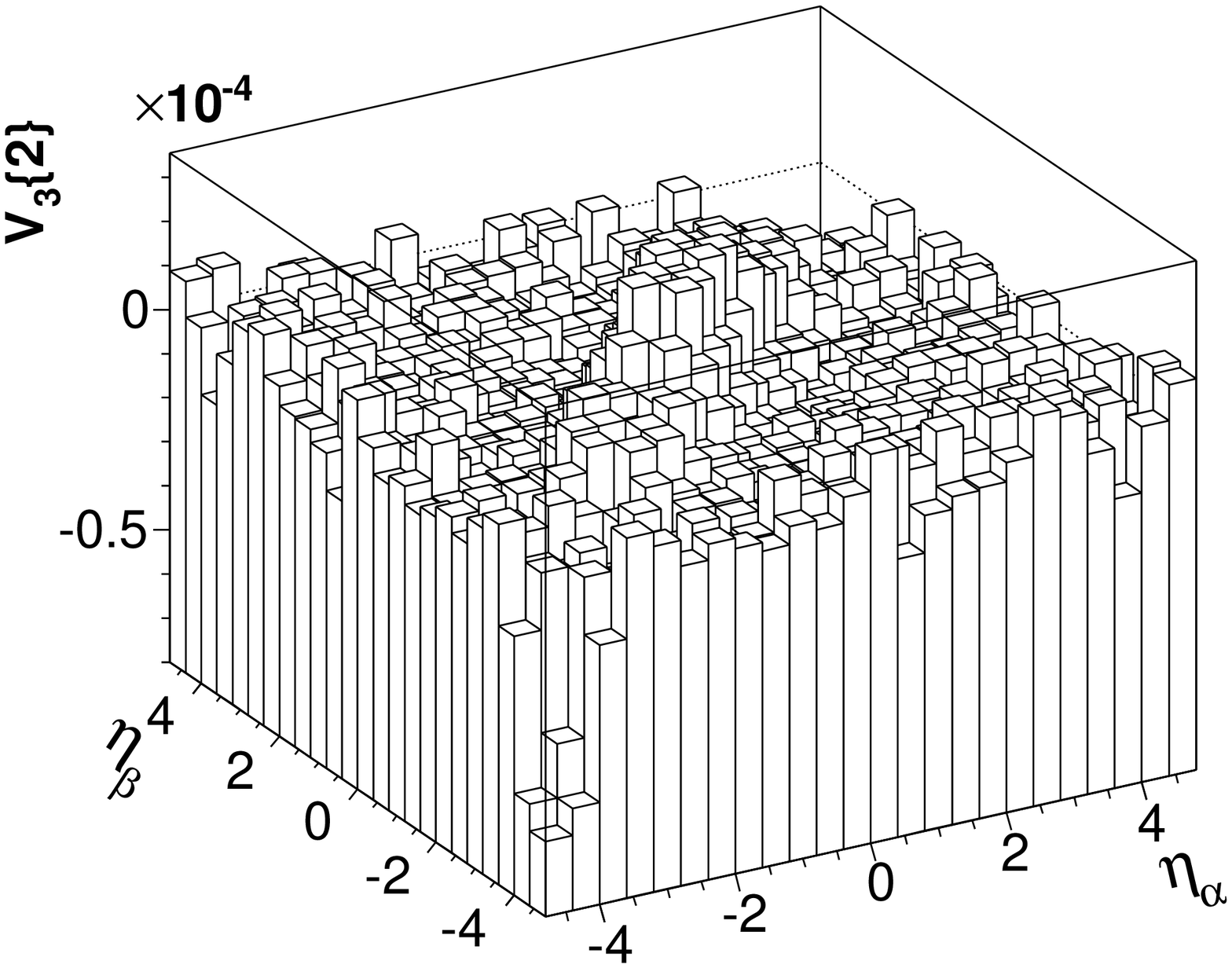}
\includegraphics[width=0.3\textwidth]{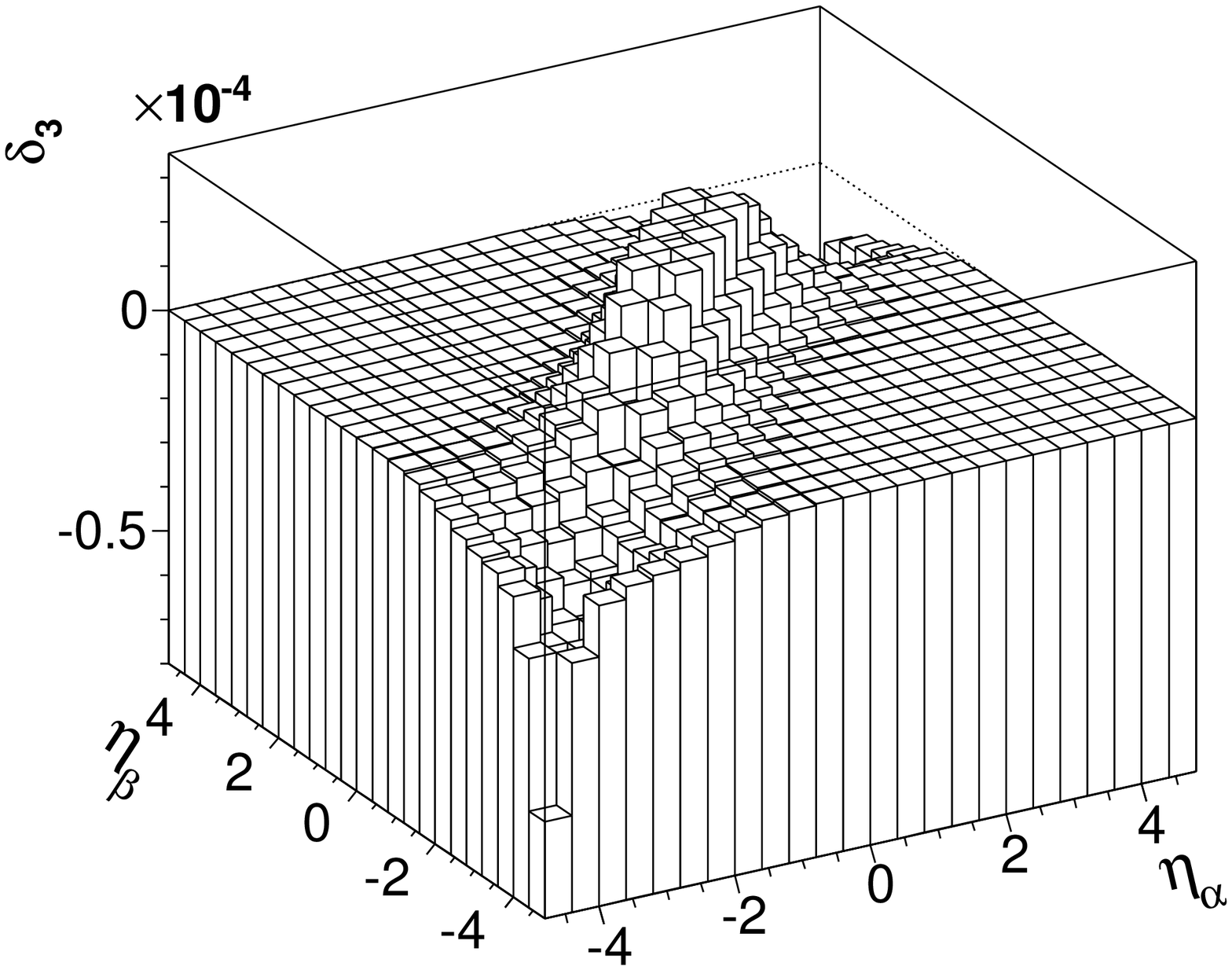}
\includegraphics[width=0.3\textwidth]{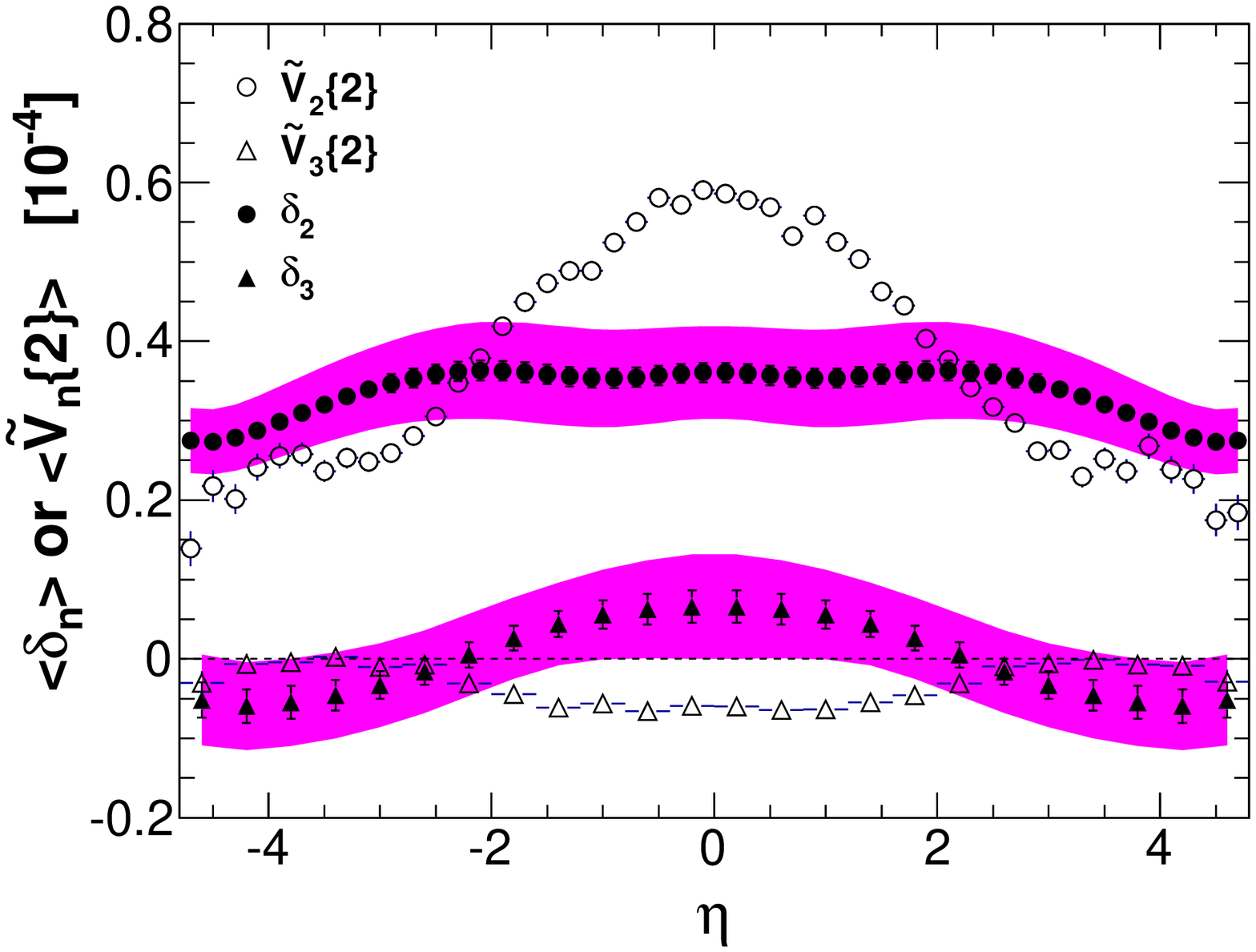}
\end{center}
\caption{Results for \hijing\ 200 GeV Au+Au collisions with \bperi\ (approximately 10-60\% centrality). Left panels: two-particle cumulants $\Vn{2}{2}$ (upper) and $\Vn{3}{2}$ (lower). Middle panels: Parameterized $\deta$-dependent nonflow correlations $\delta_2$ (upper) and $\delta_3$ (lower). Upper-right panel: $\delta_2$ subtracted two-particle cumulant $\Vt{2}{2}$. Lower-right panel: projections (averages) of $\Vt{n}{2}$ (open symbols) and $\delta_n$ (filled symbols) onto one $\eta$ axis. The bands are systematic uncertainties from the nonflow fit errors; these uncertainties also apply to the average $\Vt{n}{2}$ but anti-correlated.}
\label{fig:hijing}
\end{figure*}

Figure~\ref{fig:hijing} shows the two-particle second- and third-harmonic cumulants, $\Vn{2}{2}$ and $\Vn{3}{2}$, in the upper-left and lower-left panels, respectively. The main feature of $\Vn{2}{2}$ is the diagonal ridge along $\etaa=\etab$, characteristic of nonflow. 
In $\Vn{3}{2}$ the diagonal ridge seems to be present only in the limited mid-rapidity region. At large rapidities along the diagonal, $\Vn{3}{2}$ is negative. 

We follow the same procedure described in Sec.~\ref{sec:nonflow}, taking difference between two pairs of $\eta$-bins, $(\etaa,\etab)$ and $(\etaa,-\etab)$. 
We again fit the difference $\Delta\Vn{}{2}$ by the 2D function of Eq.~(\ref{eq:nf_fit2D}).
The fit parameters and $\chisq$ are tabulated in Table~\ref{tab}. 
We deduce the $\deta$-dependent nonflow correlations from the 2D-fit. The deduced results are depicted in the middle panels of Fig.~\ref{fig:hijing}. The second-harmonic $\delta_2$ is about an order of magnitude larger than the third-harmonic $\delta_3$, in line with the two-particle cumulants in the left panels.


Subtracting the parameterized $\deta$-dependent nonflow correlation, we obtain the remaining $\deta$-independent correlation, $\Vt{n}{2}$, as a function of $(\etaa,\etab)$. Because it is known that \hijing\ does not generate collective flow, the obtained $\Vt{n}{2}$ are in fact nonflows due to back-to-back inter-jet correlations, $\nf_n(\etaa)\nf_n(\etab)$; see Eq.~(\ref{eq:V4}). 
We show $\Vt{2}{2}$ in the upper-right panel of Fig.~\ref{fig:hijing}; 
It is peaked for particle pairs at mid-rapidity (0,0) and decreases towards large rapidities.
In the lower-right panel, we show in open circles 
the projection of $\Vt{2}{2}$ to one of the $\eta$-axes. This would be the average away-side nonflow of particle pairs with one particle at $\eta$ and the other particle from the entire $\eta$ range. 

Also superimposed in Fig.~\ref{fig:hijing} lower-right panel in solid circles are the average parameterized nonflow values of $\delta_2$. This is the $\deta$-dependent near-side nonflow, averaged over all partner particles and plotted as a function of $\eta$ of the test particle. The band indicates the uncertainties from the fit errors. This uncertainty also applies to the average $\Vt{2}{2}$ but anti-correlated.
The magnitudes of the near-side nonflow $\delta_2(\deta)$ and the away-side nonflow $\nf_2(\etaa)\nf_2(\etab)$ are similar. The near-side nonflow may largely come from intra-jet correlations, and the away-side may be dominated by inter-jet correlations. The similar magnitudes are then expected from low $\pt$ two-particle correlations from dijets due to momentum conservation. See Sec.~\ref{sec:nf} for further discussion.

The statistics of $\Vt{3}{2}$ is poor and the two-dimensional lego plot does not reveal too much information. However, we show 
the projection of
$\Vt{3}{2}$ in the solid triangles in the lower-right panel of Fig.~\ref{fig:hijing}. 
The average near-side nonflow $\delta_3$ is also superimposed with the shaded uncertainty band. Note the uncertainty band also applied to $\Vt{3}{2}$ but anti-correlated. 
The third-harmonic near- and away-side nonflows are significantly smaller than the second-harmonic nonflows.

To conclude this subsection, the features of the separated near- and away-side nonflows from the \hijing\ model seem to all make sense. This renders additional support to our decomposition method.


\section{Discussion of flow and nonflow from the models\label{discussion}}

In this section we discuss the $\deta$-dependent and independent correlations of the models separated by our method. The former is the sum of the $\deta$-dependent nonflow $\delta(\deta)$ and $\deta$-dependent flow fluctuation $\fl(\deta)$. The latter is the sum of the average flow, flow fluctuation, and the $\deta$-independent part of nonflow $\nf$. Since $\fl(\deta)$ is small as indicated by our results and $\nf$ is presumably small at our low $\pt$ region, we will simply call the $\deta$-independent correlation as ``flow'' and the $\deta$-dependent correlation as nonflow in the following discussion unless otherwise noted.

\subsection{Comparison to \ampt\ model flow}

Strong interactions among constituents in relativistic heavy-ion collisions are expected to generate hydrodynamic expansion and convert the spatial anisotropy of the initial overlap geometry into an azimuthal anisotropy in the final-state particle distribution. This phenomenon is often called hydrodynamic flow which we refer to here as ``real'' flow. This ``real'' average flow can be calculated from the \ampt\ model input because the initial geometry symmetry axis can be calculated from model information on the initial overlap geometry. In this section we compare the flow results from our decomposition method to the calculated ``real'' flow from the \ampt\ model to assess its performance. We note, however, that the spatial geometry information is inaccessible to experiment, and experimentally the flow axis has to be defined by final particle momenta. Due to particle correlations unrelated to flow, the experimental flow measurements are often contaminated by nonflow which we try to separate in this study.

\begin{figure*}[hbt]
\begin{center}
\includegraphics[width=0.3\textwidth]{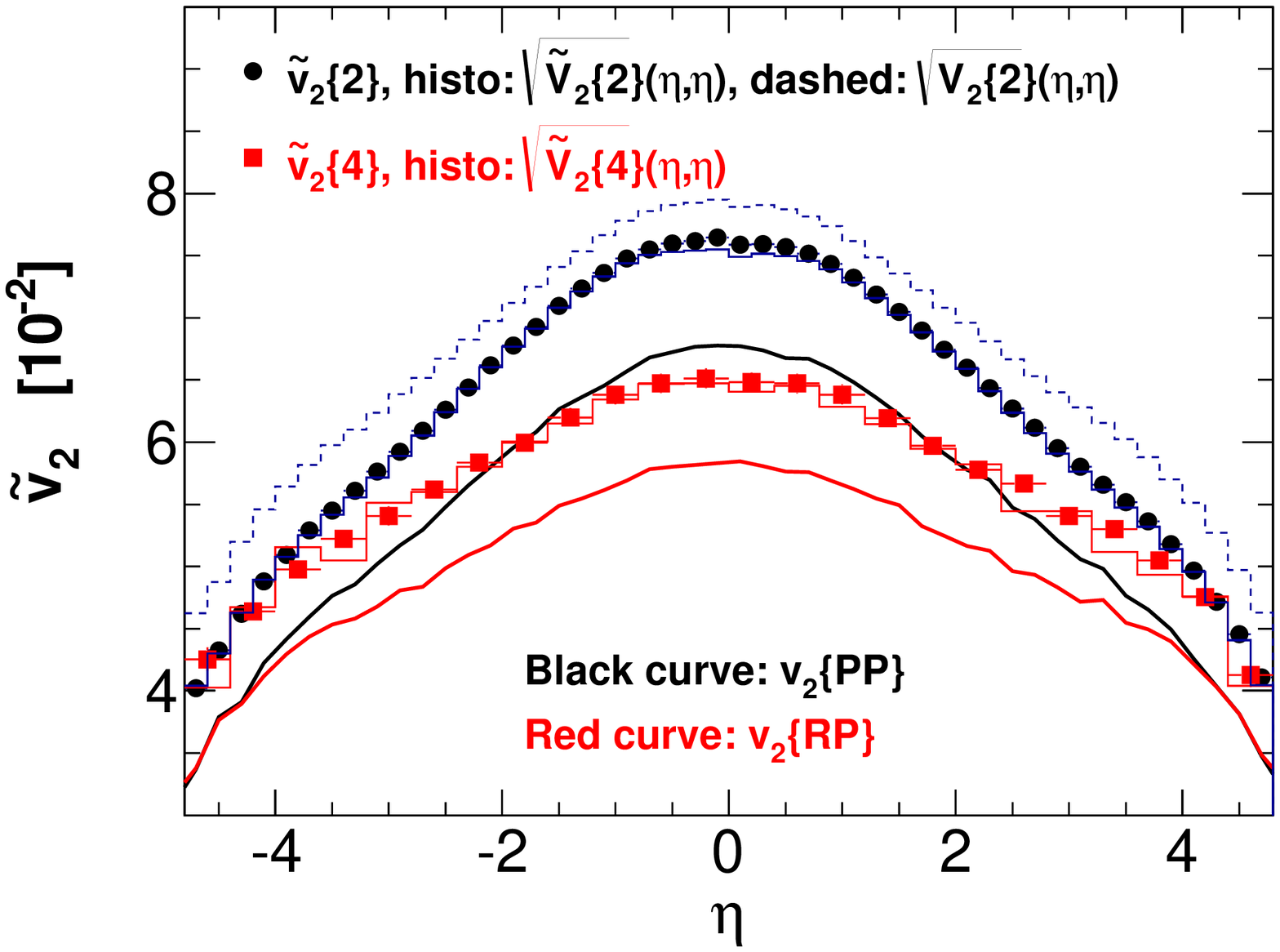}
\includegraphics[width=0.3\textwidth]{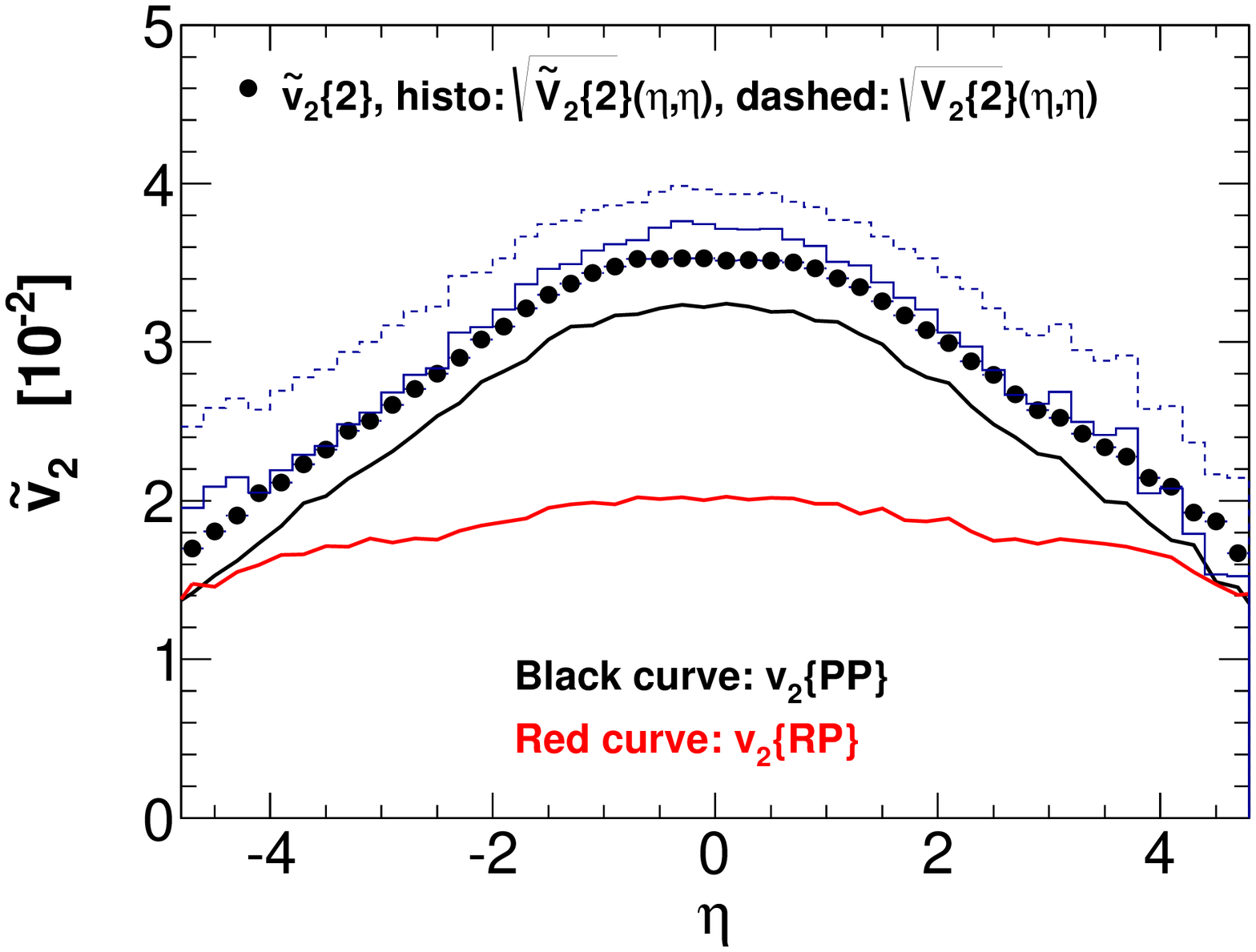}
\includegraphics[width=0.3\textwidth]{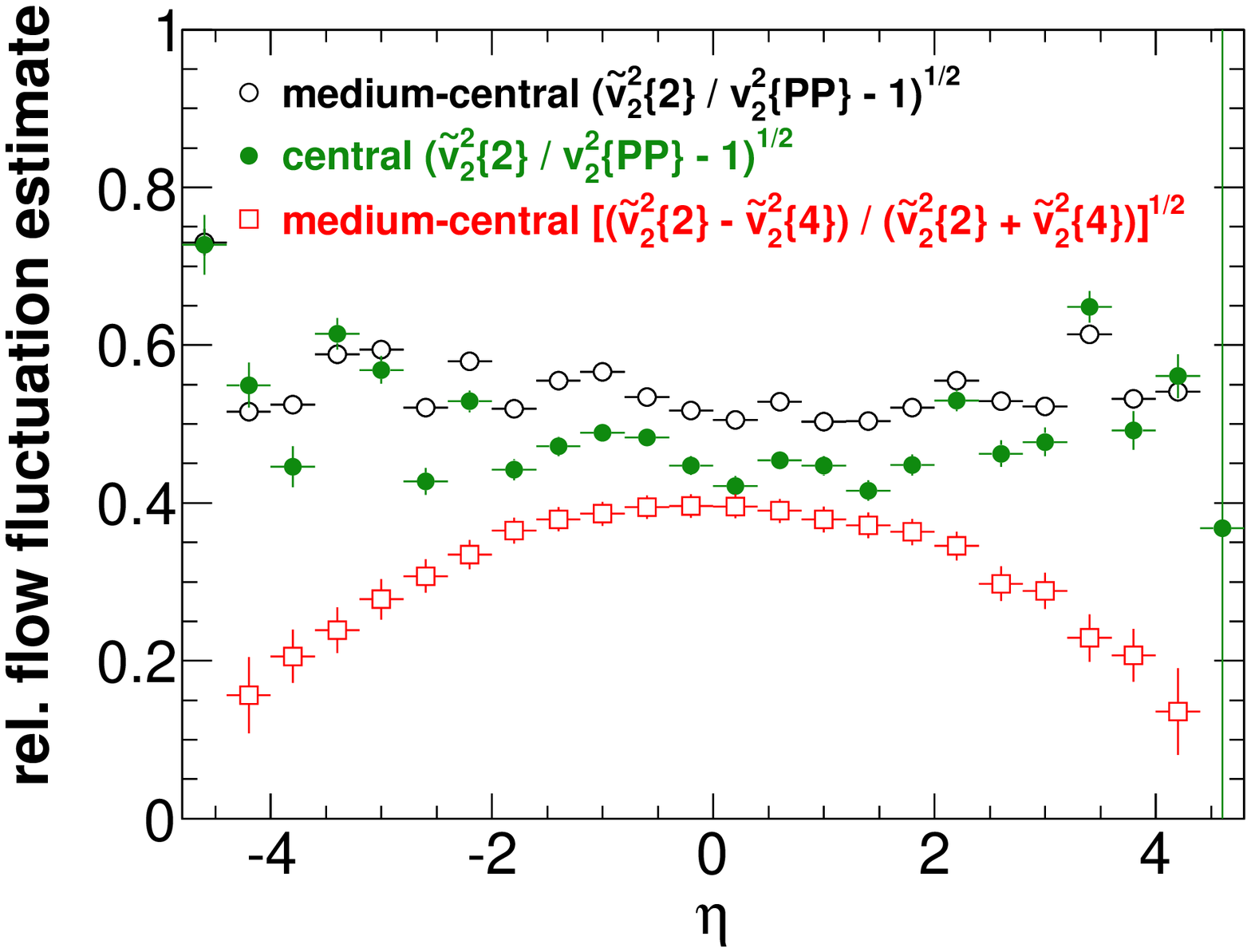}
\end{center}
\caption{(Color online) Nonflow-subtracted second-harmonic two-particle cumulant flow $\vt{2}{2}$ (circles) and $\deta$-dependent fluctuation subtracted second-harmonic four-particle cumulant flow $\vt{2}{4}$ (square) as a function of $\eta$ in \ampt\ 200 GeV medium-central (\bperi, left panel) and central (\bcent, middle panel) Au+Au collisions. Right panel: estimates of relative $v_2$ fluctuation, $\sigma_2/v_2$, in \ampt.}
\label{fig:v2fluc}
\end{figure*}

We project the $\vt{2}{4}$ and $\vt{2}{2}$ 2D plots in Fig.~\ref{fig:fact} lower-left and lower-middle panels over the entire $|\etaref|<4.8$. The result is shown in Fig.~\ref{fig:v2fluc} left panel by the filled circles and squares for second-harmonic two- and four-particle cumulant methods, respectively. These would be the two- and four-particle second-harmonic flows, largely devoid of nonflow, using all other particles as the reference particles to measure them. 
The solid histograms (almost coinciding with the data points) are, respectively, the two- and four-particle cumulant flows obtained from the diagonal elements of the nonflow-subtracted cumulants by Eq.~(\ref{eq:diag}). 
The dashed histogram shows the diagonal result obtained from the raw two-particle cumulant similar to Eq.~(\ref{eq:diag}). The difference between the dashed histogram and the solid circles is the effect of the nonflow. 

As seen from Fig.~\ref{fig:v2fluc} the two-particle cumulant $\vt{2}{2}$ is larger than the four-particle $\vt{2}{4}$ and the difference is presumably caused by flow fluctuations (as the remaining away-side nonflow is expected to be small). However, the difference becomes diminished at $|\eta|>4$. This seems hard to understand--it would imply that flow fluctuations diminish at $|\eta|>4$ if Eq.~(\ref{eq:V4}), which assumes Gaussian and relatively small flow fluctuation, is valid. 

Figure~\ref{fig:v2fluc} middle panel shows the corresponding results in central \ampt\ events of \bcent. As seen in Fig.~\ref{fig:fact}, the $\vt{2}{2}$ obtained from large $\etaref$ reference particles have large uncertainties (bed-headboard effect). We have thus restricted our projection to within $|\etaref|<3.6$ in obtaining the $\vt{2}{2}$ results in the middle panel of Fig.~\ref{fig:v2fluc}. The four-particle cumulant in central collisions is not plotted--The statistical precision is too poor to be useful. The other results are qualitatively similar between central and medium-central collisions.

\begin{figure*}[hbt]
\begin{center}
\includegraphics[width=0.3\textwidth]{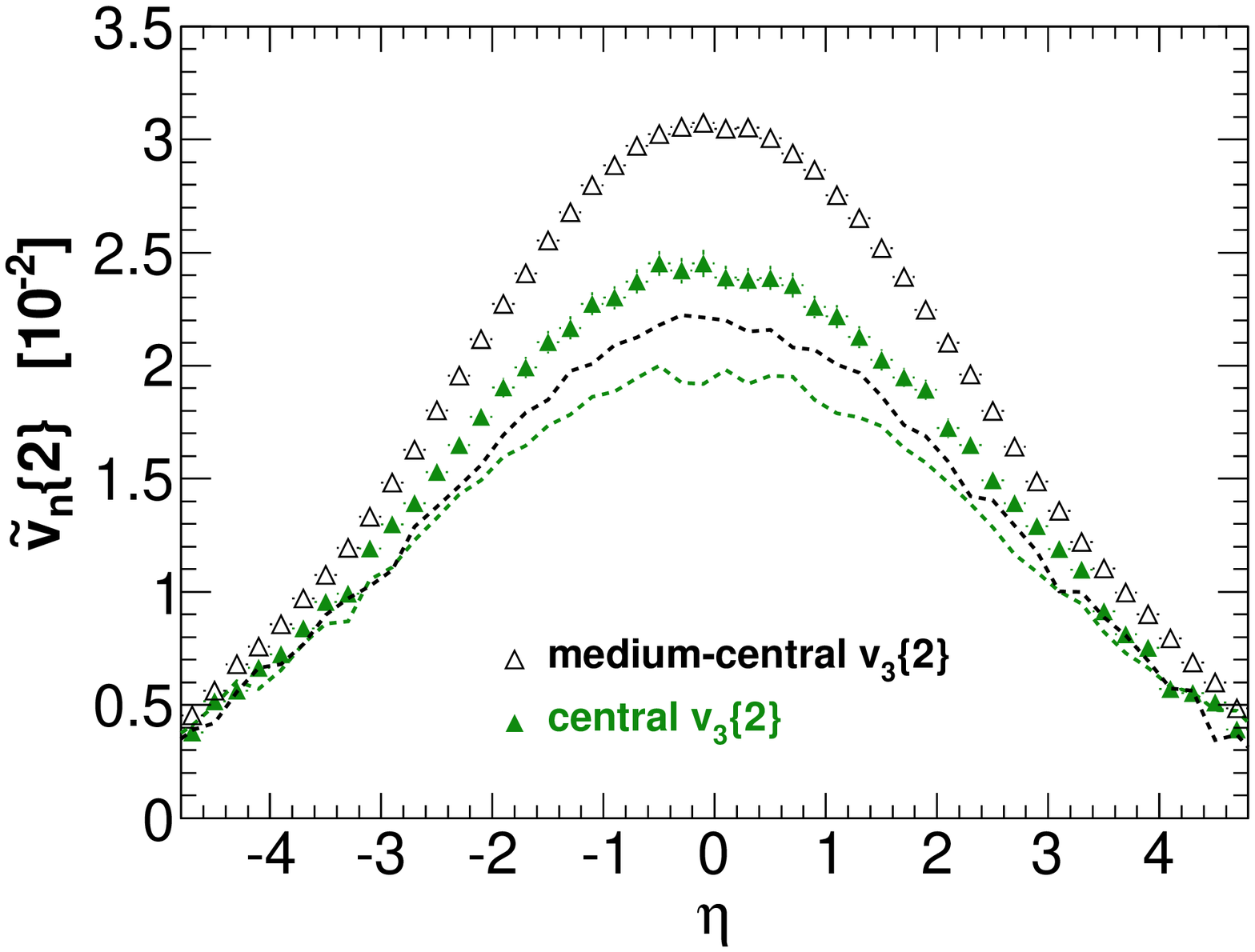}
\includegraphics[width=0.3\textwidth]{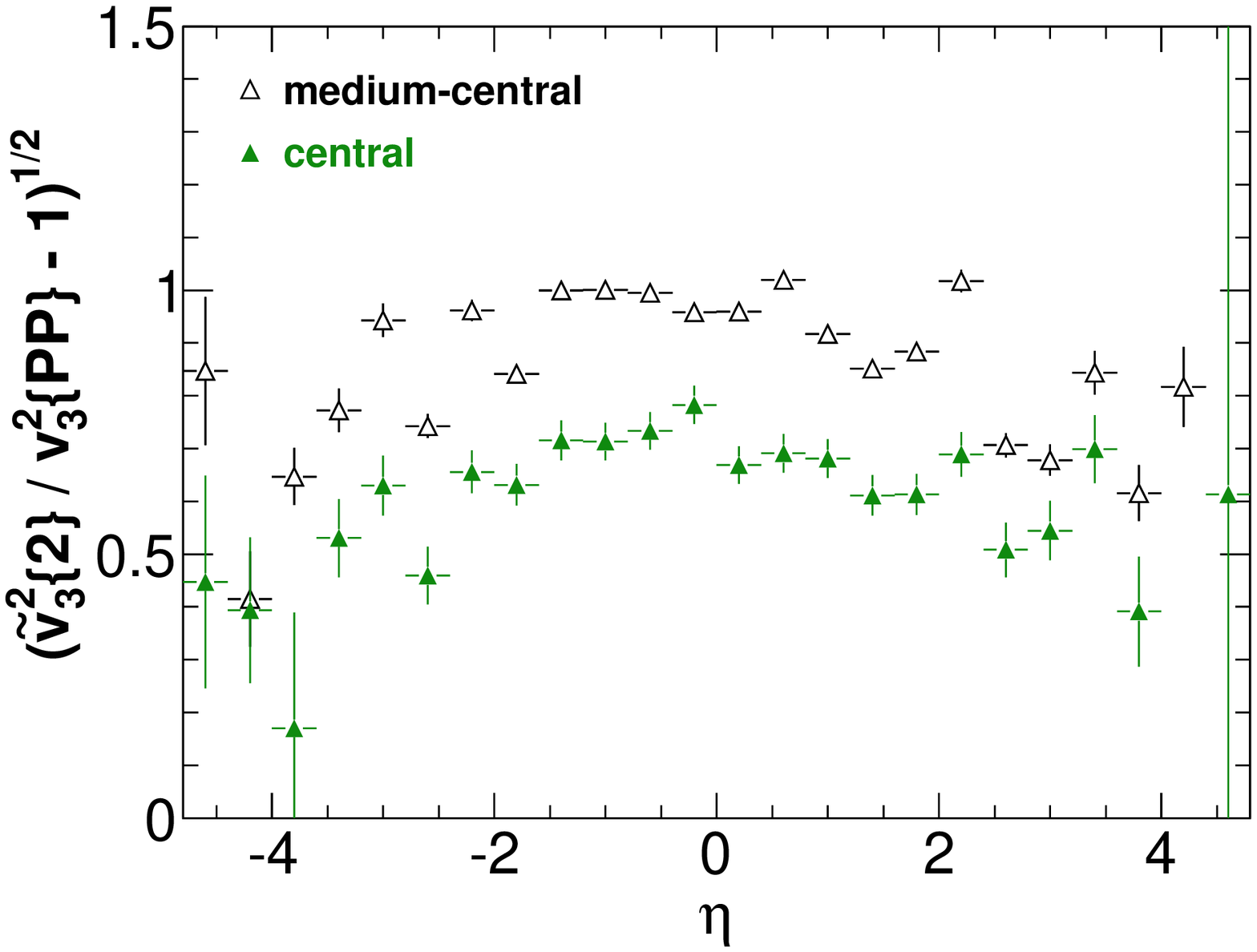}
\end{center}
\caption{(Color online) $\deta$-dependent nonflow subtracted third-harmonic two-particle cumulant flow $\vt{3}{2}$ as a function of $\eta$ in \ampt\ 200 GeV medium-central (\bperi) and central (\bcent) Au+Au collisions. Right panel: estimates of relative $v_3$ fluctuation, $\sigma_3/v_3$, in \ampt.}
\label{fig:v3fluc}
\end{figure*}

\begin{figure*}[hbt]
\begin{center}
\includegraphics[width=0.3\textwidth]{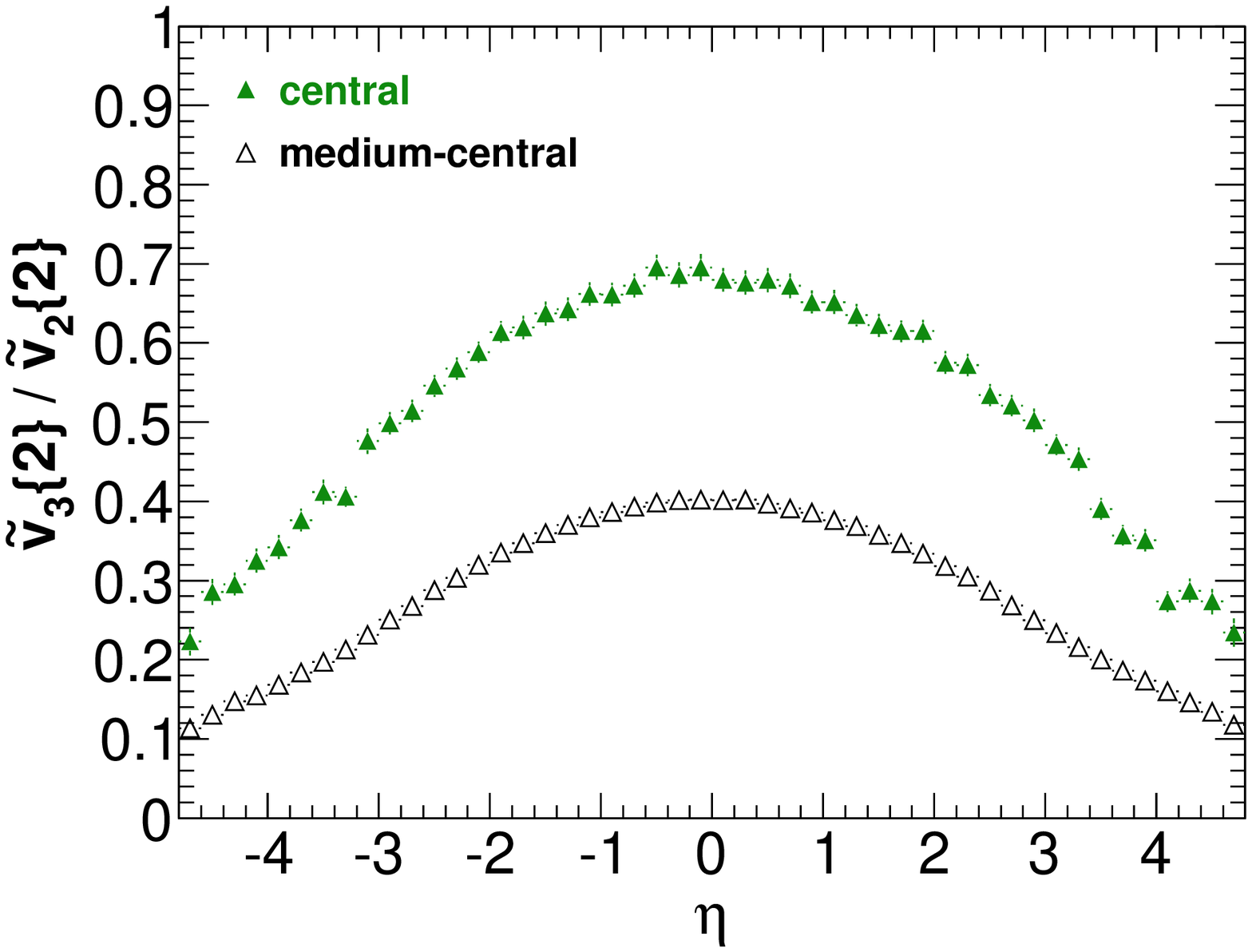}
\includegraphics[width=0.3\textwidth]{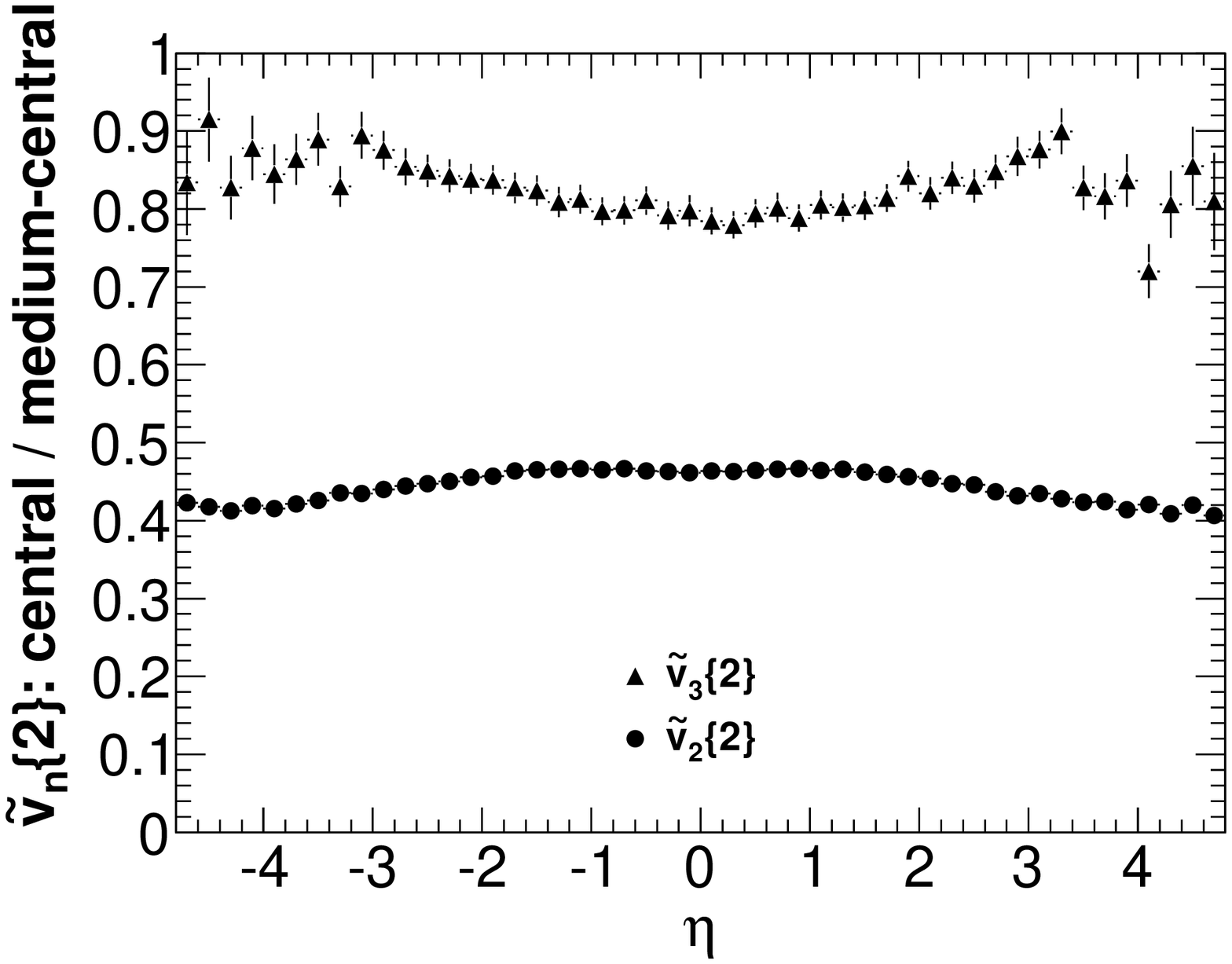}
\end{center}
\caption{(Color online) Left panel: $\vt{3}{2}/\vt{2}{2}$ ratio in \ampt\ 200 GeV medium-central (\bperi) and central (\bcent) Au+Au collisions. Right panel: Ratio of $\vt{n}{2}$ in central to that in medium-central collisions.}
\label{fig:v3v2}
\end{figure*}

We calculate the real flow in the following way. We obtain the minor axis (participant plane) of the coordinate space overlap region using the transverse coordinates (transverse radius $r$ and azimuth $\phi_r$) of all participant gluons and quarks by~\cite{Petersen}
\be\psiPP=\frac{1}{n}{\rm ATAN2}\left(\sum{r^2\sin(n\phi_r)},\sum{r^2\sin(n\phi_r)}\right)+\frac{\pi}{n}\,,\ee
where $n=2,3$ for the second- and third-harmonics, respectively. 
The function ${\rm atan2}(y, x)$ is the angle between the positive x-axis and the vector given by the coordinates $(x, y)$ on a plane; The angle is positive for $y > 0$ and negative for $y < 0$.
We calculate the participant plane resolution by using the sub-event method and find the resolution is essentially 100\%. 
We calculate the real participant-plane flow by 
\be\vn{n}{PP}=\mn{\cos n(\phi-\psiPP)}\,.\label{eq:vnPP}\ee
We also calculate the real reaction-plane flow by using the known reaction plane in \ampt\ ($\psiRP\equiv0$) by 
\be\vn{n}{RP}=\mn{\cos n(\phi-\psiRP)}\,.\label{eq:vnRP}\ee

The calculated $\vn{2}{PP}$ and $\vn{2}{RP}$ in medium-central \ampt\ collisions are shown in Fig.~\ref{fig:v2fluc} left panel as the black and red curves, respectively. Those in central \ampt\ collisions are shown in the Fig.~\ref{fig:v2fluc} middle panel. The central collision $\vn{2}{RP}$ is flatter. The $\vn{2}{PP}$ is larger than $\vn{2}{RP}$ presumably due to the effect of the fluctuating participant plane about the reaction plane. It is interesting to note that $\vn{2}{PP}$ and $\vn{2}{RP}$ also approach each other at $|\eta|>4$.

The extracted $\vt{2}{2}$ is larger than the calculated $\vn{2}{PP}$, the true average $v_2$ with respect to the participant plane from \ampt. This is presumably because the extracted $\vn{2}{2}$ should contain, besides effect from fluctuating participant plane about the reaction plane, additional flow fluctuations caused by geometry eccentricity fluctuations. Small contribution from $\deta$-independent nonflow may also play a role in $\tv_2$. 

It is commonly believed that the average two- and four-particle, $(\vt{2}{2}+\vt{2}{4})/2$, 
equals to the true average participant plane $\vn{2}{PP}$~\cite{v2method}. This does not seem to be true from our results in Fig.~\ref{fig:v2fluc}. It is also commonly believed that the four-particle $\vt{2}{4}$ should equal to the reaction plane $\vn{2}{RP}$~\cite{v2method}. This does not seem to be true either. 

It is interesting to examine the magnitude of flow fluctuation. However, it is not clear how one can reliably obtain flow fluctuation from the $\vt{}{2}$ and $\vt{}{4}$ measurements; see the previous discussion. Nevertheless, we compute $\sqrt{\frac{\vtsq{2}{2}-\vtsq{2}{4}}{\vtsq{2}{2}+\vtsq{2}{4}}}$ as extracted from our method to give a feeling about flow fluctuations in \ampt\ medium-central collisions. This is shown as the open squares in Fig.~\ref{fig:v2fluc} right panel. Note that the small $\deta$-dependent flow fluctuation effects (see Fig.~\ref{fig:d} left panel) have been already removed from the $\vt{}{2}$ and $\vt{}{4}$ measurements. One may also take the difference between the ``measured'' $\vt{2}{2}$ and the calculated $\vn{2}{PP}$ from model input as flow fluctuation, $\sqrt{\vtsq{2}{2}/\vnsq{2}{PP}-1}$. This is shown as the open and filled circles in Fig.~\ref{fig:v2fluc} right panel for medium-central and central collisions, respectively. Although experimentally inaccessible, it may give additional information regarding the size of flow fluctuation in \ampt. We note that the away-side $\deta$-independent nonflow is included in the decomposed $\vtsq{}{2}$ so the flow fluctuations are likely overestimated.

Figure~\ref{fig:v3fluc} left panel shows the third-harmonic two-particle cumulants $\vt{3}{2}$ in medium-central and central \ampt\ collisions. For $v_3$ the four-particle cumulant is consistent with zero with large statistical errors in both medium-central and central collisions. As seen in Fig.~\ref{fig:fact} and Fig.~\ref{fig:central}, the $\vt{n}{2}$ obtained from large $\etaref$ partner particles have large uncertainties (bed-headboard effect). We have thus restricted our projection to within $|\etaref|<3.6$ in obtaining the $\vt{n}{2}$ results in the left panel of Fig.~\ref{fig:v3fluc}. Figure~\ref{fig:v3fluc} left panel also shows the calculated true third-harmonic flow with respect to the participant plane by Eq.~(\ref{eq:vnPP}). The third-harmonic flow with respect to the reaction plane is found to be zero at both centralities.
 
Figure~\ref{fig:v3fluc} right panel shows an estimate of the third-harmonic flow fluctuation in \ampt\ by $\sqrt{\vtsq{3}{2}/\vnsq{3}{PP}-1}$. Similar to the second-harmonic one in Fig.~\ref{fig:v2fluc}, the estimated third-harmonic flow  is relatively independent of $\eta$, and seems smaller in central collisions than in medium-central collisions. The relative third-harmonic flow fluctuation is larger than the second-harmonic one.

It is interesting to examine the relative third-harmonic to second-harmonic flow. The left panel of Fig.~\ref{fig:v3v2} shows the ratio of $\vt{3}{2}/\vt{2}{2}$ in medium-central and central collisions. 
The $\vt{3}{2}/\vt{2}{2}$ ratio is larger at mid-rapidity and decreases significantly toward forward and backward rapidities. The ratio is larger in central collisions than in medium-central collisions. 

Figure~\ref{fig:v3v2} right panel shows the ratio of $\vt{}{2}$ in central collisions to that in medium-central collisions. It is interesting to note that the ratios are approximately independent of $\eta$; The centrality variation of flow does not depend on $\eta$. The $\vt{3}{2}$ central to medium-central ratio is close to unity--the $v_3$ magnitude has weak centrality dependence, consistent with its fluctuation nature.

\subsection{Discussion of nonflow from the models\label{sec:nf}}

In this section we discuss the extracted $\deta$-dependent nonflow $\delta(\deta)$ in the context of the models. This part of nonflow may consist primarily of effects from resonance decays and near-side, intra-jet correlations in the models. The models do not include quantum interference effects.

\begin{figure*}[hbt]
\begin{center}
\includegraphics[width=0.3\textwidth]{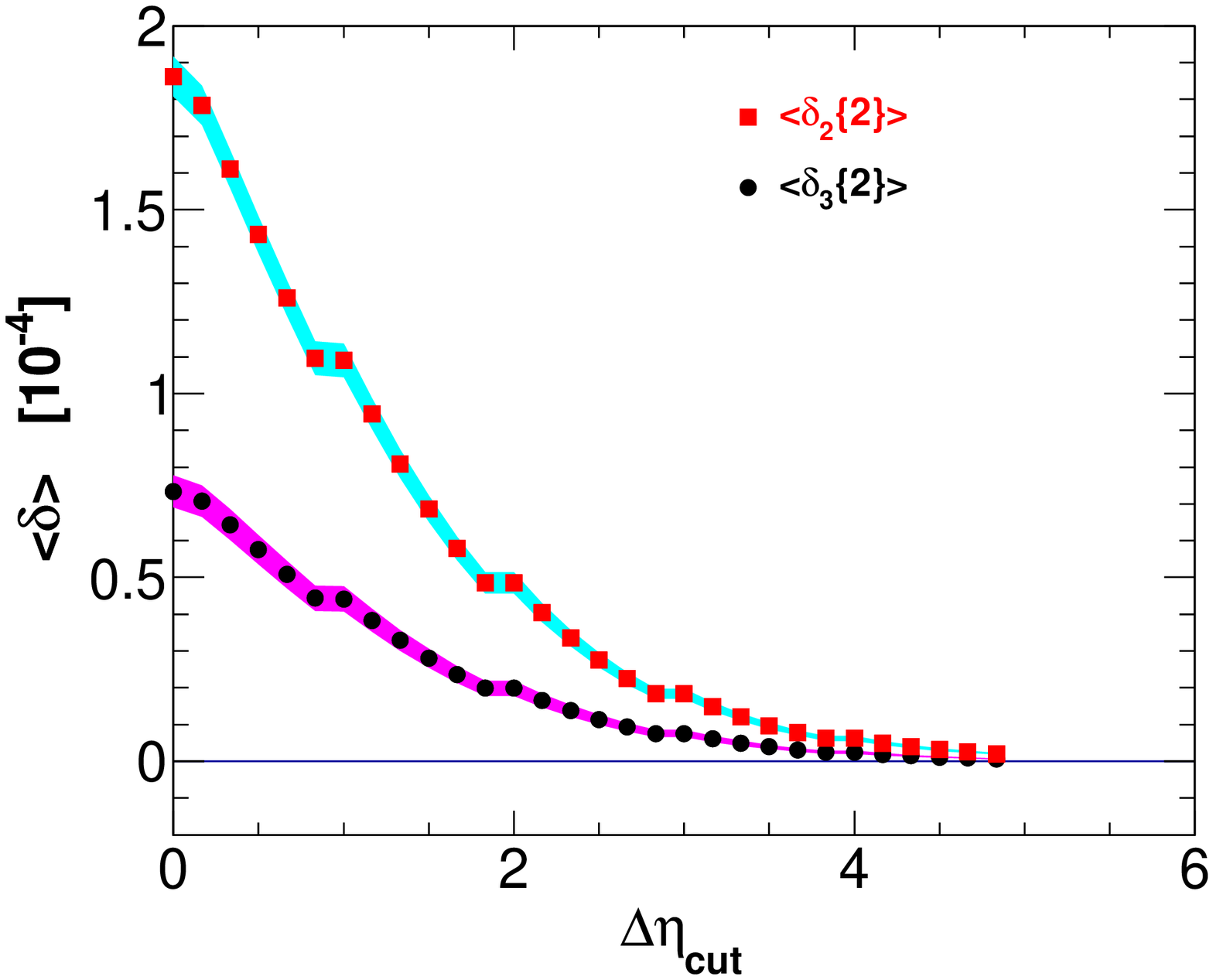}
\includegraphics[width=0.3\textwidth]{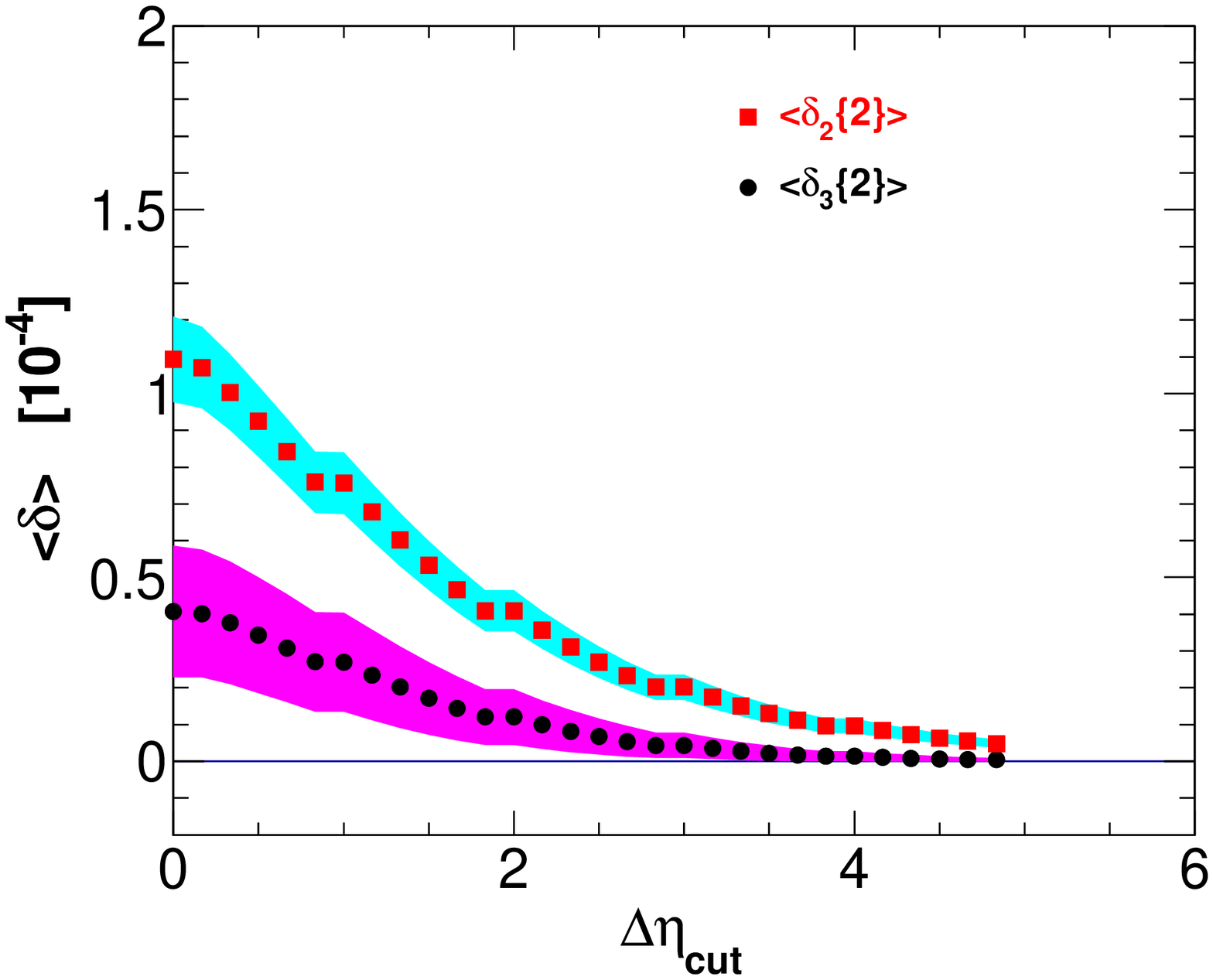}
\includegraphics[width=0.3\textwidth]{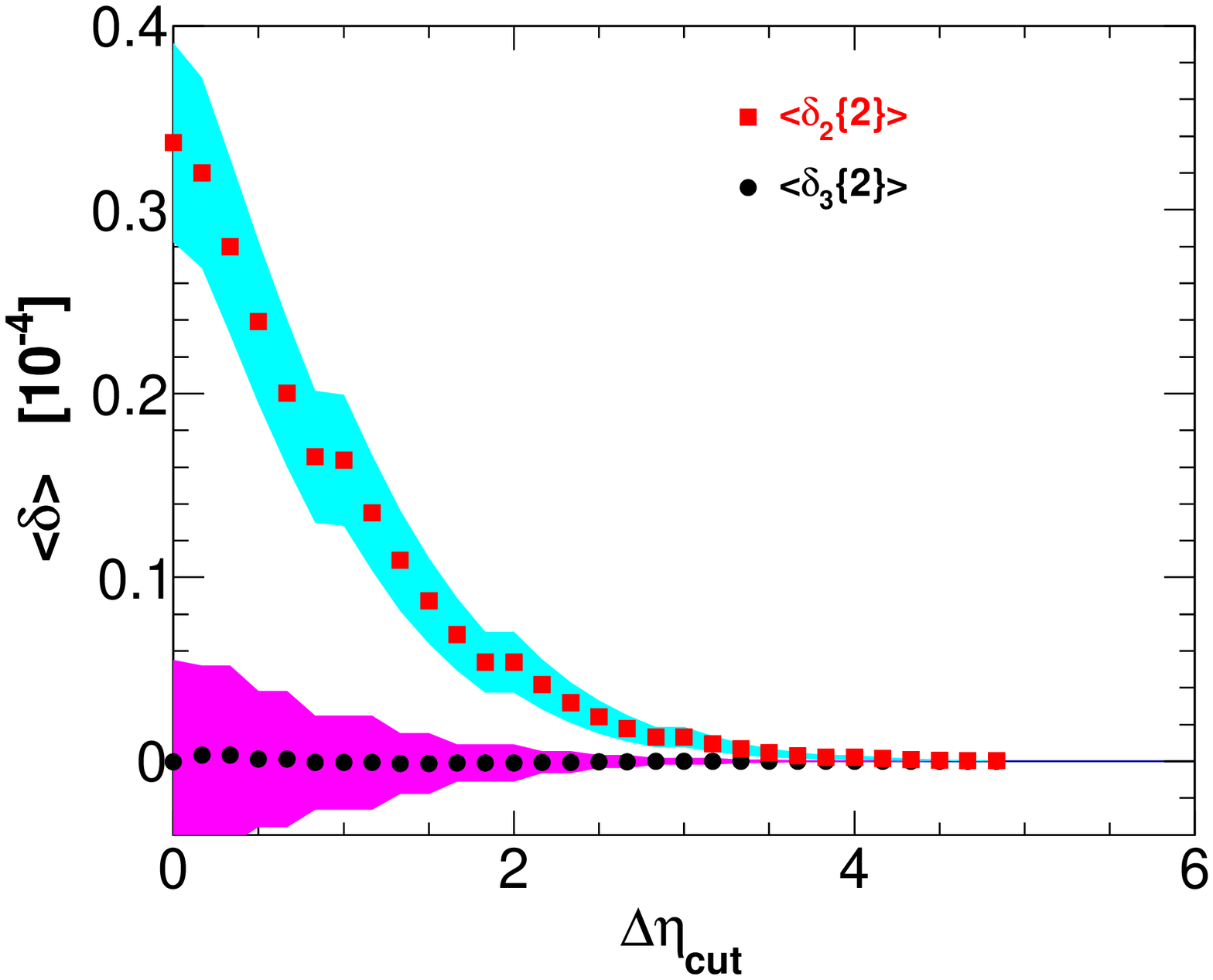}
\end{center}
\caption{(Color online) Average magnitude of $\deta$-dependent nonflow as a function of the minimum $\eta$-gap size (with a maximum $\eta$-gap of 5). Left: \ampt\ \bperi; Middle: \ampt\ \bcent; Right: \hijing\ \bperi. The bands correspond to systematic uncertainties from nonflow fit errors.}
\note{The xxx are due to binning issue when projecting onto diagonal axes.}
\label{fig:nf}
\end{figure*}

Experimentally one often applies a minimum $\eta$-gap to reduce nonflow. One difficult question has been how much nonflow still remains after a $\eta$-gap cut. We may now address this question, keeping in mind that there could be a $\deta$-independent nonflow contribution from back-to-back correlations that may not be cut out by a $\eta$-gap.
Figure~\ref{fig:nf} shows the average nonflow with $\eta$-gap between $\deta_{\rm cut}$ (the horizontal axis variable) and a maximum $\deta_{\rm max}=5$. We apply a maximum $\eta$ cut because experimentally one is limited by detector acceptance. With larger $\deta_{\rm max}$ cuts, the results are quantitatively similar. The left and middle panels show the average nonflow in 10-60\% and 0-10\% \ampt\ events, respectively. 
The second-harmonic nonflow is larger than the third-harmonic nonflow. 
The right panel shows the 10-60\% \hijing\ events; the third-harmonic $\delta_3$ is consistent with zero (see also Sec.~\ref{sec:hijing}). As expected the nonflow decreases with $\eta$-gap size because of a continuously decreasing contribution from the near-side small $\Delta\phi$ correlations. However, some finite nonflow is still present with relatively large $\eta$-gap. 

Nonflow is weaker in more central collisions. This is partially due to the large event multiplicity because nonflow is normalized by the total number of particle pairs in an event. The reduction in nonflow from medium central to central collisions is approximately a factor of 1.8. The increase in the number of participants $\Npart$ from medium central ($\Npart\sim130$) to central collisions ($\Npart\sim325$) is about a factor of 2.5. So nonflow from \ampt\ seems to scale as $1/\Npart^{2/3}$. This may indicate that nonflow in \ampt\ comes from hard-scattering sources, the magnitude of which scales with $\Npart^{4/3}$ and is then normalized by the total number of particle pairs.

It is interesting to note that \hijing\ $\delta_2$ is smaller than \ampt\ $\delta_2$ by a factor of 5. This at first glance seems counter-intuitive (if $\delta_2$ is indeed associated with nonflow as we argue). The jet-correlation effect between low-$\pt$ particles should be either similar or even stronger in \hijing\ than \ampt\ because final state scatterings in \ampt\ would likely destroy correlations. One would therefore naively expect, instead, a larger nonflow in \hijing\ than in \ampt. 
One possible reason for the observation is that 
the parton interactions in \ampt\ have moved jet-correlated particles to lower $\pt$, so that the correlation strength at low $\pt$ is relatively stronger. Experimentally it is found that the away-side jet-medium interactions soften the correlated particles~\cite{jetspec} while the near-side jet was biased toward surface emission by the high-$\pt$ trigger particles so little modification was observed. Since the particles used in our study are at low $\pt$, effects of jet-medium interactions should be also visible on the near side, which is probably what is revealed in the nonflow difference between \ampt\ and \hijing\ in Fig.~\ref{fig:nf}.

\begin{figure*}[hbt]
\begin{center}
\includegraphics[width=0.3\textwidth]{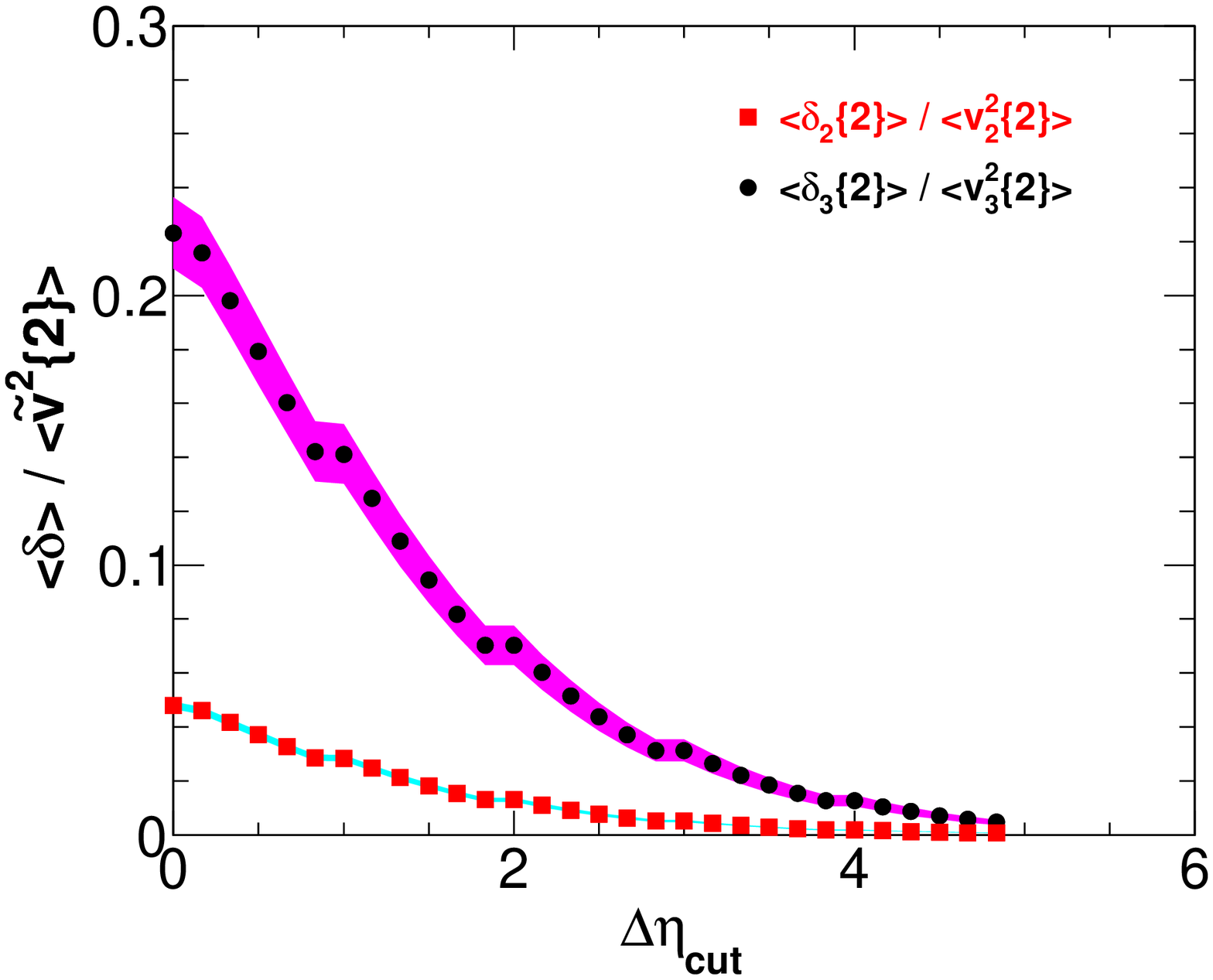}
\includegraphics[width=0.3\textwidth]{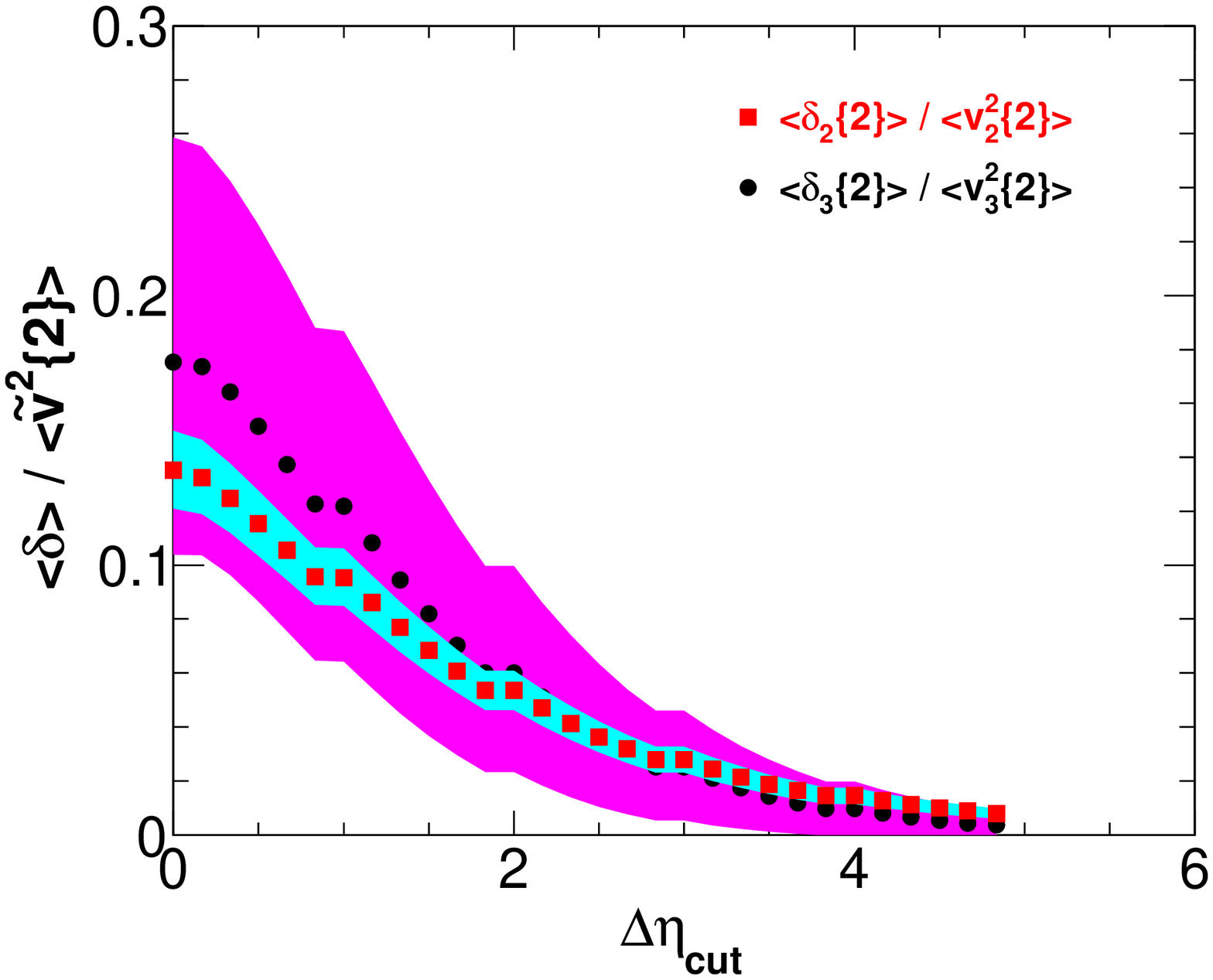}
\includegraphics[width=0.3\textwidth]{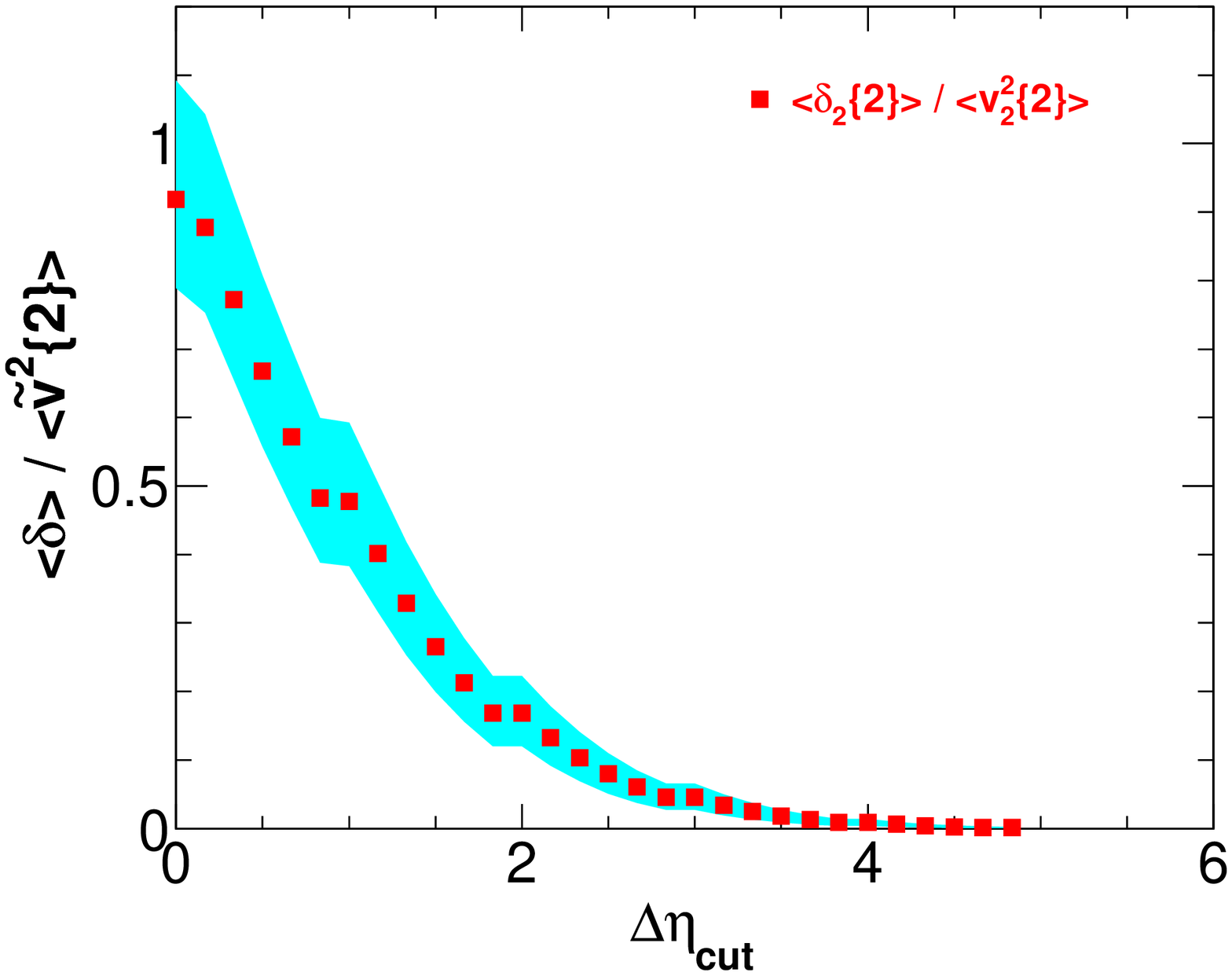}
\end{center}
\caption{(Color online) Relative strength of average nonflow to flow $\mn{\delta}/\mn{\vnsq{}{2}}$ as a function of the minimum $\eta$-gap size (with a maximum $\eta$-gap of 5). Left panel: \ampt\ \bperi. Middle panel: \ampt\ \bcent. Right panel: \hijing\ \bperi. Note that for \hijing\ the denomenator $\vnsq{}{2}$ is not flow but away-side nonflow average strength.}
\note{See Fig.~\ref{fig:nf} for explanation of the xxx.}
\label{fig:nf2v}
\end{figure*}

It is also interesting to examine the magnitude of nonflow relative to flow. Figure~\ref{fig:nf2v} left and middle panels show the ratios of average $\mn{\delta_n}$ to $\mn{\vtsq{n}{2}}$ in \ampt\ medium-central and central collisions as a function of the minimum $\eta$-gap, $\deta_{\rm cut}$. A maximum $\deta_{\rm max}=5$ is applied as before and the average $\mn{\vtsq{n}{2}}$ is calculated in the same way as the average $\mn{\delta_n}$. The relative third-harmonic nonflow to flow ratio $\mn{\delta}/\mn{\tv^2}$ is a factor of 4 larger than the second-harmonic ratio in medium-central collisions. This is mostly due to the much larger $\tv_2$ than $\tv_3$ in these collisions. In central collisions, $\mn{\delta}/\mn{\tv^2}$ becomes comparable between the second- and third-harmonics. 

The right panel of Fig.~\ref{fig:nf2v} shows the ratio of nonflow to ``flow'' from the \hijing\ model. As discussed in Sec.~\ref{sec:hijing}, \hijing\ does not have flow; the extracted flow reflects the $\deta$-independent away-side nonflow contributions. Thus the ratio reflects the relative near-side to away-side nonflow strengths. With a minimum $\eta$-gap of zero, the ratio of the second-harmonic nonflows is about unity-- this may be expected from momentum conservation of dijets resulting in similar near- and away-side nonflow correlations at low $\pt$. With increasing minimum $\eta$-gap cut, the ratio drops. This is also expected because the near-side nonflow is concentrated at narrow $\deta$ while the away-side nonflow is more evenly spreaded over a wide $\deta$ range.

\section{Conclusions\label{conclusion}}

In this article, we have proposed a data-driven method to separate particle correlations into two parts: the $\deta$-independent part and the $\deta$-dependent part. The former is likely dominated by global harmonic flows due to the common collision geometry, and the latter can be identified as nonflow due to particle correlations such as resonance decays and near-side intra-jet correlations. 

The method uses two- and four-particle cumulants between particles from two pseudo-rapidity ($\eta$) bins. It exploits the symmetry of average flow about mid-rapidity in a symmetric collision system. 
By taking the difference of four-particle cumulants between two pairs of $\eta$ bins, $(\etaa,\etab)$ and $(\etaa,-\etab)$, the average flow and the $\deta$-independent flow fluctuation effect cancel, and the $\deta$-dependent flow fluctuation can be deduced.
Similarly by taking difference of the two-particle cumulants between two pairs of $\eta$ bins, the difference in $\deta$-dependent nonflow correlation (plus $\deta$-dependent flow fluctuation effect) is obtained from which  $\deta$-dependent nonflow is deduced. 
By subtracting $\deta$-dependent nonflow correlation and $\deta$-dependent flow fluctuation, the $\deta$-independent correlation is obtained, which is dominated by flow and flow fluctuations. The $\deta$-independent nonflow contributions, such as those from back-to-back inter-jet correlations, should still remain. However, they may be negligible in our studied $\pt$ range.

We used \ampt\ to illustrate our method focusing on medium-central collisions. We find the obtained $\deta$-independent correlations satisfy factorization, commonly expected for flow. We have checked our method with central \ampt\ events, and find that the extracted flow and nonflow qualitatively agree with expectations of their centrality dependence. The obtained $\deta$-independent correlation is compared to the calculated ``real'' flow from \ampt\ and appears to be reasonable. Our \ampt\ model studies support the validity of our decomposition method.

We have also checked our method against the \hijing\ event generator. The extracted $\deta$-independent correlations are small and comparable to the $\deta$-dependent nonflow correlation. This suggests that the $\deta$-dependent nonflow and the $\deta$-independent correlation may be reflecting near- and away-side correlations of dijets, one of the main physics in \hijing. 
The \hijing\ results therefore render further support to our decomposition method.

Our method can be readily applied to real data analysis at LHC and RHIC. The large pseudo-rapidity acceptance of LHC experiments is crucial. The limited pseudo-rapidity coverage of the RHIC detectors may prevent our method from fully extracting the $\deta$-dependent nonflow correlation because the region of vanishing nonflow is likely outside the detector acceptance. Nevertheless, our method should yield valuable information on the pseudo-rapidity dependence of flow and nonflow in the limited pseudo-rapidity range.

It is important to point out that our method essentially separates $\deta$-dependent and $\deta$-independent correlations, not particularly flow and nonflow. The separated correlations may be identified as flow and nonflow, if the $\deta$-dependent flow fluctuation effect and the $\deta$-independent nonflow effect are small. These appear to be the case at low $\pt$ for the models we studied. With increasing $\pt$ the $\deta$-independent nonflow effect should increase. Therefore, it would be also interesting to study our method with high-$\pt$ particles in relativistic heavy-ion collisions. We leave such studies to future work.

\section*{Acknowledgment}

We thank Dr.~Guoliang Ma for providing us the code to calculate the participant plane angle using the initial gluon and quark transverse coordinates. 
We thank Dr.~Jiangyong Jia, Dr.~Wei Li, Dr.~Denes Molnar, and Dr.~Aihong Tang for fruitful discussions. 
We thank Dr.~Andrew Hirsch for a careful reading of the manuscript.
This work is supported by U.S.~Department of Energy under Grant No.~DE-FG02-88ER40412.

\end{document}